\pdfoutput=1

\documentclass[12pt,a4paper]{article}

\usepackage{ifthen} 
\newboolean{pdflatex}
\setboolean{pdflatex}{true} 

\newboolean{articletitles}
\setboolean{articletitles}{true} 

\newboolean{uprightparticles}
\setboolean{uprightparticles}{true} 

\newboolean{inbibliography}
\setboolean{inbibliography}{false} 

\usepackage{microtype}
\usepackage{lineno}  
\usepackage{xspace} 
\usepackage{caption} 

\usepackage{graphicx}  
\usepackage{color}
\usepackage{colortbl}
\graphicspath{{./figs/}} 

\usepackage{amsmath} 
\usepackage{amssymb}
\usepackage{amsfonts}
\usepackage{upgreek} 

\newcommand*\patchAmsMathEnvironmentForLineno[1]{%
\expandafter\let\csname old#1\expandafter\endcsname\csname #1\endcsname
\expandafter\let\csname oldend#1\expandafter\endcsname\csname
end#1\endcsname
 \renewenvironment{#1}%
   {\linenomath\csname old#1\endcsname}%
   {\csname oldend#1\endcsname\endlinenomath}%
}
\newcommand*\patchBothAmsMathEnvironmentsForLineno[1]{%
  \patchAmsMathEnvironmentForLineno{#1}%
  \patchAmsMathEnvironmentForLineno{#1*}%
}
\AtBeginDocument{%
\patchBothAmsMathEnvironmentsForLineno{equation}%
\patchBothAmsMathEnvironmentsForLineno{align}%
\patchBothAmsMathEnvironmentsForLineno{flalign}%
\patchBothAmsMathEnvironmentsForLineno{alignat}%
\patchBothAmsMathEnvironmentsForLineno{gather}%
\patchBothAmsMathEnvironmentsForLineno{multline}%
\patchBothAmsMathEnvironmentsForLineno{eqnarray}%
}

\usepackage{hyperref}    
\usepackage[all]{hypcap} 

\def\lhcb {\mbox{LHCb}\xspace}

\def\tevatron {Tevatron\xspace}

\def\MagUp {\mbox{\em Mag\kern -0.05em Up}\xspace}

\ifthenelse{\boolean{uprightparticles}}%
{

 \def\Peta        {\ensuremath{\upeta}\xspace}

 \def\Pmu         {\ensuremath{\upmu}\xspace}

 \def\Pchi        {\ensuremath{\upchi}\xspace}                 
 \def\Ppsi        {\ensuremath{\uppsi}\xspace}

 \def\PDelta      {\ensuremath{\Delta}\xspace}                 
 \def\PXi      {\ensuremath{\Xi}\xspace}                 
 \def\PLambda      {\ensuremath{\Lambda}\xspace}                 
 \def\PSigma      {\ensuremath{\Sigma}\xspace}                 
 \def\POmega      {\ensuremath{\Omega}\xspace}                 
 \def\PUpsilon      {\ensuremath{\Upsilon}\xspace}                 
 
 \def\PB      {\ensuremath{\mathrm{B}}\xspace}                 
                  
 \def\PD      {\ensuremath{\mathrm{D}}\xspace}

 \def\PJ      {\ensuremath{\mathrm{J}}\xspace}                 
 \def\PK      {\ensuremath{\mathrm{K}}\xspace}

 \def\Pb      {\ensuremath{\mathrm{b}}\xspace}                 
 \def\Pc      {\ensuremath{\mathrm{c}}\xspace}

 \def\Pi      {\ensuremath{\mathrm{i}}\xspace}

 \def\Pp      {\ensuremath{\mathrm{p}}\xspace}

}
{

 \def\Peta        {\ensuremath{\eta}\xspace}

 \def\Pmu         {\ensuremath{\mu}\xspace}

 \def\Pchi        {\ensuremath{\chi}\xspace}                 
 \def\Ppsi        {\ensuremath{\psi}\xspace}                 
                  
 \mathchardef\PDelta="7101
 \mathchardef\PXi="7104
 \mathchardef\PLambda="7103
 \mathchardef\PSigma="7106
 \mathchardef\POmega="710A
 \mathchardef\PUpsilon="7107
                  
 \def\PB      {\ensuremath{B}\xspace}                 
                  
 \def\PD      {\ensuremath{D}\xspace}

 \def\PJ      {\ensuremath{J}\xspace}                 
 \def\PK      {\ensuremath{K}\xspace}

 \def\Pb      {\ensuremath{b}\xspace}                 
 \def\Pc      {\ensuremath{c}\xspace}

 \def\Pi      {\ensuremath{i}\xspace}

 \def\Pp      {\ensuremath{p}\xspace}

}

\makeatletter
\ifcase \@ptsize \relax
  \newcommand{\miniscule}{\@setfontsize\miniscule{4}{5}}
\or
  \newcommand{\miniscule}{\@setfontsize\miniscule{5}{6}}
\or
  \newcommand{\miniscule}{\@setfontsize\miniscule{5}{6}}
\fi
\makeatother

\DeclareRobustCommand{\optbar}[1]{\shortstack{{\miniscule (\rule[.5ex]{1.25em}{.18mm})}
  \\ [-.7ex] $#1$}}

\def\mumu       {{\ensuremath{\Pmu^+\Pmu^-}}\xspace}

\def\cquark    {{\ensuremath{\Pc}}\xspace}

\def\bquark    {{\ensuremath{\Pb}}\xspace}
\def\bquarkbar {{\ensuremath{\overline \bquark}}\xspace}

  \def\Kbar    {{\kern 0.2em\overline{\kern -0.2em \PK}{}}\xspace}

\def\KorKbar    {\kern 0.18em\optbar{\kern -0.18em K}{}\xspace}

  \def\Dbar    {{\kern 0.2em\overline{\kern -0.2em \PD}{}}\xspace}

\def\DorDbar    {\kern 0.18em\optbar{\kern -0.18em D}{}\xspace}

\def\Bbar    {{\ensuremath{\kern 0.18em\overline{\kern -0.18em \PB}{}}}\xspace}

\def\BorBbar    {\kern 0.18em\optbar{\kern -0.18em B}{}\xspace}

\def\jpsi     {{\ensuremath{{\PJ\mskip -3mu/\mskip -2mu\Ppsi\mskip 2mu}}}\xspace}

\def\proton      {{\ensuremath{\Pp}}\xspace}

\def\Lbar        {{\ensuremath{\kern 0.1em\overline{\kern -0.1em\PLambda}}}\xspace}
\def\LorLbar    {\kern 0.18em\optbar{\kern -0.18em \PLambda}{}\xspace}

\def\BF         {{\ensuremath{\cal B}}\xspace}

\def\BR         {\BF}

\def\to                 {\ensuremath{\rightarrow}\xspace}

\def\AT#1     {\ensuremath{A_{\mathrm{T}}^{#1}}\xspace}           

\def\C#1      {\ensuremath{\mathcal{C}_{#1}}\xspace}                       
\def\Cp#1     {\ensuremath{\mathcal{C}_{#1}^{'}}\xspace}                    
\def\Ceff#1   {\ensuremath{\mathcal{C}_{#1}^{\mathrm{(eff)}}}\xspace}        
\def\Cpeff#1  {\ensuremath{\mathcal{C}_{#1}^{'\mathrm{(eff)}}}\xspace}       
\def\Ope#1    {\ensuremath{\mathcal{O}_{#1}}\xspace}                       
\def\Opep#1   {\ensuremath{\mathcal{O}_{#1}^{'}}\xspace}                    

\newcommand{\tev}{\ifthenelse{\boolean{inbibliography}}{\ensuremath{~T\kern -0.05em eV}\xspace}{\ensuremath{\mathrm{\,Te\kern -0.1em V}}}\xspace}
\newcommand{\gev}{\ensuremath{\mathrm{\,Ge\kern -0.1em V}}\xspace}
\newcommand{\mev}{\ensuremath{\mathrm{\,Me\kern -0.1em V}}\xspace}
\newcommand{\kev}{\ensuremath{\mathrm{\,ke\kern -0.1em V}}\xspace}
\newcommand{\ev}{\ensuremath{\mathrm{\,e\kern -0.1em V}}\xspace}
\newcommand{\gevc}{\ensuremath{{\mathrm{\,Ge\kern -0.1em V\!/}c}}\xspace}
\newcommand{\mevc}{\ensuremath{{\mathrm{\,Me\kern -0.1em V\!/}c}}\xspace}
\newcommand{\gevcc}{\ensuremath{{\mathrm{\,Ge\kern -0.1em V\!/}c^2}}\xspace}
\newcommand{\gevgevcccc}{\ensuremath{{\mathrm{\,Ge\kern -0.1em V^2\!/}c^4}}\xspace}
\newcommand{\mevcc}{\ensuremath{{\mathrm{\,Me\kern -0.1em V\!/}c^2}}\xspace}

\def\mum  {\ensuremath{{\,\upmu\rm m}}\xspace}

\def\nb {\ensuremath{\rm \,nb}\xspace}

\def\pb {\ensuremath{\rm \,pb}\xspace}
\def\invpb {\ensuremath{\mbox{\,pb}^{-1}}\xspace}

\def\invfb   {\ensuremath{\mbox{\,fb}^{-1}}\xspace}

\def\deriv {\ensuremath{\mathrm{d}}}

\def\gsim{{~\raise.15em\hbox{$>$}\kern-.85em
          \lower.35em\hbox{$\sim$}~}\xspace}
\def\lsim{{~\raise.15em\hbox{$<$}\kern-.85em
          \lower.35em\hbox{$\sim$}~}\xspace}

\def\sPlot{\mbox{\em sPlot}\xspace}

\def\sqs   {\ensuremath{\protect\sqrt{s}}\xspace}

\def\ptot       {\ensuremath{p}\xspace}
\def\pt         {\ensuremath{p_{\mathrm T}}\xspace}

\newcommand{\lum} {\ensuremath{\mathcal{L}}\xspace}

\def\evtgen     {\mbox{\textsc{EvtGen}}\xspace}

\def\geant      {\mbox{\textsc{Geant4}}\xspace}

\def\photos     {\mbox{\textsc{Photos}}\xspace}

\def\pythia     {\mbox{\textsc{Pythia}}\xspace}

\def\tell1  {TELL1\xspace}
\def\ukl1   {UKL1\xspace}

\newcommand{\ups}    {\PUpsilon}

\newcommand{\YiS}    {\ensuremath{\ups\mathrm{(iS)}}\xspace}
\newcommand{\YoneS}  {\ensuremath{\ups\mathrm{(1S)}}\xspace}
\newcommand{\YtwoS}  {\ensuremath{\ups\mathrm{(2S)}}\xspace}
\newcommand{\YthreeS}{\ensuremath{\ups\mathrm{(3S)}}\xspace}

\usepackage{cite} 
\usepackage{mciteplus}

\usepackage{longtable} 
\usepackage{mathtools}
\usepackage{rotating}  
\usepackage{graphpap}  
\usepackage{multirow}  
\usepackage{mathrsfs}  
\usepackage{pifont}    
\usepackage{cancel}    
\usepackage{afterpage} 
\usepackage{pdflscape} 
\begin{document}

\renewcommand{\thefootnote}{\fnsymbol{footnote}}
\setcounter{footnote}{1}


\begin{titlepage}
\pagenumbering{roman}

\vspace*{-1.5cm}
\centerline{\large EUROPEAN ORGANIZATION FOR NUCLEAR RESEARCH (CERN)}
\vspace*{1.5cm}
\noindent
\begin{tabular*}{\linewidth}{lc@{\extracolsep{\fill}}r@{\extracolsep{0pt}}}
\ifthenelse{\boolean{pdflatex}}
{\vspace*{-2.7cm}\mbox{\!\!\!\includegraphics[width=.14\textwidth]{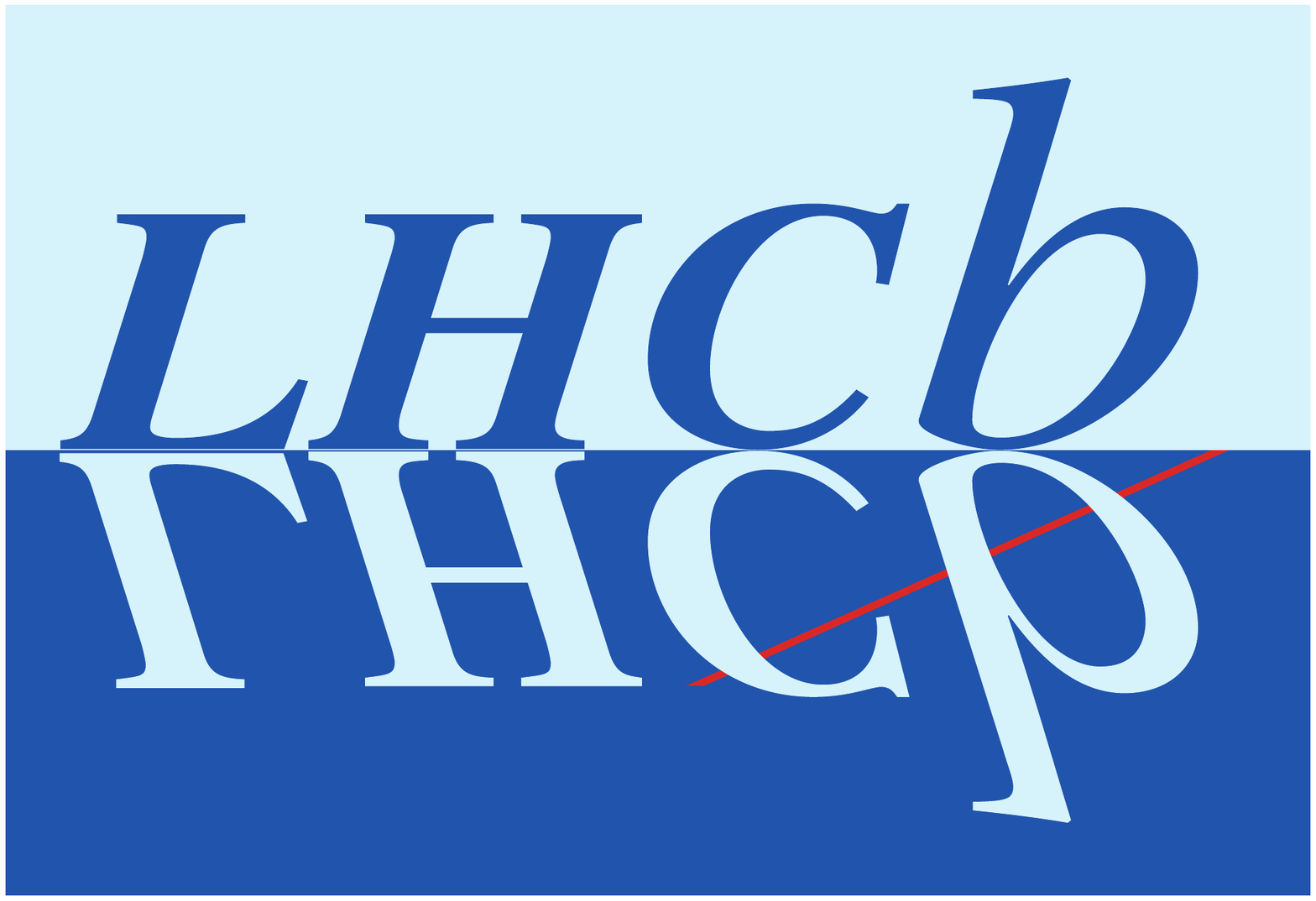}} & &}%
{\vspace*{-1.2cm}\mbox{\!\!\!\includegraphics[width=.12\textwidth]{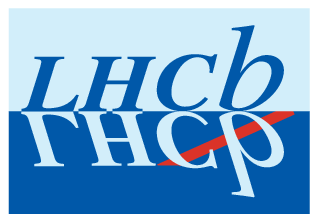}} & &}%
\\
 & & CERN-PH-EP-2015-232 \\  
 & & LHCb-PAPER-2015-045 \\  
 & & September 8, 2015\\ 
 & & \\
\end{tabular*}

\vspace*{4.0cm}

{\bf\boldmath\huge
  \begin{center}
    Forward production 
    of~\ups~mesons in $\proton\proton$~collisions 
    at~$\sqs=7$~and~$8\tev$
  \end{center}
}

\vspace*{2.0cm}

\begin{center}
The LHCb collaboration\footnote{Authors are listed at the end of this paper.}
\end{center}

\vspace{\fill}

\begin{abstract}
  \noindent
  The production of \ups~mesons in $\proton\proton$~collisions 
  at \mbox{$\sqs=7$}~and~\mbox{$8\tev$}
  is studied with the~LHCb detector
  using data samples corresponding 
  to an~integrated 
  \mbox{luminosity} of 1\invfb and 2\invfb respectively.
  The~production cross\nobreakdash-sections and ratios of
  cross\nobreakdash-sections are measured as functions 
  of the~meson transverse momentum~\pt
  and rapidity~$y$,
  for~\mbox{$\pt<30\gevc$}~and~\mbox{$2.0<y<4.5$}.
\end{abstract}

\vspace*{2.0cm}

\begin{center}
  Published in JHEP 1511(2015) 103
\end{center}

\vspace{\fill}

{\footnotesize 
\centerline{\copyright~CERN on behalf of the \lhcb collaboration, licence \href{http://creativecommons.org/licenses/by/4.0/}{CC-BY-4.0}.}}
\vspace*{2mm}

\end{titlepage}


\newpage
\setcounter{page}{2}
\mbox{~}

\cleardoublepage


\renewcommand{\thefootnote}{\arabic{footnote}}
\setcounter{footnote}{0}



\pagestyle{plain} 
\setcounter{page}{1}
\pagenumbering{arabic}


%

\section{Introduction}
\label{sec:Introduction}

In high energy hadron collisions, 
the~production of heavy quarkonium systems
such as the~$\bquark\bquarkbar$~states\,(\YoneS, \YtwoS and \YthreeS, 
represented generically as \ups~in the following)
probes the~dynamics of the~colliding partons
and provides insight into
the~non-perturbative regime of quantum chromodynamics\,(QCD).
Despite many models that have been proposed, a~complete
description of heavy quarkonium production is still not available.

The effective theory of 
non-relativistic~QCD\,(NRQCD)~\cite{CaswellLepage1986PL,PhysRevD.51.1125}
provides the foundation for much of the~current theoretical work. 
According to NRQCD, the production of heavy quarkonium 
factorises into two steps:  a~heavy quark-antiquark pair is first created 
at short distances, and subsequently evolves
non-perturbatively into a~quarkonium state.
The~NRQCD calculations include
the~colour-singlet\,(CS) 
and colour-octet\,(CO) matrix elements
for the~pertubative stage.
The~CS model~\cite{Kartvelishvili:1978id,Baier:1981uk}, 
which provides a~leading\nobreakdash-order description of quarkonium production, 
underestimates the~cross-section 
for single \jpsi~production 
at the~\tevatron~\cite{Abe:1992ww} at high~\pt,
where \pt~is the~component of the~meson momentum transverse to the~beam. 
To~resolve this
discrepancy, the~CO mechanism was introduced~\cite{Braaten:1994vv}. 
The~corresponding matrix elements were determined from the~high-\pt data, 
as the~CO cross-section decreases  
more slowly with \pt than that predicted by the~CS model. 
More~recent higher-order 
calculations~\cite{Campbell:2007ws,Gong:2008sn,
artoisenet:2008,lansberg:2009,Han:2014kxa}
show better agreement between CS~predictions and 
the~experimental data~\cite{Brambilla:2010cs},
reducing the~need for large CO contributions.
The~production of \ups~mesons in 
proton-proton\,($\proton\proton$)~collisions 
can occur either directly in parton scattering or via feed down from 
the~decay of heavier bottomonium states, such as  
$\Pchi_{\bquark}$~\cite{Aad:2011ih,Abazov:2012gh,LHCB-PAPER-2012-015,
LHCb-PAPER-2014-031,Mazurov:1693482,LHCb-PAPER-2014-040},  
or higher-mass $\ups$ states,
which complicates the theoretical description 
of bottomonium production~\cite{Likhoded:2012hw,Wang:2012is}.

The production of the~\ups~mesons has been studied using $\proton\proton$~collision 
data taken at~\mbox{$\sqs=2.76$}, 7 and~\mbox{$8\tev$} 
by the~LHCb~\cite{LHCb-PAPER-2013-066,LHCb-PAPER-2011-036,LHCb-PAPER-2013-016},
ALICE~\cite{Abelev:2014qha}, ATLAS~\cite{AtlasUpsilon} 
and CMS~\mbox{\cite{CmsUpsilon,CMSUpsilon2}} experiments in 
different kinematic regions.
The~existing LHCb measurements of these quantities were performed 
at \mbox{$\sqs=7\tev$} with a~data sample collected 
in~2010 corresponding to an~integrated luminosity of $25\invpb$, 
and at \mbox{$\sqs=8\tev$} for early 2012~data using~$50\invpb$. 
Both~measurements were differential in $\pt$ and $y$ of 
the~\ups~mesons in the~ranges \mbox{$2.0<y<4.5$}
and \mbox{$\pt<15\gevc$}.   
Based on these measurements, an~increase of 
the~production cross-section 
in excess of 30\% between  
\mbox{$\sqs=7$}~and \mbox{$8\tev$} was observed, 
which is larger than the increase observed for other quarkonium states 
such as the~$\jpsi$~\cite{LHCb-PAPER-2011-003,LHCb-PAPER-2013-016}
and larger than the~expectations from NRQCD~\cite{Han:2014kxa}.

In this paper we report on the~measurement of 
the~inclusive production cross-sections of the $\ups$ states 
at \mbox{$\sqs=7$}~and~$8\tev$ and 
the~ratios of these cross-sections.
The~\ups~cross-section measurement 
is performed using a~data sample
corresponding to the~complete LHCb Run~1
data set with integrated luminosities 
of~$1\invfb$ and~$2\invfb$, accumulated 
at \mbox{$\sqs=7$}~and~$8\tev$, respectively.
These~samples are independent from those used in the~previous
analyses~\cite{LHCb-PAPER-2011-036,LHCb-PAPER-2013-016}.
The~increased size of the~data sample 
results in a~better statistical
precision and allows the~measurements to be 
extended up~to $\pt$~values of~$30\gevc$.

\bigskip
\section{Detector and simulation}
\label{sec:Detector}

The~\lhcb detector~\cite{Alves:2008zz,LHCb-DP-2014-002} is a~single-arm forward
spectrometer covering the~\mbox{pseudorapidity} range \mbox{$2<\Peta<5$},
designed for the study of particles containing \bquark or \cquark~quarks. 
The~detector includes a~high-precision tracking system
consisting of a silicon-strip vertex detector surrounding 
the~$\proton\proton$~interaction region, 
a~large-area silicon-strip detector located upstream 
of a~dipole magnet with a bending power of about $4{\rm\,Tm}$, 
and three stations of silicon-strip detectors and straw
drift tubes placed downstream of the~magnet.
The tracking system provides a~measurement of momentum, 
\ptot, of charged particles with a~relative uncertainty 
that varies from 0.5\% at low momentum to 1.0\% at 200\gevc.
The~minimum distance of a~track to a~primary vertex, 
the~impact parameter, is measured with a~resolution 
of \mbox{$(15+29/\pt)\mum$},
where \pt~is in\,\gevc.
Different types of charged hadrons are distinguished using information
from two ring-imaging Cherenkov detectors. 
Photons, electrons and hadrons are identified 
by a~calorimeter system consisting of
scintillating-pad and preshower detectors, an electromagnetic
calorimeter and a hadronic calorimeter. 
Muons~are identified by 
a~system composed of alternating layers of iron and multiwire
proportional chambers~\cite{LHCb-DP-2012-002}.
The~online event selection is performed by a~trigger~\cite{LHCb-DP-2012-004}, 
which consists of a~hardware stage, based on information from 
the~calorimeter and muon
systems, followed by a~software stage, which applies a~full event
reconstruction.
At~the~hardware stage, events for this analysis 
are selected by requiring
dimuon candidates with a~product of their \pt~values exceeding
\mbox{$1.7\,(2.6)\,(\!\gevc)^2$} for data 
collected at~\mbox{$\sqs=7\,(8)\tev$}. 
In~the~subsequent software trigger, two 
well-reconstructed tracks are required to have hits 
in the~muon system,
\mbox{$\pt>500\mevc$}, 
\mbox{$\ptot>6\gevc$} and to form a~common vertex.  
Only~events with a~dimuon candidate with 
a~mass~\mbox{$m_{\mumu}>4.7\gevcc$} 
are retained for further analysis.
In~the~offline selection, trigger decisions 
are associated with reconstructed particles.
Selection requirements can therefore be made 
on the~trigger selection itself
and on whether the~decision was due to the~signal candidate, 
the~other particles produced in the~$\proton\proton$~collision, 
or a~combination of~both.

In the~simulation,
$\proton\proton$~collisions are generated using
\mbox{$\pythia\,6$}~\cite{Sjostrand:2006za} 
with a~specific \lhcb~configuration~\cite{LHCb-PROC-2010-056}. 
Decays of hadronic particles are described by \evtgen~\cite{Lange:2001uf}, 
in~which final-state radiation is generated using \photos~\cite{Golonka:2005pn}. 
The~interaction of the~generated particles with the detector, 
and its response, are implemented using the \geant
toolkit~\cite{Allison:2006ve, *Agostinelli:2002hh} as described in
Ref.~\cite{LHCb-PROC-2011-006}.

\section{Selection and cross-section determination} 
\label{sec:cross}
The~event selection is based on the~criteria described in
the~previous LHCb 
\ups~analyses~\mbox{\cite{LHCb-PAPER-2011-036,LHCb-PAPER-2013-016,LHCb-PAPER-2013-066}} but slightly modified to improve the~signal-to-background ratio. 
It~includes selection criteria that ensure good quality
track reconstruction~\cite{LHCb-DP-2013-002}, 
muon identification~\cite{LHCb-DP-2013-001}, 
and the~requirement of a~good fit quality for the~dimuon vertex, 
where the~associated primary vertex position is used as a~constraint 
in the~fit~\cite{Hulsbergen:2005pu}.
In~addition, the~muon candidates are required to 
have
\mbox{$1<\pt<25\gevc$}, 
\mbox{$10<\ptot<400\gevc$}
and pseudorapidity within the~region~\mbox{$2.0<\Peta<4.5$}.

The~differential cross-section  for the~production 
of an~$\ups$~meson 
decaying into a~muon pair is 
\begin{equation}
\BR_{\ups} \times 
\dfrac { \deriv^2}
       { \deriv \pt\, \deriv y}
\upsigma ( \proton\proton \to \ups \mathrm{X})
\equiv 
\dfrac{ 1}
      { \Delta \pt\Delta y }
\upsigma^{\ups\to\mumu}_{\mathrm{bin}}
= 
\dfrac{ 1}
      { \Delta \pt\Delta y }
\dfrac{ N_{\ups\to\mumu}}
      { \lum},                \label{eq:sigma}
\end{equation}
where $\BR_{\ups}$ is the~branching fraction 
of the~\mbox{$\ups\to\mumu$}~decay, 
$\Delta y$~and $\Delta \pt$~are 
the~rapidity and \pt~bin sizes,
$\upsigma^{\ups\to\mumu}_{\mathrm{bin}}$~is a production cross-section 
for $\ups\to\mumu$~events in the~given $(\pt,y)$~bin, 
$N_{\ups \to \mumu}$~is  the~efficiency-corrected number 
of $\ups \to \mumu$~decays and \lum~is the~integrated luminosity.
Given the~sizeable uncertainty on the~dimuon branching fractions 
of the~$\ups$ mesons~\cite{PDG2014}, the~measurement of 
the~production cross-section multiplied 
by the~dimuon branching fraction is presented,
as in previous LHCb 
measurements~\mbox{\cite{LHCb-PAPER-2011-036,LHCb-PAPER-2013-016,LHCb-PAPER-2013-066}}. 

A~large part of the~theoretical and experimental uncertainties cancel 
in the~ratios of production cross-sections of various \ups~mesons, 
defined for a~given \mbox{$(\pt,y)$}~bin as  
\begin{equation}
\mathscr{R}_{\mathrm{i,j}}\equiv 
\dfrac 
{\upsigma^{\YiS\to\mumu}_{\mathrm{bin}}}
{\upsigma^{\ups\mathrm{(jS)}\to\mumu}_{\mathrm{bin}}} = 
\dfrac{ N_{\YiS\to\mumu}}
      { N_{\ups\mathrm{(jS)}\to\mumu}}.  \label{eq:ryns}
\end{equation}
The~evolution of the~production cross-sections as a~function of 
$\proton\proton$~collision energy is studied using 
the~ratio 
\begin{equation}
\mathscr{R}_{8/7} \equiv
\dfrac 
{ \left.\upsigma^{\ups\to\mumu}_{\mathrm{bin}}\right|_{\sqs=8\tev}}
{ \left.\upsigma^{\ups\to\mumu}_{\mathrm{bin}}\right|_{\sqs=7\tev}}. 
\end{equation}

The~signal yields 
$N_{\ups\to\mumu}$~in each $(\pt,y)$~bin are 
determined from an~unbinned extended maximum likelihood fit to 
the~dimuon mass spectrum of 
the~selected candidates within the~range~\mbox{$8.5<m_{\mumu}<12.5~\gevcc$}. 
The correction for efficiency is embedded in the~fit procedure.
Each dimuon~candidate is given a~weight calculated as
$1/\varepsilon^{\mathrm{tot}}$,
where $\varepsilon^{\mathrm{tot}}$~is the~total efficiency,
which is determined for each $\ups\to\mumu$~candidate as
\begin{equation}
  \varepsilon^{\mathrm{tot}} = 
  \varepsilon^{\mathrm{rec\&sel}} \times 
  \varepsilon^{\mathrm{trg}}      \times 
  \varepsilon^{\Pmu\mathrm{ID}},    \label{eq:effic}
\end{equation}
where 
$\varepsilon^{\mathrm{rec\&sel}}$ 
is the reconstruction and selection efficiency,
$\varepsilon^{\mathrm{trg}}$~is the~trigger efficiency
and $\varepsilon^{\Pmu\mathrm{ID}}$~is the~efficiency of the muon identification
criteria.
The~efficiencies 
$\varepsilon^{\mathrm{rec\&sel}}$~and $\varepsilon^{\mathrm{trg}}$~are 
determined using simulation, and corrected using data-driven
techniques to account for small differences in 
the~muon reconstruction efficiency between data and 
simulation~\mbox{\cite{LHCb-DP-2013-002,LHCb-DP-2013-001}}.
The~efficiency of the muon identification 
criteria $\varepsilon^{\Pmu\mathrm{ID}}$ 
is measured directly from data using 
a~large sample of low-background~\mbox{$\jpsi\to\mumu$}~events.
All~efficiencies are evaluated as functions of 
the~muon and dimuon kinematics.
The~mean total 
efficiency~$\left\langle\varepsilon^{\mathrm{tot}}\right\rangle$
reaches a~maximum of about~$45\%$ 
for the~region
\mbox{$15<\pt<20\gevc$}, 
\mbox{$3.0<y<3.5$}, 
and drops down to~$10\%$ 
at high~\pt and large~$y$,
with the~average efficiency being about~$30\%$.

In each $(\pt,y)$~bin, 
the~dimuon mass distribution 
is described by the~sum of three Crystal~Ball 
functions~\cite{Skwarnicki:1986xj}, one for 
each of the~\YoneS, \YtwoS~and~\YthreeS~signals, 
and the~product of an~exponential function with 
a~second-order polynomial for the~combinatorial background. 
The~mean value and the~resolution 
of the~Crystal~Ball function 
describing the~mass distribution of the~\YoneS~meson
are free fit parameters.
For the~\YtwoS~and \YthreeS~mesons
the~mass differences 
\mbox{$m(\YtwoS)-m(\YoneS)$} and 
\mbox{$m(\YthreeS)-m(\YoneS)$}
are fixed to the~known values~\cite{PDG2014}, 
while 
the~resolutions 
are fixed to the~value of 
the~resolution of the~\YoneS~signal, scaled by
the~ratio of the~masses of the~\YtwoS and \YthreeS~to the~\YoneS~meson.
The~tail parameters of the~Crystal~Ball function 
describing the~radiative tail are fixed 
from studies of simulated samples.

\begin{figure}[t]
  \setlength{\unitlength}{1mm}
  \centering
  \begin{picture}(150,60)
    \put(  0,  0){ 
      \includegraphics*[width=75mm,height=60mm,%
      ]{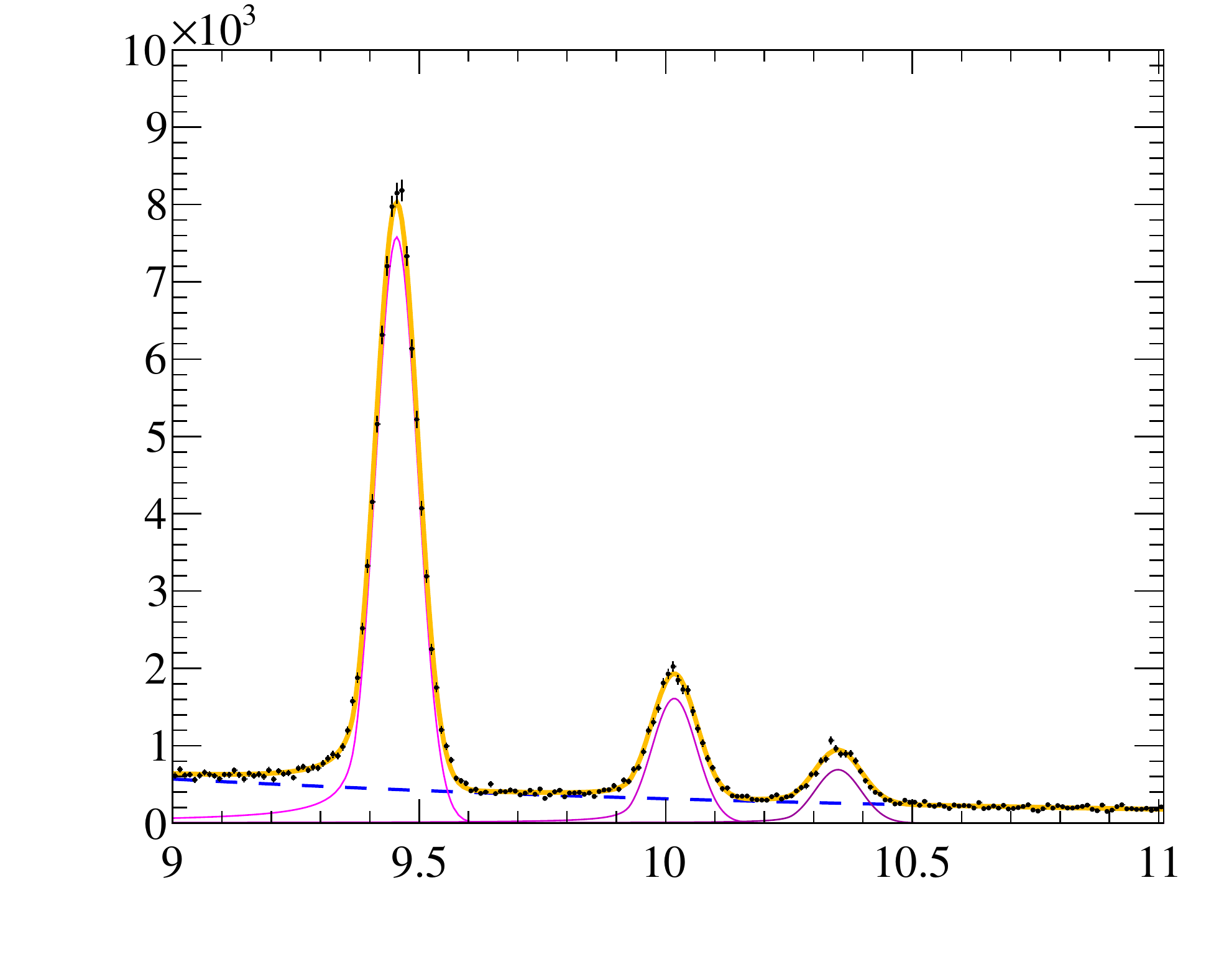}
    }
    \put( 75,  0){ 
      \includegraphics*[width=75mm,height=60mm,%
      ]{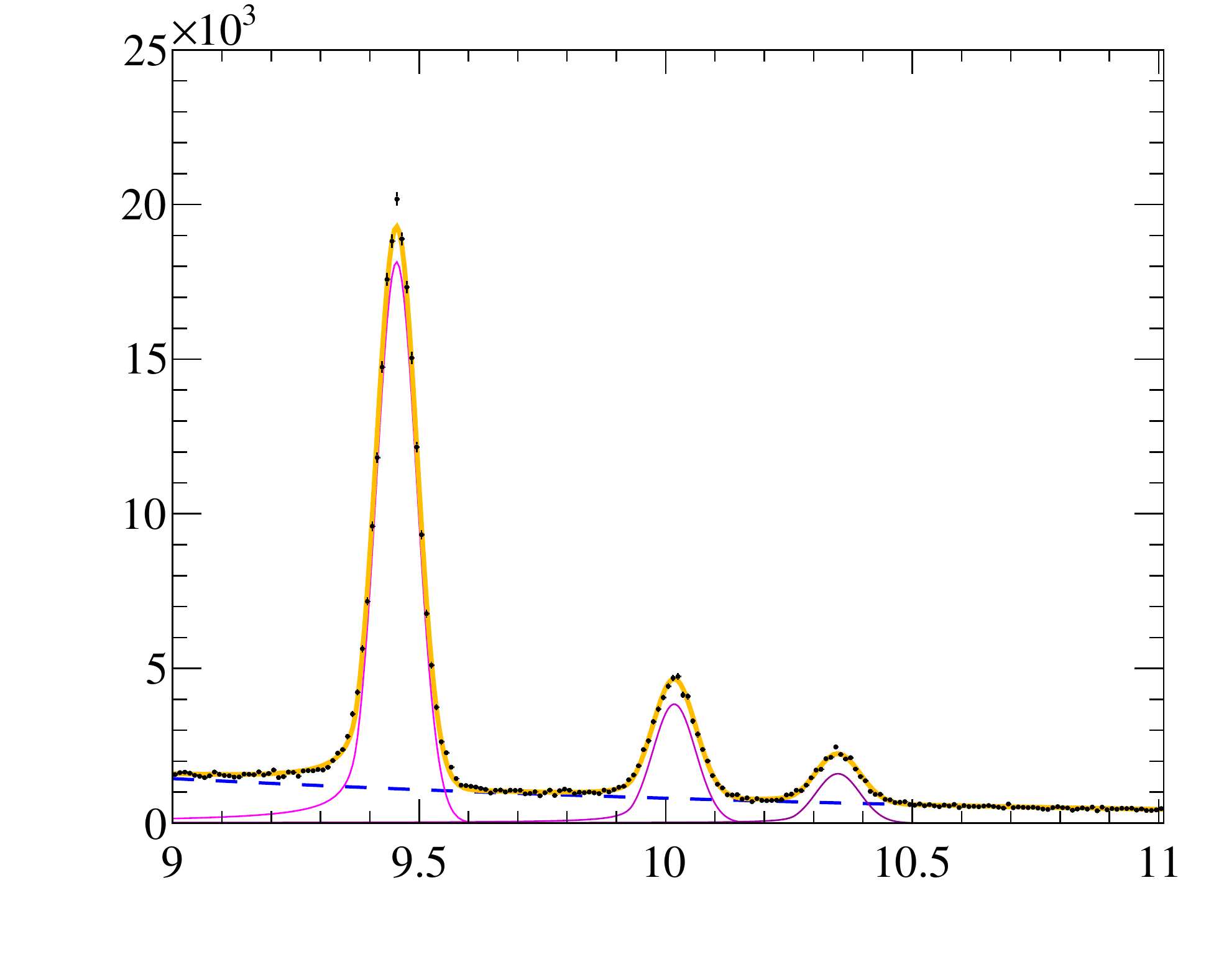}
    }
    \put( 32, 45) { \small { {$\begin{array}{l}\lhcb~\sqs=7\tev \\ 3<\pt<4\gevc \\ 3.0<y<3.5\end{array}$}}}
    \put(106, 45) { \small { {$\begin{array}{l}\lhcb~\sqs=8\tev \\ 3<\pt<4\gevc \\ 3.0<y<3.5\end{array}$}}}
    \put(  2, 14) { \begin{sideways}Candidates/(10\mevcc)\end{sideways}}
    \put( 77, 14) { \begin{sideways}Candidates/(10\mevcc)\end{sideways}}
    \put( 38,  2) { $m_{\mumu}$}  \put( 57.5,2) { $\left[\!\gevcc\right]$ }
    \put(113,  2) { $m_{\mumu}$}  \put(132.5,2) { $\left[\!\gevcc\right]$}  
  \end{picture}
  \caption { \small
    Efficiency-corrected dimuon mass distributions 
    for 
    (left)~\mbox{$\sqrt{s}=7\,\mathrm{TeV}$} and 
    (right)~\mbox{$\sqrt{s}=8\,\mathrm{TeV}$} samples
    in the~region \mbox{$3<\pt<4\gevc$}, \mbox{$3.0<y<3.5$}.
    The~thick dark yellow solid curves show the~result of the~fits, 
    as described in the~text.
    The~three peaks, shown with thin magenta solid lines, 
    correspond to the~\YoneS, \YtwoS and \YthreeS~signals\,(left to right).
    The~background component is indicated with a~blue dashed line.
    To~show the~signal peaks clearly, 
    the~range of the~dimuon mass shown is narrower 
    than that used in the~fit.
  } 
  \label{fig:onebin}
\end{figure}

The~fits are performed independently on the~efficiency-corrected 
dimuon~mass distributions 
in each (\pt,$y$)~bin.
As an~example, 
Fig.~\ref{fig:onebin} shows the~results of 
the~fits in the~region \mbox{$3<\pt<4\gevc$} and~\mbox{$3.0<y<3.5$}.
For each bin the~position and the~resolution of the~\YoneS~signal
is found to be consistent 
between~\mbox{$\sqs=7$} and \mbox{$8\tev$}~data sets.
The~resolution varies between $33\mevcc$ in 
the~region of low~\pt and small~rapidity
and~$90\mevcc$~for the~high~\pt and large~$y$~region,
with the~average~value being close to~$42\mevcc$. 
The~total signal yields are obtained by summing the  
signal yields over  all $(\pt,y)$~bins and are summarised in 
Table~\ref{tab:yields}.

\begin{table}[t]
  \centering
  \caption{ \small 
    Efficiency-corrected signal yields
    for data samples accumulated at \mbox{$\sqs=7$} and~\mbox{$8\,\mathrm{TeV}$}
    summed over the~full kinematic range~\mbox{$\pt<30\gevc$}, 
    \mbox{$2.0<y<4.5$}.
    The~uncertainties are statistical only.
  } \label{tab:yields}
  \vspace*{3mm}
  \begin{tabular*}{0.75\textwidth}{@{\hspace{2mm}}c@{\extracolsep{\fill}}cc@{\hspace{2mm}}}
    & $\sqs=7\tev$
    & $\sqs=8\tev$
    \\[1mm]
    \hline 
    \\[-2mm]    
    $N_{\YoneS\to\mumu}$
    & $(2639.8 \pm 3.7)\cdot10^{3}$ 
    & $(6563.1 \pm 6.3)\cdot10^3$
    \\
    $N_{\YtwoS\to\mumu}$
    & $\phantom{0}(667.3 \pm 2.2)\cdot10^3$ 
    & $(          1674.3 \pm 3.5)\cdot10^3 $
    \\
    $N_{\YthreeS\to\mumu}$
    & $\phantom{0}(328.8 \pm 1.5)\cdot10^3$ 
    & $\phantom{0}(786.6 \pm 2.6)\cdot10^3$ 
  \end{tabular*}   
\end{table}

\section{Systematic uncertainties}
\label{sec:syst}

The~systematic uncertainties are summarised in~Table~\ref{tab:syst},
separately for the~measurement of the~cross-sections  and of their ratios.

\begin{table}[t]
  \centering
  \caption{\small 
    Summary of relative systematic uncertainties\,(in \%) for 
    the~differential 
    production cross-sections, their ratios, 
    integrated cross-sections and 
    the~ratios $\mathscr{R}_{8/7}$.
    The ranges indicate variations depending on the~$(\pt,y)$~bin 
    and the~\ups~state.
  }\label{tab:syst}
  \vspace*{3mm}
  \begin{tabular*}{0.97\textwidth}{@{\hspace{2mm}}l@{\extracolsep{\fill}}cccc@{\hspace{2mm}}}
    Source                           
    &  $\upsigma^{\ups\to\mumu}_{\mathrm{bin}}$ 
    &  $\mathscr{R}_{\mathrm{i,j}}$ 
    &  $\upsigma^{\ups\to\mumu}$ 
    &  $\mathscr{R}_{8/7}$ 
    \\[1mm]
    \hline 
    \\[-2mm]    
    Fit model and range         & $0.1-4.8$                & $0.1-2.9$        & $0.1$ &  ---    \\ 
    Efficiency correction       & $0.2-0.6$                & $0.1-1.1$        & $0.4$ &  ---    \\ 
    Efficiency uncertainty      & $0.2-0.3$                &  ---             & $0.2$ &  $0.3$  \\ 
    Muon identification         & $0.3-0.5$                &  ---             & $0.3$ &  $0.2$  \\ 
    Data-simulation agreement   &                          &                  & &        \\
    ~~~Radiative tails          & $1.0$                    &  ---             & $1.0$ & ---     \\ 
    ~~~Selection efficiency     & $1.0$                    &  $0.5$           & $1.0$ &  $0.5$  \\
    ~~~Tracking efficiency      & $0.5\oplus\left(2\times0.4\right)$ & ---   & $0.5\oplus\left(2\times0.4\right)$ & ---     \\ 
    ~~~Trigger efficiency       & $2.0$                    &  ---             & $2.0$ &  $1.0$  \\ 
    \multirow{2}{*}{Luminosity} & $1.7\,(\sqrt{s}=7\tev)$  &  \multirow{2}{*}{---} & $1.7\,(\sqrt{s}=7\tev)$ & \multirow{2}{*}{$1.4$} \\ 
                                & $1.2\,(\sqrt{s}=8\tev)$  &                       & $1.2\,(\sqrt{s}=8\tev)$ &              
  \end{tabular*}   
\end{table}


The uncertainty related to the~mass model describing the~shape of
the~dimuon mass distribution is studied by varying the fit~range and 
the signal and background parametrisation used in the~fit model.
The~fit range is varied by moving the~upper edge
from~$12.5$~to~$11.5\gevcc$; 
the~degree of the~polynomial function used
in the~estimation of the~background is varied 
between zeroth and the~third order.
Also the~tail parameters 
of the~Crystal Ball function are allowed to vary 
in the~fit.
In~addition, the~constraints on the~difference
in the~\ups~signal peak positions
are removed for all bins with high signal yields.
The~maximum relative difference in the~number of 
signal events is taken as a~systematic uncertainty
arising from the~choice of the~fit model.

As an~alternative to the~determination of 
the~signal yields from
efficiency-corrected data, the~method 
employed in 
Ref.~\cite{LHCb-PAPER-2013-066} is used.
In~this method the~efficiency-corrected yields
for each $(\pt,y)$~bin  are calculated
using the~\sPlot~technique~\cite{Pivk:2004ty}.
The~difference between this method and the~nominal one is taken 
as a~systematic uncertainty on the~efficiency correction.

Reconstruction, selection and trigger~efficiencies in Eq.~\eqref{eq:effic} 
are obtained using simulated samples. 
The uncertainties due to the~finite size of these samples 
are propagated to the~measurement using a~large number of 
pseudoexperiments. The~same technique is used for the~propagation of 
the uncertainties on the~muon identification efficiency determined 
from large low-background samples of \mbox{$\jpsi\to\mumu$}~decays.

Several systematic uncertainties are assigned 
to account for possible  imperfections in the~simulated samples.
The~possible mismodelling of the bremsstrahlung simulation for 
the~radiative tail and its effect on the~signal shape has been 
estimated in previous LHCb analyses~\cite{LHCb-PAPER-2013-016} 
and leads to an~additional uncertainty of 1.0\% on the~cross-section.

Good agreement between data and simulation is observed for 
all variables used in the~selection. 
The~small differences seen would affect the~efficiencies 
by less than~1.0\%, which is conservatively taken as 
the~systematic uncertainty to account for 
the~disagreement between data and simulation.

The~efficiency is corrected using data-driven
techniques to account for small differences in 
the~tracking efficiency between data and 
\mbox{simulation~\cite{LHCb-DP-2013-002,LHCb-DP-2013-001}}.
The~uncertainty in the~correction factor is propagated 
to the cross-section measurement using pseudoexperiments 
and results in a global 0.5\%~systematic uncertainty plus 
an~additional uncertainty of 0.4\%~per track.

The systematic uncertainty associated with the~trigger requirements 
is assessed by studying the~performance of the~dimuon trigger
for \YoneS~events selected using the~single muon 
high-\pt~trigger~\cite{LHCb-DP-2012-004}
in data and simulation. The comparison is perfomed 
in bins of the \YoneS~meson  transverse momentum and rapidity 
and the~largest observed difference of 2.0\% is assigned 
as the~systematic uncertainty associated with 
the~imperfection of trigger simulation.

The~luminosity measurement 
was calibrated during dedicated data taking periods, 
using both van~der~Meer 
scans~\cite{vandermeer}
and a~beam-gas imaging method~\cite{FerroLuzzi:2005em,LHCb-PAPER-2011-015}.
The absolute luminosity scale is determined 
with \mbox{$1.7\,(1.2)\%$}~uncertainty 
for the~sample collected at~\mbox{$\sqs=7\,(8)\tev$},
of which the~beam-gas resolution, 
the~spread of the~measurements
and the~detector alignment
are the largest 
contributions~\cite{LHCb-PAPER-2014-047,LHCb-PAPER-2011-015,Barschel:1693671}.
The~ratio of absolute luminosities for samples accumulated 
at $\sqs=7$~and~$8\tev$~is known with a 1.4\%~uncertainty.

The total systematic uncertainty in each $(\pt,y)$~bin  is  
the~sum in quadrature of the~individual components described above. 
For the~integrated production cross-section the~systematic uncertainty 
is estimated by taking into account bin-to-bin correlations.
Several systematic uncertainties cancel or significantly reduce
in the~measurement of the~ratios
 $\mathscr{R}_{\mathrm{i,j}}$~and $\mathscr{R}_{8/7}$, 
as shown in Table~\ref{tab:syst}.

The~production cross-sections are  measured at centre-of-mass energies 
of~$7$~and $8\tev$, where the~actual beam energy for $\proton\proton$~collisions 
is known with a~precision of 0.65\%~\cite{CERN-ATS-2013-040}.
Assuming a~linear dependence of the~production cross-section on
the~$\proton\proton$~collision energy, 
and using the~measured production cross-sections 
at~\mbox{$\sqs=7\,(8)\tev$}, 
the~change in the~production cross-section
due to the~imprecise knowledge of 
the~beam energy is estimated to be~\mbox{$1.4\,(1.2)\%$}.
The~effect is strongly correlated
between $\sqs=7$ and~$8\tev$~data
and will therefore mostly cancel in 
the~measurement of the~ratio of 
cross-sections at the~two energies.

The efficiency is dependent on the~polarisation of the \ups~mesons.
The~polarisation of the~\ups~mesons  produced in $\proton\proton$~collisions
at \mbox{$\sqs=7\tev$} at high~\pt and central rapidity 
has been studied by the~CMS collaboration~\cite{Chatrchyan:2012woa}
in the~centre-of-mass helicity, 
Collins\nobreakdash-Soper~\cite{Collins:1977iv}
and the~perpendicular helicity frames.
No evidence of significant  transverse or longitudinal polarisation
has been observed for 
the~region~\mbox{$10<\pt<50\gevc$}, 
\mbox{$\left|y\right|<1.2$}.
Therefore, results are quoted under the~assumption of 
unpolarised production of \ups~mesons 
and no corresponding systematic uncertainty is assigned on 
the~cross-section. 
Under~the~assumption of transversely polarised \ups~mesons with \mbox{$\uplambda_{\vartheta}=0.2$} 
in the~\lhcb~kinematic region,\footnote{The CMS~measurements for \YoneS~mesons are consistent 
with small transverse polarisation in the~helicity frame with 
the~central values for the~polarisation parameter~\mbox{$0\lesssim\uplambda_{\vartheta}\lesssim0.2$}~\cite{Chatrchyan:2012woa}.} 
the~total production cross-section 
would result in an~increase of~3\%, with 
the~largest local increase of
around~6\% 
occuring in the~low~\pt region\,\mbox{$(\pt<3\gevc)$}, 
both for small\,\mbox{$(y<2.5)$} 
and large\,\mbox{$(y>4.0)$}~rapidities.

\section{Results} 
\label{sec:results}

\begin{figure}[p]
  \setlength{\unitlength}{1mm}
  \centering
  \begin{picture}(150,180)
    \put( 0,120){ 
      \includegraphics*[width=75mm,height=60mm,%
      ]{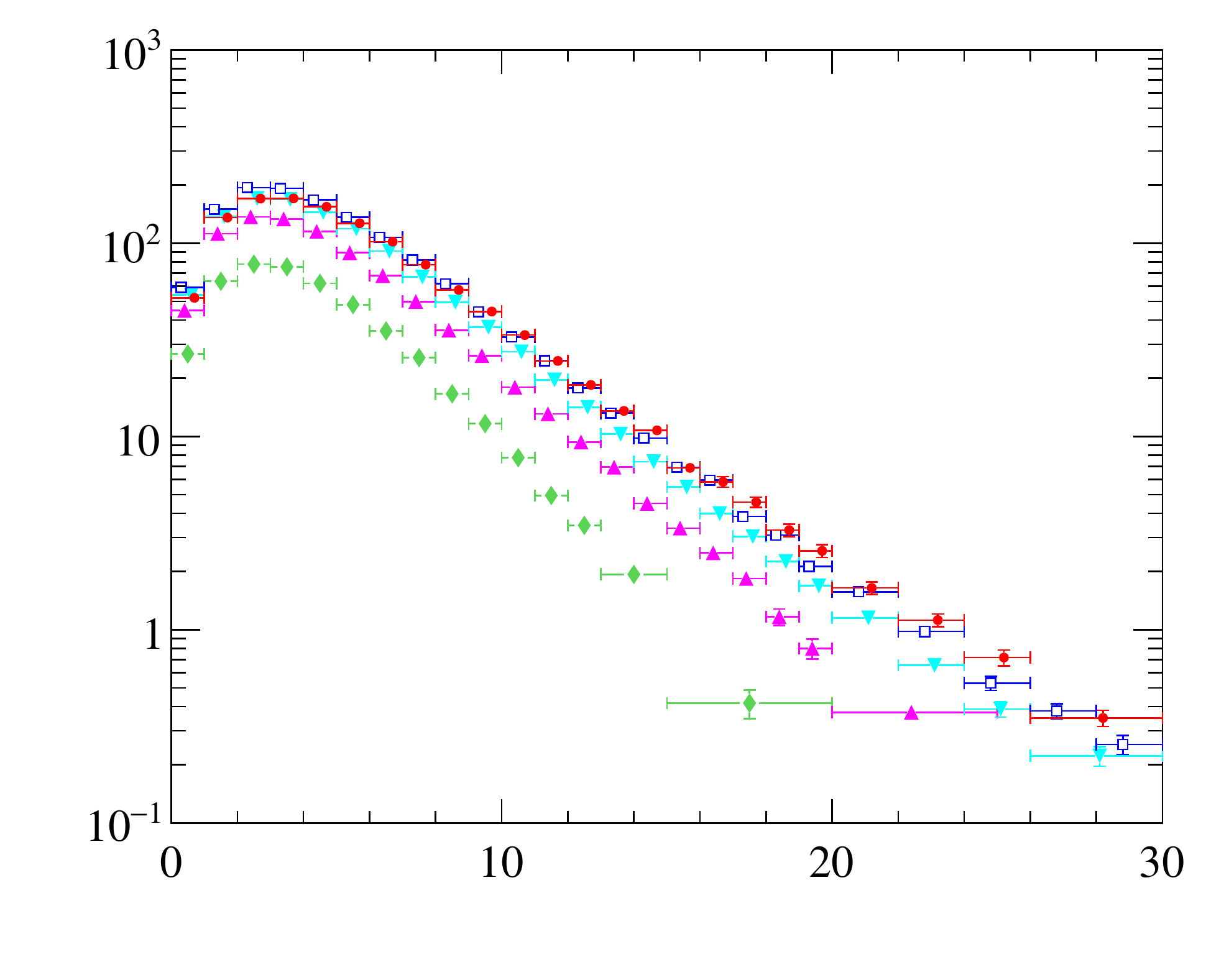}
    }
    \put(75,120){ 
      \includegraphics*[width=75mm,height=60mm,%
      ]{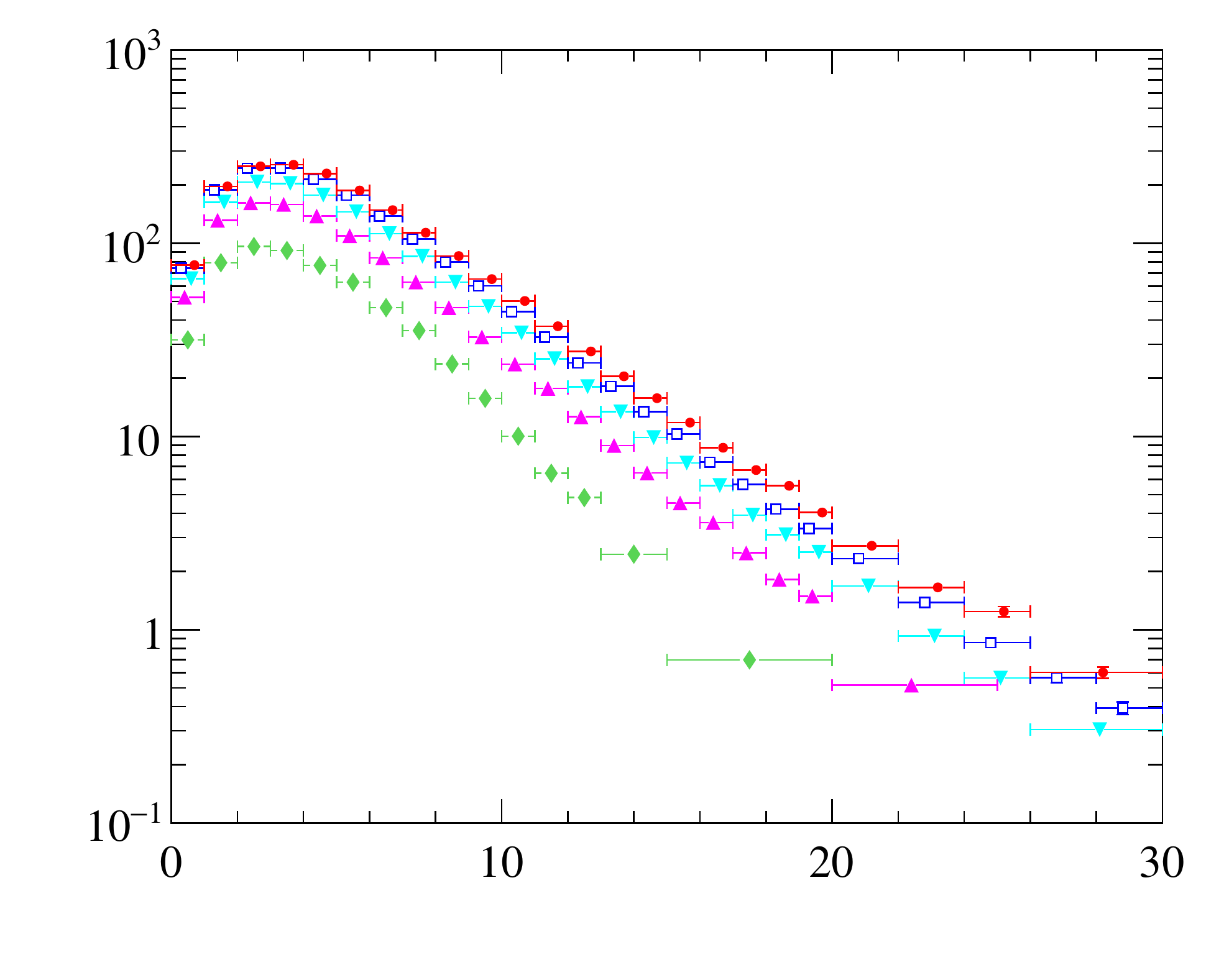}
    }
    \put( 0, 60){ 
      \includegraphics*[width=75mm,height=60mm,%
      ]{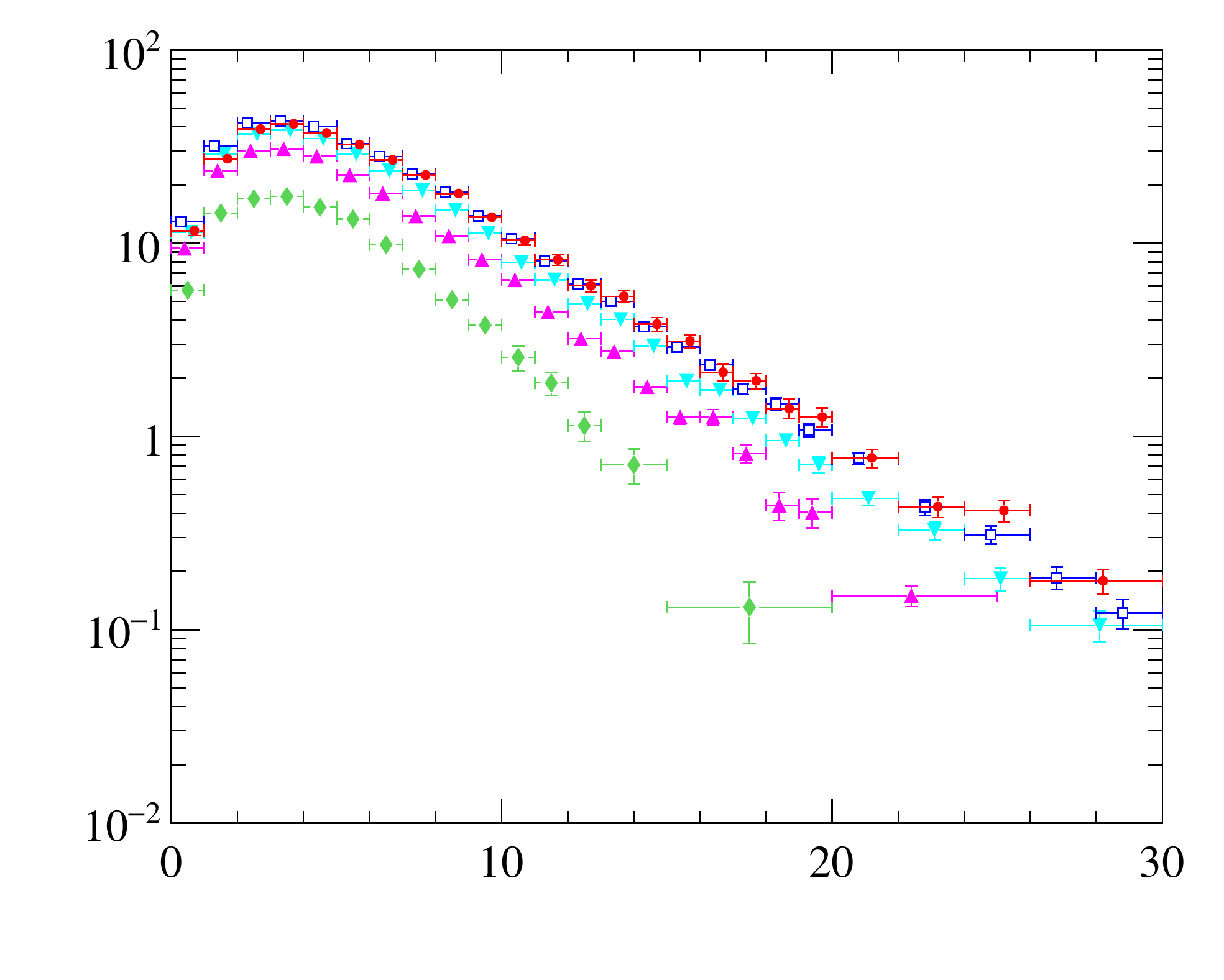}
    }
    \put(75, 60){ 
      \includegraphics*[width=75mm,height=60mm,%
      ]{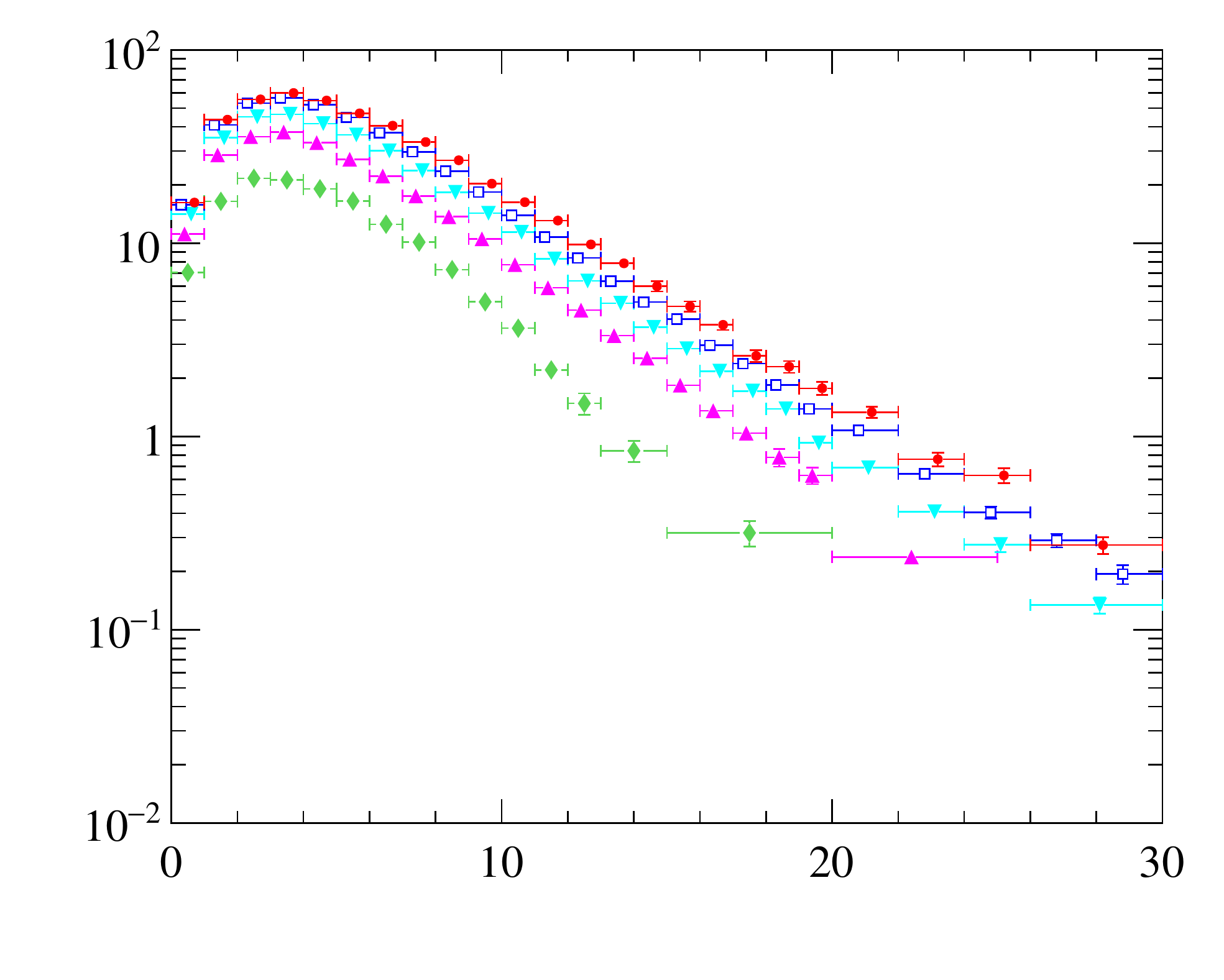}
    }
    \put( 0,  0){ 
      \includegraphics*[width=75mm,height=60mm,%
      ]{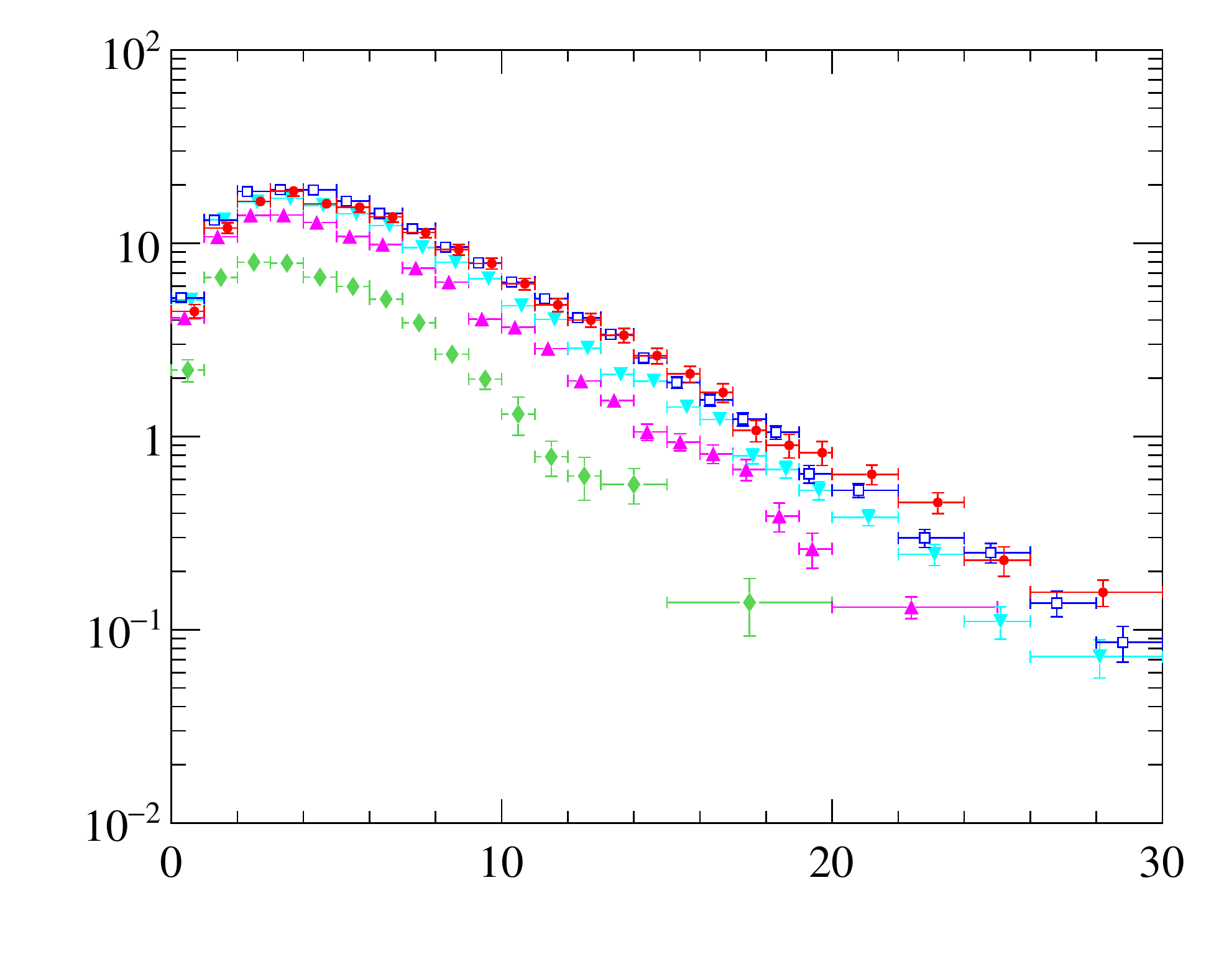}
    }
    \put(75,  0){ 
      \includegraphics*[width=75mm,height=60mm,%
      ]{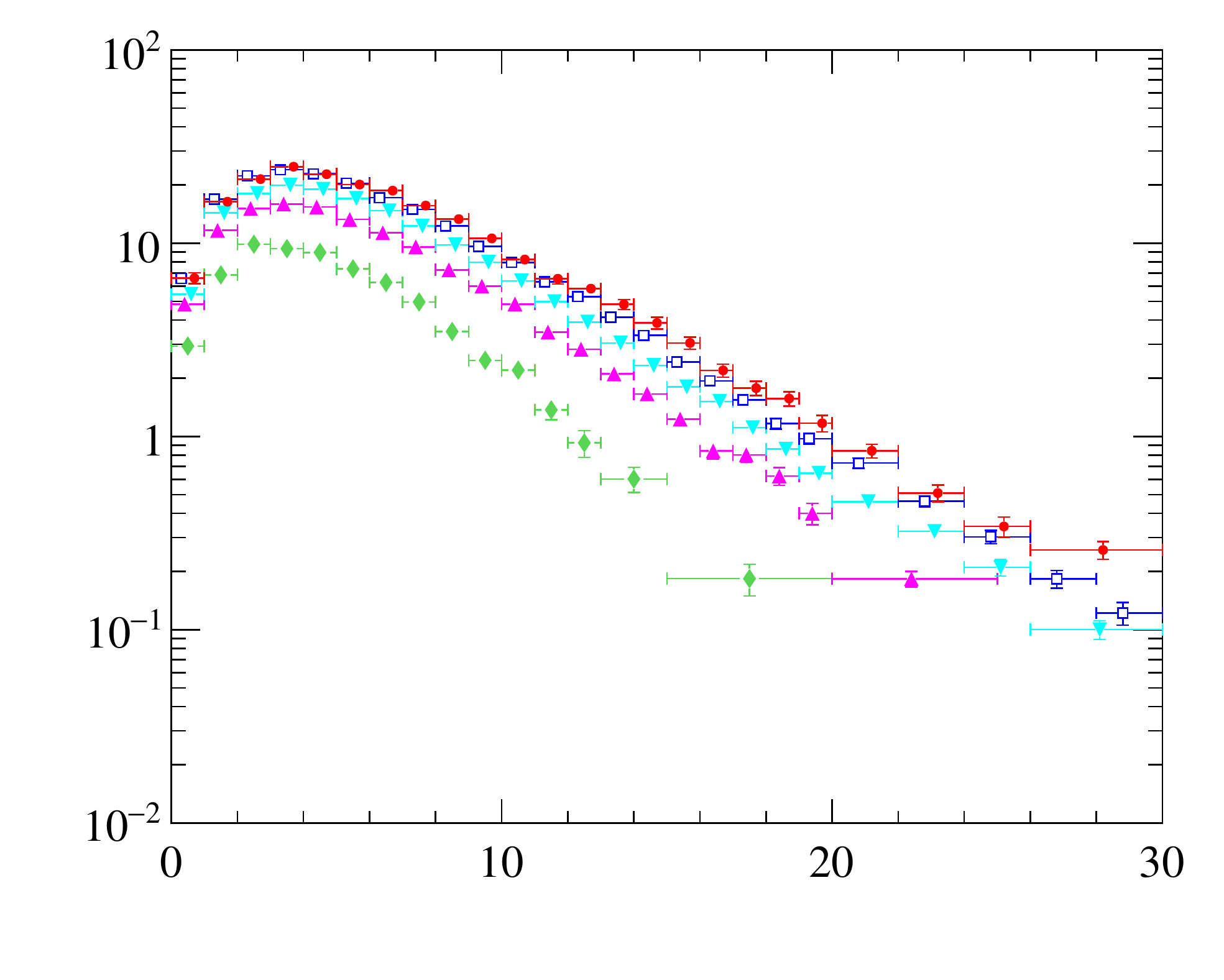}
    }
    \put( -2,133) { \begin{sideways}
        $\tfrac{\deriv^2}{\deriv\pt \deriv y}\upsigma^{\YoneS\to\mumu}~\left[\tfrac{\pb}{\!\gevc}\right]$
      \end{sideways}}
    \put( 73,133) { \begin{sideways}
         $\tfrac{\deriv^2}{\deriv\pt \deriv y}\upsigma^{\YoneS\to\mumu}~\left[\tfrac{\pb}{\!\gevc}\right]$
      \end{sideways}}
    \put( -2, 73) { \begin{sideways}
        $\tfrac{\deriv^2}{\deriv\pt \deriv y}\upsigma^{\YtwoS\to\mumu}~\left[\tfrac{\pb}{\!\gevc}\right]$
      \end{sideways}}
    \put( 73, 73) { \begin{sideways}
        $\tfrac{\deriv^2}{\deriv\pt \deriv y}\upsigma^{\YtwoS\to\mumu}~\left[\tfrac{\pb}{\!\gevc}\right]$
      \end{sideways}}
    \put( -2, 13) { \begin{sideways}
        $\tfrac{\deriv^2}{\deriv\pt \deriv y}\upsigma^{\YthreeS\to\mumu}~\left[\tfrac{\pb}{\!\gevc}\right]$
      \end{sideways}}
    \put( 73, 13) { \begin{sideways}
        $\tfrac{\deriv^2}{\deriv\pt \deriv y}\upsigma^{\YthreeS\to\mumu}~\left[\tfrac{\pb}{\!\gevc}\right]$
      \end{sideways}}
    \put (12,138){\scriptsize $\begin{array}{cc}  
        {\color{red}\bullet}                       & 2.0<y<2.5 \\
        {\color{blue}\square}                      & 2.5<y<3.0 \\
        {\color[rgb]{0,1,1}\blacktriangledown}     & 3.0<y<3.5 \\ 
        {\color[rgb]{1,0,1}\blacktriangle}         & 3.5<y<4.0 \\ 
        {\color[rgb]{0.33,0.83,0.33}\blacklozenge} & 4.0<y<4.5 \\ \end{array}$}
    \put (87.5,138){\scriptsize $\begin{array}{cc}  
        {\color{red}\bullet}                       & 2.0<y<2.5 \\
        {\color{blue}\square}                      & 2.5<y<3.0 \\
        {\color[rgb]{0,1,1}\blacktriangledown}     & 3.0<y<3.5 \\ 
        {\color[rgb]{1,0,1}\blacktriangle}         & 3.5<y<4.0 \\ 
        {\color[rgb]{0.33,0.83,0.33}\blacklozenge} & 4.0<y<4.5 \\ \end{array}$}
    \put (12.5, 78){\scriptsize $\begin{array}{cc}  
        {\color{red}\bullet}                       & 2.0<y<2.5 \\
        {\color{blue}\square}                      & 2.5<y<3.0 \\
        {\color[rgb]{0,1,1}\blacktriangledown}     & 3.0<y<3.5 \\ 
        {\color[rgb]{1,0,1}\blacktriangle}         & 3.5<y<4.0 \\ 
        {\color[rgb]{0.33,0.83,0.33}\blacklozenge} & 4.0<y<4.5 \\ \end{array}$}
    \put (87.5, 78){\scriptsize $\begin{array}{cc}  
        {\color{red}\bullet}                       & 2.0<y<2.5 \\
        {\color{blue}\square}                      & 2.5<y<3.0 \\
        {\color[rgb]{0,1,1}\blacktriangledown}     & 3.0<y<3.5 \\ 
        {\color[rgb]{1,0,1}\blacktriangle}         & 3.5<y<4.0 \\ 
        {\color[rgb]{0.33,0.83,0.33}\blacklozenge} & 4.0<y<4.5 \\ \end{array}$}
    \put (12.5, 18){\scriptsize $\begin{array}{cc}  
        {\color{red}\bullet}                       & 2.0<y<2.5 \\
        {\color{blue}\square}                      & 2.5<y<3.0 \\
        {\color[rgb]{0,1,1}\blacktriangledown}     & 3.0<y<3.5 \\ 
        {\color[rgb]{1,0,1}\blacktriangle}         & 3.5<y<4.0 \\ 
        {\color[rgb]{0.33,0.83,0.33}\blacklozenge} & 4.0<y<4.5 \\ \end{array}$}
    \put (87.5, 18){\scriptsize $\begin{array}{cc}  
        {\color{red}\bullet}                       & 2.0<y<2.5 \\
        {\color{blue}\square}                      & 2.5<y<3.0 \\
        {\color[rgb]{0,1,1}\blacktriangledown}     & 3.0<y<3.5 \\ 
        {\color[rgb]{1,0,1}\blacktriangle}         & 3.5<y<4.0 \\ 
        {\color[rgb]{0.33,0.83,0.33}\blacklozenge} & 4.0<y<4.5 \\ \end{array}$}
    \put( 45,168) { $\begin{array}{l} \text{LHCb}  \\ \sqrt{s}=7\tev \end{array}$}
    \put(120,168) { $\begin{array}{l} \text{LHCb}  \\ \sqrt{s}=8\tev \end{array}$}
    \put( 45,108) { $\begin{array}{l} \text{LHCb}  \\ \sqrt{s}=7\tev \end{array}$}
    \put(120,108) { $\begin{array}{l} \text{LHCb}  \\ \sqrt{s}=8\tev \end{array}$}
    \put( 45, 48) { $\begin{array}{l} \text{LHCb}  \\ \sqrt{s}=7\tev \end{array}$}
    \put(120, 48) { $\begin{array}{l} \text{LHCb}  \\ \sqrt{s}=8\tev \end{array}$}
    \put( 40,122) { $\pt$} \put( 59,122){ $\left[\!\gevc\right]$}
    \put(115,122) { $\pt$} \put(134,122){ $\left[\!\gevc\right]$}
    \put( 40, 62) { $\pt$} \put( 59, 62){ $\left[\!\gevc\right]$}
    \put(115, 62) { $\pt$} \put(134, 62){ $\left[\!\gevc\right]$}
    \put( 40,  2) { $\pt$} \put( 59,  2){ $\left[\!\gevc\right]$}
    \put(115,  2) { $\pt$} \put(134,  2){ $\left[\!\gevc\right]$}
  \end{picture}
  \caption { \small
    Double differential cross-sections  
    \mbox{$\dfrac{\deriv^2}{\deriv p_{\mathrm{T}}\,\deriv y}\upsigma^{\ups\to\mumu}$}
    for 
    (top)~\YoneS, 
    (middle)~\YtwoS and  
    \mbox{(bottom)}~\YthreeS 
    at 
    (left)~\mbox{$\sqrt{s}=7\,\mathrm{TeV}$} and 
    (right)~\mbox{$\sqrt{s}=8\,\mathrm{TeV}$}.
    The~error bars indicate the~sum in quadrature 
    of the~statistical and systematic uncertainties.
    The~rapidity ranges 
    \mbox{$2.0<y<2.5$},
    \mbox{$2.5<y<3.0$},
    \mbox{$3.0<y<3.5$},
    \mbox{$3.5<y<4.0$} and 
    \mbox{$4.0<y<4.5$} are shown with 
    red filled circles, 
    blue open squares, 
    cyan downward triangles,  
    magenta upward triangles 
    and green diamonds, respectively.
    Some data points are displaced from 
    the~bin centres to improve visibility.
  }
  \label{fig:diff_xsec}
\end{figure}

\begin{sidewaystable}[p]
  \centering
  \caption{\small 
    Production cross-section $\upsigma^{\YoneS\to\mumu}_{\mathrm{bin}}\left[\!\pb\right]$~in $(\pt,y)$~bins
    for $\sqrt{s}=7\,\mathrm{TeV}$. The~first uncertainties are statistical and the~second 
    are the~uncorrelated component of the~systematic uncertainties. 
    The~overall correlated systematic uncertainty is 3.1\% 
    and is not included in the~numbers in the~table.
    The~horizontal lines indicate bin boundaries.
  }\label{tab:y1s7TeV}
  \begin{small}
    \begin{tabular*}{0.99\textwidth}{@{\hspace{1mm}}c@{\extracolsep{\fill}}ccccc@{\hspace{1mm}}}
      $\pt\left[\!\gevc\right]$  
      &  $2.0<y<2.5$
      &  $2.5<y<3.0$
      &  $3.0<y<3.5$
      &  $3.5<y<4.0$
      &  $4.0<y<4.5$  
      \\
      \hline 
$0-1$
 & $26.1 \pm 0.5 \pm 0.3$  & $29.55 \pm 0.30 \pm 0.11$  & $27.0 \pm 0.3 \pm 0.4$  & $22.5 \pm 0.3 \pm 0.7$  & $13.4 \pm 0.4 \pm 0.2$ \\
$1-2$
 & $67.9 \pm 0.8 \pm 1.0$  & $74.9 \pm 0.5 \pm 0.4$  & $68.8 \pm 0.4 \pm 0.5$  & $56.0 \pm 0.4 \pm 0.3$  & $31.8 \pm 0.6 \pm 0.1$ \\
$2-3$
 & $85.0 \pm 0.8 \pm 0.7$  & $97.0 \pm 0.6 \pm 0.4$  & $85.2 \pm 0.5 \pm 0.3$  & $68.5 \pm 0.5 \pm 0.8$  & $38.9 \pm 0.6 \pm 1.0$ \\
$3-4$
 & $85.3 \pm 0.8 \pm 1.7$  & $96.0 \pm 0.6 \pm 0.4$  & $84.2 \pm 0.5 \pm 0.1$  & $66.7 \pm 0.5 \pm 0.4$  & $37.7 \pm 0.6 \pm 0.3$ \\
$4-5$
 & $77.2 \pm 0.8 \pm 0.3$  & $83.7 \pm 0.5 \pm 0.2$  & $72.2 \pm 0.4 \pm 0.3$  & $57.6 \pm 0.4 \pm 0.8$  & $31.0 \pm 0.5 \pm 0.2$ \\
$5-6$
 & $63.4 \pm 0.7 \pm 1.1$  & $68.1 \pm 0.5 \pm 0.3$  & $59.4 \pm 0.4 \pm 0.4$  & $44.6 \pm 0.4 \pm 0.3$  & $24.0 \pm 0.5 \pm 0.1$ \\
$6-7$
 & $50.9 \pm 0.6 \pm 0.8$  & $53.6 \pm 0.4 \pm 0.4$  & $45.5 \pm 0.4 \pm 0.4$  & $34.0 \pm 0.3 \pm 0.2$  & $17.6 \pm 0.4 \pm 0.4$ \\
$7-8$
 & $38.7 \pm 0.5 \pm 0.6$  & $40.9 \pm 0.4 \pm 0.4$  & $33.4 \pm 0.3 \pm 0.2$  & $25.0 \pm 0.3 \pm 0.2$  & $12.78 \pm 0.33 \pm 0.04$ \\
$8-9$
 & $28.6 \pm 0.5 \pm 0.4$  & $30.8 \pm 0.3 \pm 0.3$  & $24.76 \pm 0.25 \pm 0.25$  & $17.74 \pm 0.24 \pm 0.12$  & $8.31 \pm 0.27 \pm 0.14$ \\
$9-10$
 & $22.2 \pm 0.4 \pm 0.3$  & $22.05 \pm 0.26 \pm 0.13$  & $18.39 \pm 0.22 \pm 0.14$  & $13.10 \pm 0.21 \pm 0.12$  & $5.83 \pm 0.23 \pm 0.06$ \\
$10-11$
 & $16.7 \pm 0.4 \pm 0.2$  & $16.35 \pm 0.22 \pm 0.06$  & $13.71 \pm 0.18 \pm 0.03$  & $8.99 \pm 0.17 \pm 0.04$  & $3.9 \pm 0.2 \pm 0.3$ \\
$11-12$
 & $12.3 \pm 0.3 \pm 0.2$  & $12.32 \pm 0.19 \pm 0.16$  & $9.81 \pm 0.16 \pm 0.02$  & $6.55 \pm 0.14 \pm 0.08$  & $2.48 \pm 0.17 \pm 0.02$ \\
$12-13$
 & $9.24 \pm 0.26 \pm 0.15$  & $8.92 \pm 0.16 \pm 0.05$  & $7.08 \pm 0.13 \pm 0.01$  & $4.68 \pm 0.12 \pm 0.03$  & $1.73 \pm 0.16 \pm 0.04$ \\  \cline{6-6}
$13-14$
 & $6.78 \pm 0.22 \pm 0.09$  & $6.60 \pm 0.13 \pm 0.08$  & $5.14 \pm 0.11 \pm 0.03$  & $3.47 \pm 0.10 \pm 0.02$  & \multirow{2}{*}{$1.93 \pm 0.19 \pm 0.07$} \\
$14-15$
 & $5.38 \pm 0.19 \pm 0.10$  & $4.91 \pm 0.11 \pm 0.04$  & $3.70 \pm 0.09 \pm 0.07$  & $2.25 \pm 0.08 \pm 0.04$  &  \\  \cline{6-6}
$15-16$
 & $3.44 \pm 0.15 \pm 0.02$  & $3.46 \pm 0.10 \pm 0.04$  & $2.74 \pm 0.08 \pm 0.01$  & $1.68 \pm 0.07 \pm 0.02$  & \multirow{5}{*}{$1.04 \pm 0.17 \pm 0.03$} \\
$16-17$
 & $2.91 \pm 0.14 \pm 0.07$  & $2.97 \pm 0.09 \pm 0.03$  & $1.99 \pm 0.07 \pm 0.01$  & $1.25 \pm 0.06 \pm 0.03$  &  \\
$17-18$
 & $2.29 \pm 0.12 \pm 0.02$  & $1.93 \pm 0.07 \pm 0.01$  & $1.52 \pm 0.06 \pm 0.01$  & $0.92 \pm 0.06 \pm 0.01$  &  \\
$18-19$
 & $1.64 \pm 0.10 \pm 0.04$  & $1.54 \pm 0.06 \pm 0.01$  & $1.13 \pm 0.05 \pm 0.01$  & $0.58 \pm 0.05 \pm 0.03$  &  \\
$19-20$
 & $1.28 \pm 0.08 \pm 0.02$  & $1.06 \pm 0.05 \pm 0.01$  & $0.84 \pm 0.05 \pm 0.01$  & $0.40 \pm 0.04 \pm 0.02$  &  \\  \cline{2-6}
$20-21$
 & \multirow{2}{*}{$1.65 \pm 0.10 \pm 0.05$}  & \multirow{2}{*}{$1.57 \pm 0.06 \pm 0.01$}  & \multirow{2}{*}{$1.15 \pm 0.05 \pm 0.01$}  & \multirow{5}{*}{$0.94 \pm 0.06 \pm 0.01$} &  \\
$21-22$
 &   &   &   &  &  \\  \cline{2-4}
$22-23$
 & \multirow{2}{*}{$1.12 \pm 0.08 \pm 0.01$}  & \multirow{2}{*}{$0.98 \pm 0.05 \pm 0.01$}  & \multirow{2}{*}{$0.65 \pm 0.04 \pm 0.02$}  &  &  \\
$23-24$
 &   &   &   &  &  \\  \cline{2-4}
$24-25$
 & \multirow{2}{*}{$0.72 \pm 0.06 \pm 0.01$}  & \multirow{2}{*}{$0.53 \pm 0.04 \pm 0.01$}  & \multirow{2}{*}{$0.39 \pm 0.03 \pm 0.01$}  &  &  \\  \cline{5-5}
$25-26$
 &   &   &  &  &  \\  \cline{2-4}
$26-27$
 & \multirow{4}{*}{$0.70 \pm 0.06 \pm 0.01$}  & \multirow{2}{*}{$0.38 \pm 0.03 \pm 0.01$}  & \multirow{4}{*}{$0.45 \pm 0.04 \pm 0.03$} &  &  \\
$27-28$
 &   &   &  &  &  \\  \cline{3-3}
$28-29$
 &   & \multirow{2}{*}{$0.26 \pm 0.03 \pm 0.01$}  &  &  &  \\
$29-30$
 &   &   &  &  &  
\end{tabular*}   
\end{small}
\end{sidewaystable}

\begin{sidewaystable}[p]
  \centering
  \caption{\small 
    Production cross-section $\upsigma^{\YtwoS\to\mumu}_{\mathrm{bin}}~\left[\!\pb\right]$~in $(\pt,y)$~bins
    for $\sqrt{s}=7\,\mathrm{TeV}$. The~first uncertainties are statistical and the~second 
    are the~uncorrelated component of the~systematic uncertainties. 
    The~overall correlated systematic uncertainty is 3.1\% 
    and is not included in the~numbers in the~table.
    The~horizontal lines indicate bin boundaries.
  }\label{tab:y2s7TeV}
  \begin{small}
    \begin{tabular*}{0.99\textwidth}{@{\hspace{1mm}}c@{\extracolsep{\fill}}ccccc@{\hspace{1mm}}}
      $\pt\left[\!\gevc\right]$  
      &  $2.0<y<2.5$
      &  $2.5<y<3.0$
      &  $3.0<y<3.5$
      &  $3.5<y<4.0$
      &  $4.0<y<4.5$  
      \\
      \hline 
$0-1$
 & $5.80 \pm 0.25 \pm 0.12$  & $6.44 \pm 0.16 \pm 0.04$  & $5.69 \pm 0.14 \pm 0.12$  & $4.71 \pm 0.14 \pm 0.24$  & $2.86 \pm 0.19 \pm 0.10$ \\
$1-2$
 & $13.7 \pm 0.4 \pm 0.4$  & $15.95 \pm 0.25 \pm 0.13$  & $14.40 \pm 0.23 \pm 0.14$  & $11.87 \pm 0.22 \pm 0.10$  & $7.16 \pm 0.30 \pm 0.03$ \\
$2-3$
 & $19.5 \pm 0.4 \pm 0.3$  & $20.98 \pm 0.29 \pm 0.13$  & $18.35 \pm 0.25 \pm 0.10$  & $15.1 \pm 0.2 \pm 0.3$  & $8.5 \pm 0.3 \pm 0.4$ \\
$3-4$
 & $20.7 \pm 0.5 \pm 0.7$  & $21.49 \pm 0.29 \pm 0.13$  & $19.22 \pm 0.26 \pm 0.06$  & $15.42 \pm 0.25 \pm 0.13$  & $8.72 \pm 0.31 \pm 0.12$ \\
$4-5$
 & $18.6 \pm 0.4 \pm 0.1$  & $20.16 \pm 0.28 \pm 0.07$  & $17.40 \pm 0.24 \pm 0.14$  & $14.1 \pm 0.2 \pm 0.4$  & $7.67 \pm 0.28 \pm 0.13$ \\
$5-6$
 & $16.2 \pm 0.4 \pm 0.4$  & $16.37 \pm 0.26 \pm 0.11$  & $14.47 \pm 0.22 \pm 0.18$  & $11.26 \pm 0.21 \pm 0.11$  & $6.67 \pm 0.26 \pm 0.04$ \\
$6-7$
 & $13.5 \pm 0.4 \pm 0.4$  & $14.04 \pm 0.24 \pm 0.13$  & $11.84 \pm 0.20 \pm 0.17$  & $9.07 \pm 0.19 \pm 0.10$  & $4.91 \pm 0.22 \pm 0.19$ \\
$7-8$
 & $11.3 \pm 0.3 \pm 0.3$  & $11.42 \pm 0.21 \pm 0.20$  & $9.37 \pm 0.17 \pm 0.11$  & $6.92 \pm 0.17 \pm 0.13$  & $3.67 \pm 0.19 \pm 0.04$ \\
$8-9$
 & $9.04 \pm 0.29 \pm 0.17$  & $9.17 \pm 0.19 \pm 0.12$  & $7.43 \pm 0.15 \pm 0.12$  & $5.46 \pm 0.15 \pm 0.07$  & $2.55 \pm 0.16 \pm 0.08$ \\
$9-10$
 & $6.82 \pm 0.25 \pm 0.15$  & $6.91 \pm 0.16 \pm 0.05$  & $5.64 \pm 0.13 \pm 0.09$  & $4.12 \pm 0.13 \pm 0.09$  & $1.88 \pm 0.15 \pm 0.03$ \\
$10-11$
 & $5.17 \pm 0.22 \pm 0.11$  & $5.28 \pm 0.13 \pm 0.04$  & $3.96 \pm 0.11 \pm 0.01$  & $3.23 \pm 0.11 \pm 0.02$  & $1.28 \pm 0.13 \pm 0.13$ \\
$11-12$
 & $4.10 \pm 0.19 \pm 0.11$  & $4.03 \pm 0.12 \pm 0.08$  & $3.23 \pm 0.10 \pm 0.02$  & $2.20 \pm 0.09 \pm 0.04$  & $0.95 \pm 0.13 \pm 0.01$ \\
$12-13$
 & $3.02 \pm 0.16 \pm 0.09$  & $3.07 \pm 0.10 \pm 0.04$  & $2.43 \pm 0.08 \pm 0.01$  & $1.60 \pm 0.08 \pm 0.01$  & $0.57 \pm 0.10 \pm 0.02$ \\  \cline{6-6}
$13-14$
 & $2.66 \pm 0.15 \pm 0.07$  & $2.50 \pm 0.09 \pm 0.04$  & $2.02 \pm 0.07 \pm 0.02$  & $1.38 \pm 0.07 \pm 0.02$  & \multirow{2}{*}{$0.71 \pm 0.14 \pm 0.03$} \\
$14-15$
 & $1.90 \pm 0.12 \pm 0.07$  & $1.85 \pm 0.07 \pm 0.03$  & $1.47 \pm 0.06 \pm 0.04$  & $0.90 \pm 0.06 \pm 0.03$  &  \\  \cline{6-6}
$15-16$
 & $1.56 \pm 0.11 \pm 0.01$  & $1.45 \pm 0.07 \pm 0.02$  & $0.97 \pm 0.05 \pm 0.01$  & $0.63 \pm 0.05 \pm 0.01$  & \multirow{5}{*}{$0.33 \pm 0.11 \pm 0.01$} \\
$16-17$
 & $1.08 \pm 0.09 \pm 0.05$  & $1.17 \pm 0.06 \pm 0.02$  & $0.87 \pm 0.05 \pm 0.01$  & $0.63 \pm 0.05 \pm 0.03$  &  \\
$17-18$
 & $0.97 \pm 0.08 \pm 0.01$  & $0.88 \pm 0.05 \pm 0.01$  & $0.62 \pm 0.04 \pm 0.01$  & $0.41 \pm 0.04 \pm 0.01$  &  \\
$18-19$
 & $0.70 \pm 0.07 \pm 0.03$  & $0.74 \pm 0.05 \pm 0.01$  & $0.48 \pm 0.04 \pm 0.01$  & $0.22 \pm 0.03 \pm 0.02$  &  \\
$19-20$
 & $0.63 \pm 0.07 \pm 0.02$  & $0.54 \pm 0.04 \pm 0.01$  & $0.36 \pm 0.03 \pm 0.01$  & $0.20 \pm 0.03 \pm 0.01$  &  \\  \cline{2-6}
$20-21$
 & \multirow{2}{*}{$0.77 \pm 0.07 \pm 0.04$}  & \multirow{2}{*}{$0.77 \pm 0.05 \pm 0.01$}  & \multirow{2}{*}{$0.48 \pm 0.04 \pm 0.01$}  & \multirow{5}{*}{$0.38 \pm 0.04 \pm 0.01$} &  \\
$21-22$
 &   &   &   &  &  \\  \cline{2-4}
$22-23$
 & \multirow{2}{*}{$0.43 \pm 0.05 \pm 0.01$}  & \multirow{2}{*}{$0.43 \pm 0.04 \pm 0.01$}  & \multirow{2}{*}{$0.33 \pm 0.03 \pm 0.01$}  &  &  \\
$23-24$
 &   &   &   &  &  \\  \cline{2-4}
$24-25$
 & \multirow{2}{*}{$0.41 \pm 0.05 \pm 0.01$}  & \multirow{2}{*}{$0.31 \pm 0.03 \pm 0.01$}  & \multirow{2}{*}{$0.18 \pm 0.02 \pm 0.01$}  &  &  \\  \cline{5-5}
$25-26$
 &   &   &  &  &  \\  \cline{2-4}
$26-27$
 & \multirow{4}{*}{$0.36 \pm 0.05 \pm 0.01$}  & \multirow{2}{*}{$0.19 \pm 0.02 \pm 0.01$}  & \multirow{4}{*}{$0.21 \pm 0.03 \pm 0.02$} &  &  \\
$27-28$
 &   &   &  &  &  \\  \cline{3-3}
$28-29$
 &   & \multirow{2}{*}{$0.12 \pm 0.02 \pm 0.01$}  &  &  &  \\
$29-30$
 &   &   &  &  &  
\end{tabular*}   
\end{small}
\end{sidewaystable}

\begin{sidewaystable}[p]
  \centering
  \caption{\small 
    Production cross-section $\upsigma^{\YthreeS\to\mumu}_{\mathrm{bin}}~\left[\!\pb\right]$~in $(\pt,y)$~bins
    for $\sqrt{s}=7\,\mathrm{TeV}$. The~first uncertainties are statistical and the~second
    are the~uncorrelated component of the~systematic uncertainties. 
    The~overall correlated systematic uncertainty is 3.1\% 
    and is not included in the~numbers in the~table.
    The~horizontal lines indicate bin boundaries.
  }\label{tab:y3s7TeV}
  \begin{small}
    \begin{tabular*}{0.99\textwidth}{@{\hspace{1mm}}c@{\extracolsep{\fill}}ccccc@{\hspace{1mm}}}
      $\pt\left[\!\gevc\right]$  
      &  $2.0<y<2.5$
      &  $2.5<y<3.0$
      &  $3.0<y<3.5$
      &  $3.5<y<4.0$
      &  $4.0<y<4.5$  
      \\
      \hline 
$0-1$
 & $2.22 \pm 0.17 \pm 0.05$  & $2.61 \pm 0.11 \pm 0.02$  & $2.54 \pm 0.11 \pm 0.08$  & $2.05 \pm 0.10 \pm 0.12$  & $1.10 \pm 0.13 \pm 0.05$ \\
$1-2$
 & $5.99 \pm 0.28 \pm 0.17$  & $6.58 \pm 0.18 \pm 0.08$  & $6.61 \pm 0.17 \pm 0.09$  & $5.40 \pm 0.16 \pm 0.07$  & $3.33 \pm 0.22 \pm 0.01$ \\
$2-3$
 & $8.22 \pm 0.31 \pm 0.11$  & $9.25 \pm 0.21 \pm 0.08$  & $8.17 \pm 0.19 \pm 0.03$  & $6.98 \pm 0.18 \pm 0.17$  & $3.99 \pm 0.23 \pm 0.24$ \\
$3-4$
 & $9.3 \pm 0.3 \pm 0.3$  & $9.46 \pm 0.22 \pm 0.06$  & $8.53 \pm 0.19 \pm 0.01$  & $6.99 \pm 0.18 \pm 0.05$  & $3.94 \pm 0.22 \pm 0.06$ \\
$4-5$
 & $8.00 \pm 0.31 \pm 0.02$  & $9.42 \pm 0.21 \pm 0.05$  & $7.84 \pm 0.18 \pm 0.03$  & $6.40 \pm 0.17 \pm 0.21$  & $3.33 \pm 0.20 \pm 0.05$ \\
$5-6$
 & $7.7 \pm 0.3 \pm 0.2$  & $8.25 \pm 0.20 \pm 0.05$  & $7.10 \pm 0.17 \pm 0.10$  & $5.42 \pm 0.16 \pm 0.08$  & $2.98 \pm 0.18 \pm 0.02$ \\
$6-7$
 & $6.84 \pm 0.27 \pm 0.20$  & $7.13 \pm 0.18 \pm 0.09$  & $6.17 \pm 0.15 \pm 0.10$  & $4.93 \pm 0.15 \pm 0.07$  & $2.57 \pm 0.17 \pm 0.14$ \\
$7-8$
 & $5.68 \pm 0.25 \pm 0.15$  & $5.93 \pm 0.16 \pm 0.11$  & $4.75 \pm 0.13 \pm 0.07$  & $3.72 \pm 0.13 \pm 0.08$  & $1.94 \pm 0.15 \pm 0.03$ \\
$8-9$
 & $4.63 \pm 0.22 \pm 0.11$  & $4.77 \pm 0.14 \pm 0.07$  & $3.99 \pm 0.12 \pm 0.08$  & $3.15 \pm 0.12 \pm 0.04$  & $1.33 \pm 0.13 \pm 0.04$ \\
$9-10$
 & $3.93 \pm 0.20 \pm 0.11$  & $3.96 \pm 0.13 \pm 0.04$  & $3.28 \pm 0.10 \pm 0.05$  & $2.03 \pm 0.09 \pm 0.05$  & $0.99 \pm 0.11 \pm 0.01$ \\
$10-11$
 & $3.08 \pm 0.18 \pm 0.06$  & $3.15 \pm 0.11 \pm 0.03$  & $2.38 \pm 0.09 \pm 0.01$  & $1.84 \pm 0.09 \pm 0.01$  & $0.65 \pm 0.10 \pm 0.10$ \\
$11-12$
 & $2.40 \pm 0.15 \pm 0.07$  & $2.58 \pm 0.10 \pm 0.06$  & $2.02 \pm 0.08 \pm 0.01$  & $1.43 \pm 0.08 \pm 0.03$  & $0.39 \pm 0.08 \pm 0.01$ \\
$12-13$
 & $2.00 \pm 0.14 \pm 0.06$  & $2.06 \pm 0.08 \pm 0.02$  & $1.43 \pm 0.07 \pm 0.01$  & $0.97 \pm 0.06 \pm 0.01$  & $0.31 \pm 0.08 \pm 0.01$ \\  \cline{6-6}
$13-14$
 & $1.67 \pm 0.12 \pm 0.06$  & $1.69 \pm 0.07 \pm 0.03$  & $1.05 \pm 0.06 \pm 0.01$  & $0.77 \pm 0.05 \pm 0.01$  & \multirow{2}{*}{$0.56 \pm 0.11 \pm 0.02$} \\
$14-15$
 & $1.31 \pm 0.11 \pm 0.04$  & $1.28 \pm 0.06 \pm 0.02$  & $0.97 \pm 0.05 \pm 0.02$  & $0.53 \pm 0.05 \pm 0.01$  &  \\  \cline{6-6}
$15-16$
 & $1.05 \pm 0.09 \pm 0.01$  & $0.95 \pm 0.06 \pm 0.01$  & $0.71 \pm 0.04 \pm 0.01$  & $0.47 \pm 0.04 \pm 0.01$  & \multirow{5}{*}{$0.35 \pm 0.11 \pm 0.01$} \\
$16-17$
 & $0.84 \pm 0.08 \pm 0.03$  & $0.77 \pm 0.05 \pm 0.01$  & $0.61 \pm 0.04 \pm 0.01$  & $0.41 \pm 0.04 \pm 0.01$  &  \\
$17-18$
 & $0.54 \pm 0.07 \pm 0.01$  & $0.61 \pm 0.04 \pm 0.01$  & $0.40 \pm 0.03 \pm 0.01$  & $0.34 \pm 0.04 \pm 0.01$  &  \\
$18-19$
 & $0.45 \pm 0.06 \pm 0.02$  & $0.53 \pm 0.04 \pm 0.01$  & $0.34 \pm 0.03 \pm 0.01$  & $0.19 \pm 0.03 \pm 0.01$  &  \\
$19-20$
 & $0.41 \pm 0.05 \pm 0.01$  & $0.32 \pm 0.03 \pm 0.01$  & $0.26 \pm 0.03 \pm 0.01$  & $0.13 \pm 0.03 \pm 0.01$  &  \\  \cline{2-6}
$20-21$
 & \multirow{2}{*}{$0.64 \pm 0.07 \pm 0.02$}  & \multirow{2}{*}{$0.53 \pm 0.04 \pm 0.01$}  & \multirow{2}{*}{$0.38 \pm 0.03 \pm 0.01$}  & \multirow{5}{*}{$0.33 \pm 0.04 \pm 0.01$} &  \\
$21-22$
 &   &   &   &  &  \\  \cline{2-4}
$22-23$
 & \multirow{2}{*}{$0.46 \pm 0.05 \pm 0.01$}  & \multirow{2}{*}{$0.20 \pm 0.03 \pm 0.01$}  & \multirow{2}{*}{$0.25 \pm 0.03 \pm 0.01$}  &  &  \\
$23-24$
 &   &   &   &  &  \\  \cline{2-4}
$24-25$
 & \multirow{2}{*}{$0.23 \pm 0.04 \pm 0.01$}  & \multirow{2}{*}{$0.25  \pm 0.03 \pm 0.01$}  & \multirow{2}{*}{$0.11 \pm 0.02 \pm 0.01$}  &  &  \\  \cline{5-5}
$25-26$
 &   &   &  &  &  \\  \cline{2-4}
$26-27$
 & \multirow{4}{*}{$0.31 \pm 0.05 \pm 0.01$}  & \multirow{2}{*}{$0.14 \pm 0.02 \pm 0.01$}  & \multirow{4}{*}{$0.15 \pm 0.03 \pm 0.02$} &  &  \\
$27-28$
 &   &   &  &  &  \\  \cline{3-3}
$28-29$
 &   & \multirow{2}{*}{$ 0.09 \pm 0.02 \pm 0.01$}  &  &  &  \\
$29-30$
 &   &   &  &  &  
\end{tabular*}   
\end{small}
\end{sidewaystable}

\begin{sidewaystable}[p]
  \centering
  \caption{\small 
    Production cross-section $\upsigma^{\YoneS\to\mumu}_{\mathrm{bin}}~\left[\!\pb\right]$~in $(\pt,y)$~bins
    for $\sqrt{s}=8\,\mathrm{TeV}$. The~first uncertainties are statistical and the~second 
    are the~uncorrelated component of the~systematic uncertainties. 
    The~overall correlated systematic uncertainty is 2.8\% 
    and is not included in the~numbers in the~table.
    The~horizontal lines indicate bin boundaries.
  }\label{tab:y1s8TeV}
  \begin{small}
    \begin{tabular*}{0.99\textwidth}{@{\hspace{1mm}}c@{\extracolsep{\fill}}ccccc@{\hspace{1mm}}}
      $\pt\left[\!\gevc\right]$  
      &  $2.0<y<2.5$
      &  $2.5<y<3.0$
      &  $3.0<y<3.5$
      &  $3.5<y<4.0$
      &  $4.0<y<4.5$  
      \\
      \hline 
$0-1$
 & $38.5 \pm 0.5 \pm 0.6$  & $37.2 \pm 0.3 \pm 0.3$  & $32.7 \pm 0.2 \pm 0.3$  & $26.28 \pm 0.22 \pm 0.12$  & $15.8 \pm 0.3 \pm 0.2$ \\
$1-2$
 & $98.4 \pm 0.8 \pm 0.5$  & $94.3 \pm 0.4 \pm 0.3$  & $81.5 \pm 0.4 \pm 0.4$  & $65.7 \pm 0.4 \pm 0.7$  & $39.6 \pm 0.5 \pm 0.4$ \\
$2-3$
 & $124.9 \pm 0.8 \pm 0.8$  & $122.1 \pm 0.5 \pm 0.7$  & $103.7 \pm 0.4 \pm 0.8$  & $80.9 \pm 0.4 \pm 0.2$  & $48.0 \pm 0.5 \pm 0.3$ \\
$3-4$
 & $127.3 \pm 0.8 \pm 0.9$  & $122.4 \pm 0.5 \pm 0.3$  & $101.9 \pm 0.4 \pm 0.6$  & $79.4 \pm 0.4 \pm 0.5$  & $45.8 \pm 0.5 \pm 0.3$ \\
$4-5$
 & $114.7 \pm 0.8 \pm 0.5$  & $107.1 \pm 0.4 \pm 0.3$  & $88.7 \pm 0.4 \pm 0.3$  & $69.2 \pm 0.4 \pm 0.3$  & $38.4 \pm 0.4 \pm 0.6$ \\
$5-6$
 & $93.7 \pm 0.7 \pm 0.7$  & $88.6 \pm 0.4 \pm 0.5$  & $72.7 \pm 0.3 \pm 0.5$  & $54.7 \pm 0.3 \pm 0.5$  & $31.4 \pm 0.4 \pm 0.1$ \\
$6-7$
 & $74.1 \pm 0.6 \pm 0.8$  & $69.1 \pm 0.4 \pm 0.3$  & $56.0 \pm 0.3 \pm 0.3$  & $42.0 \pm 0.3 \pm 0.4$  & $23.20 \pm 0.33 \pm 0.08$ \\
$7-8$
 & $56.7 \pm 0.5 \pm 0.5$  & $52.7 \pm 0.3 \pm 0.3$  & $42.8 \pm 0.2 \pm 0.2$  & $31.48 \pm 0.24 \pm 0.16$  & $17.65 \pm 0.29 \pm 0.15$ \\
$8-9$
 & $42.9 \pm 0.5 \pm 0.3$  & $39.9 \pm 0.3 \pm 0.3$  & $31.41 \pm 0.21 \pm 0.13$  & $23.19 \pm 0.20 \pm 0.04$  & $11.86 \pm 0.24 \pm 0.19$ \\
$9-10$
 & $32.6 \pm 0.4 \pm 0.3$  & $30.04 \pm 0.23 \pm 0.14$  & $23.53 \pm 0.18 \pm 0.07$  & $16.36 \pm 0.17 \pm 0.14$  & $7.87 \pm 0.20 \pm 0.05$ \\
$10-11$
 & $25.1 \pm 0.4 \pm 0.4$  & $22.10 \pm 0.19 \pm 0.25$  & $17.17 \pm 0.15 \pm 0.07$  & $11.85 \pm 0.14 \pm 0.14$  & $5.02 \pm 0.17 \pm 0.13$ \\
$11-12$
 & $18.6 \pm 0.3 \pm 0.2$  & $16.32 \pm 0.16 \pm 0.12$  & $12.62 \pm 0.13 \pm 0.14$  & $8.87 \pm 0.12 \pm 0.06$  & $3.23 \pm 0.14 \pm 0.05$ \\
$12-13$
 & $13.77 \pm 0.25 \pm 0.12$  & $12.00 \pm 0.14 \pm 0.12$  & $9.05 \pm 0.11 \pm 0.05$  & $6.32 \pm 0.10 \pm 0.06$  & $2.41 \pm 0.14 \pm 0.05$ \\  \cline{6-6}
$13-14$
 & $10.24 \pm 0.22 \pm 0.13$  & $9.09 \pm 0.12 \pm 0.08$  & $6.70 \pm 0.09 \pm 0.09$  & $4.49 \pm 0.09 \pm 0.05$  & \multirow{2}{*}{$2.46 \pm 0.15 \pm 0.07$} \\
$14-15$
 & $7.89 \pm 0.19 \pm 0.17$  & $6.71 \pm 0.10 \pm 0.06$  & $4.93 \pm 0.08 \pm 0.01$  & $3.24 \pm 0.07 \pm 0.07$  &  \\  \cline{6-6}
$15-16$
 & $5.90 \pm 0.16 \pm 0.09$  & $5.15 \pm 0.09 \pm 0.04$  & $3.64 \pm 0.07 \pm 0.03$  & $2.27 \pm 0.06 \pm 0.02$  & \multirow{5}{*}{$1.75 \pm 0.15 \pm 0.05$} \\
$16-17$
 & $4.37 \pm 0.13 \pm 0.02$  & $3.68 \pm 0.07 \pm 0.03$  & $2.79 \pm 0.06 \pm 0.01$  & $1.79 \pm 0.06 \pm 0.03$  &  \\
$17-18$
 & $3.35 \pm 0.12 \pm 0.05$  & $2.83 \pm 0.06 \pm 0.01$  & $1.96 \pm 0.05 \pm 0.02$  & $1.25 \pm 0.05 \pm 0.02$  &  \\
$18-19$
 & $2.78 \pm 0.10 \pm 0.03$  & $2.10 \pm 0.05 \pm 0.01$  & $1.55 \pm 0.05 \pm 0.01$  & $0.91 \pm 0.04 \pm 0.03$  &  \\
$19-20$
 & $2.02 \pm 0.09 \pm 0.01$  & $1.67 \pm 0.05 \pm 0.01$  & $1.26 \pm 0.04 \pm 0.02$  & $0.75 \pm 0.04 \pm 0.01$  &  \\  \cline{2-6}
$20-21$
 & \multirow{2}{*}{$2.72 \pm 0.10 \pm 0.02$}  & \multirow{2}{*}{$2.34 \pm 0.06 \pm 0.03$}  & \multirow{2}{*}{$1.69 \pm 0.05 \pm 0.03$}  & \multirow{5}{*}{$1.29 \pm 0.05 \pm 0.02$} &  \\
$21-22$
 &   &   &   &  &  \\  \cline{2-4}
$22-23$
 & \multirow{2}{*}{$1.66 \pm 0.08 \pm 0.01$}  & \multirow{2}{*}{$1.38 \pm 0.04 \pm 0.01$}  & \multirow{2}{*}{$0.93 \pm 0.04 \pm 0.01$}  &  &  \\
$23-24$
 &   &   &   &  &  \\  \cline{2-4}
$24-25$
 & \multirow{2}{*}{$1.24 \pm 0.07 \pm 0.02$}  & \multirow{2}{*}{$0.86 \pm 0.04 \pm 0.02$}  & \multirow{2}{*}{$0.56 \pm 0.03 \pm 0.01$}  &  &  \\  \cline{5-5}
$25-26$
 &   &   &  &  &  \\  \cline{2-4}
$26-27$
 & \multirow{4}{*}{$1.20 \pm 0.07 \pm 0.02$}  & \multirow{2}{*}{$0.56 \pm 0.03 \pm 0.01$}  & \multirow{4}{*}{$0.61 \pm 0.03 \pm 0.01$} &  &  \\
$27-28$
 &   &   &  &  &  \\  \cline{3-3}
$28-29$
 &   & \multirow{2}{*}{$0.39 \pm 0.03 \pm 0.01$}  &  &  &  \\
$29-30$
 &   &   &  &  &  
\end{tabular*}   
\end{small}
\end{sidewaystable}

\begin{sidewaystable}[p]
  \centering
  \caption{\small 
    Production cross-section $\upsigma^{\YtwoS\to\mumu}_{\mathrm{bin}}~\left[\!\pb\right]$~in $(\pt,y)$~bins
    for $\sqrt{s}=8\,\mathrm{TeV}$. The~first uncertainties are statistical and the~second 
    are the~uncorrelated component of the~systematic uncertainties. 
    The~overall correlated systematic uncertainty is 2.8\% 
    and is not included in the~numbers in the~table.
    The~horizontal lines indicate bin boundaries.
  }\label{tab:y2s8TeV}
  \begin{small}
    \begin{tabular*}{0.99\textwidth}{@{\hspace{1mm}}c@{\extracolsep{\fill}}ccccc@{\hspace{1mm}}}
      $\pt\left[\!\gevc\right]$  
      &  $2.0<y<2.5$
      &  $2.5<y<3.0$
      &  $3.0<y<3.5$
      &  $3.5<y<4.0$
      &  $4.0<y<4.5$  
      \\
      \hline 
$0-1$
 & $8.11 \pm 0.24 \pm 0.22$  & $7.90 \pm 0.13 \pm 0.13$  & $7.07 \pm 0.12 \pm 0.09$  & $5.58 \pm 0.11 \pm 0.05$  & $3.53 \pm 0.15 \pm 0.07$ \\
$1-2$
 & $21.8 \pm 0.4 \pm 0.2$  & $20.44 \pm 0.22 \pm 0.09$  & $17.55 \pm 0.19 \pm 0.12$  & $14.30 \pm 0.18 \pm 0.22$  & $8.23 \pm 0.24 \pm 0.18$ \\
$2-3$
 & $27.7 \pm 0.4 \pm 0.2$  & $26.53 \pm 0.25 \pm 0.23$  & $22.55 \pm 0.21 \pm 0.24$  & $17.80 \pm 0.20 \pm 0.08$  & $10.83 \pm 0.26 \pm 0.10$ \\
$3-4$
 & $29.9 \pm 0.4 \pm 0.4$  & $28.24 \pm 0.26 \pm 0.12$  & $23.24 \pm 0.21 \pm 0.27$  & $18.80 \pm 0.20 \pm 0.25$  & $10.62 \pm 0.25 \pm 0.17$ \\
$4-5$
 & $27.4 \pm 0.4 \pm 0.2$  & $26.00 \pm 0.25 \pm 0.13$  & $20.79 \pm 0.20 \pm 0.11$  & $16.57 \pm 0.19 \pm 0.15$  & $9.6 \pm 0.2 \pm 0.3$ \\
$5-6$
 & $23.5 \pm 0.4 \pm 0.2$  & $22.39 \pm 0.23 \pm 0.18$  & $18.16 \pm 0.19 \pm 0.20$  & $13.59 \pm 0.17 \pm 0.19$  & $8.26 \pm 0.21 \pm 0.03$ \\
$6-7$
 & $20.3 \pm 0.4 \pm 0.4$  & $18.62 \pm 0.21 \pm 0.13$  & $15.02 \pm 0.17 \pm 0.17$  & $11.13 \pm 0.16 \pm 0.20$  & $6.27 \pm 0.18 \pm 0.04$ \\
$7-8$
 & $16.7 \pm 0.3 \pm 0.2$  & $14.85 \pm 0.18 \pm 0.16$  & $11.87 \pm 0.15 \pm 0.14$  & $8.78 \pm 0.14 \pm 0.10$  & $5.06 \pm 0.16 \pm 0.08$ \\
$8-9$
 & $13.43 \pm 0.28 \pm 0.17$  & $11.79 \pm 0.16 \pm 0.16$  & $9.16 \pm 0.13 \pm 0.05$  & $6.86 \pm 0.12 \pm 0.03$  & $3.65 \pm 0.15 \pm 0.12$ \\
$9-10$
 & $10.16 \pm 0.24 \pm 0.15$  & $9.20 \pm 0.14 \pm 0.07$  & $7.14 \pm 0.11 \pm 0.04$  & $5.26 \pm 0.10 \pm 0.07$  & $2.49 \pm 0.13 \pm 0.02$ \\
$10-11$
 & $8.15 \pm 0.22 \pm 0.16$  & $6.97 \pm 0.12 \pm 0.11$  & $5.70 \pm 0.09 \pm 0.05$  & $3.87 \pm 0.09 \pm 0.07$  & $1.82 \pm 0.11 \pm 0.06$ \\
$11-12$
 & $6.55 \pm 0.20 \pm 0.15$  & $5.37 \pm 0.10 \pm 0.07$  & $4.15 \pm 0.08 \pm 0.09$  & $2.94 \pm 0.08 \pm 0.04$  & $1.11 \pm 0.09 \pm 0.03$ \\
$12-13$
 & $4.93 \pm 0.16 \pm 0.08$  & $4.19 \pm 0.09 \pm 0.05$  & $3.19 \pm 0.07 \pm 0.03$  & $2.25 \pm 0.07 \pm 0.03$  & $0.74 \pm 0.08 \pm 0.04$ \\  \cline{6-6}
$13-14$
 & $3.93 \pm 0.15 \pm 0.08$  & $3.18 \pm 0.08 \pm 0.04$  & $2.45 \pm 0.06 \pm 0.06$  & $1.66 \pm 0.06 \pm 0.03$  & \multirow{2}{*}{$0.84 \pm 0.09 \pm 0.04$} \\
$14-15$
 & $2.99 \pm 0.13 \pm 0.10$  & $2.48 \pm 0.07 \pm 0.04$  & $1.83 \pm 0.05 \pm 0.01$  & $1.27 \pm 0.05 \pm 0.05$  &  \\  \cline{6-6}
$15-16$
 & $2.36 \pm 0.11 \pm 0.07$  & $2.03 \pm 0.06 \pm 0.02$  & $1.42 \pm 0.05 \pm 0.03$  & $0.92 \pm 0.04 \pm 0.01$  & \multirow{5}{*}{$0.79 \pm 0.11 \pm 0.03$} \\
$16-17$
 & $1.89 \pm 0.10 \pm 0.02$  & $1.48 \pm 0.05 \pm 0.02$  & $1.09 \pm 0.04 \pm 0.01$  & $0.68 \pm 0.04 \pm 0.03$  &  \\
$17-18$
 & $1.31 \pm 0.08 \pm 0.03$  & $1.19 \pm 0.04 \pm 0.01$  & $0.86 \pm 0.04 \pm 0.01$  & $0.52 \pm 0.04 \pm 0.01$  &  \\
$18-19$
 & $1.15 \pm 0.07 \pm 0.02$  & $0.92 \pm 0.04 \pm 0.01$  & $0.69 \pm 0.03 \pm 0.01$  & $0.39 \pm 0.03 \pm 0.02$  &  \\
$19-20$
 & $0.89 \pm 0.06 \pm 0.01$  & $0.70 \pm 0.03 \pm 0.01$  & $0.46 \pm 0.03 \pm 0.01$  & $0.32 \pm 0.02 \pm 0.01$  &  \\  \cline{2-6}
$20-21$
 & \multirow{2}{*}{$1.34 \pm 0.08 \pm 0.01$}  & \multirow{2}{*}{$1.08 \pm 0.04 \pm 0.02$}  & \multirow{2}{*}{$0.69 \pm 0.03 \pm 0.02$}  & \multirow{5}{*}{$0.59 \pm 0.04 \pm 0.02$} &  \\
$21-22$
 &   &   &   &  &  \\  \cline{2-4}
$22-23$
 & \multirow{2}{*}{$0.76 \pm 0.06 \pm 0.01$}  & \multirow{2}{*}{$0.64 \pm 0.03 \pm 0.01$}  & \multirow{2}{*}{$0.41 \pm 0.03 \pm 0.01$}  &  &  \\
$23-24$
 &   &   &   &  &  \\  \cline{2-4}
$24-25$
 & \multirow{2}{*}{$0.63 \pm 0.05 \pm 0.01$}  & \multirow{2}{*}{$0.41 \pm 0.03 \pm 0.01$}  & \multirow{2}{*}{$0.28 \pm 0.02 \pm 0.01$}  &  &  \\  \cline{5-5}
$25-26$
 &   &   &  &  &  \\  \cline{2-4}
$26-27$
 & \multirow{4}{*}{$0.55 \pm 0.05 \pm 0.02$}  & \multirow{2}{*}{$0.29 \pm 0.02 \pm 0.01$}  & \multirow{4}{*}{$0.27 \pm 0.02 \pm 0.01$} &  &  \\
$27-28$
 &   &   &  &  &  \\  \cline{3-3}
$28-29$
 &   & \multirow{2}{*}{$0.19 \pm 0.02 \pm 0.01$}  &  &  &  \\
$29-30$
 &   &   &  &  &  
    \end{tabular*}   
  \end{small}
\end{sidewaystable}

\begin{sidewaystable}[p]
  \centering
  \caption{\small 
    Production cross-section $\upsigma^{\YthreeS\to\mumu}_{\mathrm{bin}}~\left[\!\pb\right]$~in $(\pt,y)$~bins
    for $\sqrt{s}=8\,\mathrm{TeV}$. The~first uncertainties are statistical and the~second 
    are the~uncorrelated component of the~systematic uncertainty. 
    The~overall correlated systematic uncertainty is 2.8\% 
    and is not included in the~numbers in the~table.
    The horizontal lines indicate the~bin boundaries.
  }\label{tab:y3s8TeV}
  \begin{small}
    \begin{tabular*}{0.99\textwidth}{@{\hspace{1mm}}c@{\extracolsep{\fill}}ccccc@{\hspace{1mm}}}
      $\pt\left[\!\gevc\right]$  
      &  $2.0<y<2.5$
      &  $2.5<y<3.0$
      &  $3.0<y<3.5$
      &  $3.5<y<4.0$
      &  $4.0<y<4.5$  
      \\
      \hline 
$0-1$
 & $3.30 \pm 0.17 \pm 0.09$  & $3.29 \pm 0.10 \pm 0.04$  & $2.72 \pm 0.09 \pm 0.05$  & $2.42 \pm 0.08 \pm 0.02$  & $1.47 \pm 0.11 \pm 0.03$ \\
$1-2$
 & $8.19 \pm 0.27 \pm 0.01$  & $8.45 \pm 0.16 \pm 0.06$  & $7.18 \pm 0.14 \pm 0.02$  & $5.83 \pm 0.13 \pm 0.13$  & $3.43 \pm 0.17 \pm 0.11$ \\
$2-3$
 & $10.73 \pm 0.30 \pm 0.14$  & $11.16 \pm 0.18 \pm 0.07$  & $9.05 \pm 0.15 \pm 0.14$  & $7.56 \pm 0.14 \pm 0.03$  & $4.94 \pm 0.19 \pm 0.05$ \\
$3-4$
 & $12.44 \pm 0.31 \pm 0.07$  & $12.00 \pm 0.18 \pm 0.04$  & $9.99 \pm 0.16 \pm 0.08$  & $7.98 \pm 0.15 \pm 0.10$  & $4.69 \pm 0.18 \pm 0.08$ \\
$4-5$
 & $11.37 \pm 0.30 \pm 0.07$  & $11.42 \pm 0.18 \pm 0.01$  & $9.51 \pm 0.15 \pm 0.05$  & $7.70 \pm 0.14 \pm 0.05$  & $4.48 \pm 0.17 \pm 0.20$ \\
$5-6$
 & $10.06 \pm 0.27 \pm 0.04$  & $10.21 \pm 0.17 \pm 0.07$  & $8.53 \pm 0.14 \pm 0.09$  & $6.64 \pm 0.13 \pm 0.12$  & $3.68 \pm 0.15 \pm 0.01$ \\
$6-7$
 & $9.35 \pm 0.26 \pm 0.16$  & $8.60 \pm 0.15 \pm 0.03$  & $7.36 \pm 0.13 \pm 0.07$  & $5.66 \pm 0.12 \pm 0.12$  & $3.13 \pm 0.14 \pm 0.01$ \\
$7-8$
 & $7.83 \pm 0.23 \pm 0.06$  & $7.48 \pm 0.14 \pm 0.05$  & $6.14 \pm 0.11 \pm 0.08$  & $4.79 \pm 0.11 \pm 0.04$  & $2.48 \pm 0.12 \pm 0.04$ \\
$8-9$
 & $6.66 \pm 0.21 \pm 0.05$  & $6.13 \pm 0.12 \pm 0.08$  & $4.91 \pm 0.10 \pm 0.03$  & $3.64 \pm 0.09 \pm 0.01$  & $1.75 \pm 0.11 \pm 0.05$ \\
$9-10$
 & $5.29 \pm 0.19 \pm 0.07$  & $4.81 \pm 0.11 \pm 0.04$  & $3.99 \pm 0.09 \pm 0.02$  & $3.00 \pm 0.08 \pm 0.05$  & $1.24 \pm 0.09 \pm 0.01$ \\
$10-11$
 & $4.11 \pm 0.17 \pm 0.08$  & $3.98 \pm 0.09 \pm 0.08$  & $3.19 \pm 0.07 \pm 0.03$  & $2.42 \pm 0.07 \pm 0.05$  & $1.10 \pm 0.09 \pm 0.07$ \\
$11-12$
 & $3.27 \pm 0.15 \pm 0.09$  & $3.16 \pm 0.08 \pm 0.04$  & $2.49 \pm 0.06 \pm 0.07$  & $1.73 \pm 0.06 \pm 0.02$  & $0.69 \pm 0.07 \pm 0.02$ \\
$12-13$
 & $2.91 \pm 0.13 \pm 0.04$  & $2.65 \pm 0.07 \pm 0.04$  & $1.95 \pm 0.06 \pm 0.02$  & $1.41 \pm 0.06 \pm 0.02$  & $0.46 \pm 0.07 \pm 0.01$ \\  \cline{6-6}
$13-14$
 & $2.41 \pm 0.12 \pm 0.04$  & $2.07 \pm 0.06 \pm 0.03$  & $1.52 \pm 0.05 \pm 0.04$  & $1.05 \pm 0.05 \pm 0.02$  & \multirow{2}{*}{$0.60 \pm 0.09 \pm 0.02$} \\
$14-15$
 & $1.93 \pm 0.11 \pm 0.07$  & $1.67 \pm 0.06 \pm 0.04$  & $1.17 \pm 0.04 \pm 0.01$  & $0.83 \pm 0.04 \pm 0.03$  &  \\  \cline{6-6}
$15-16$
 & $1.52 \pm 0.09 \pm 0.04$  & $1.21 \pm 0.05 \pm 0.02$  & $0.90 \pm 0.04 \pm 0.02$  & $0.61 \pm 0.04 \pm 0.01$  & \multirow{5}{*}{$0.46 \pm 0.08 \pm 0.01$} \\
$16-17$
 & $1.10 \pm 0.08 \pm 0.02$  & $0.97 \pm 0.04 \pm 0.01$  & $0.76 \pm 0.04 \pm 0.01$  & $0.42 \pm 0.03 \pm 0.02$  &  \\
$17-18$
 & $0.89 \pm 0.07 \pm 0.02$  & $0.77 \pm 0.04 \pm 0.01$  & $0.56 \pm 0.03 \pm 0.01$  & $0.40 \pm 0.032 \pm 0.01$  &  \\
$18-19$
 & $0.79 \pm 0.06 \pm 0.01$  & $0.58 \pm 0.03 \pm 0.01$  & $0.43 \pm 0.03 \pm 0.01$  & $0.31 \pm 0.029 \pm 0.01$  &  \\
$19-20$
 & $0.59 \pm 0.05 \pm 0.01$  & $0.49 \pm 0.03 \pm 0.01$  & $0.32 \pm 0.02 \pm 0.01$  & $0.20 \pm 0.02 \pm 0.01$  &  \\  \cline{2-6}
$20-21$
 & \multirow{2}{*}{$0.84 \pm 0.06 \pm 0.01$}  & \multirow{2}{*}{$0.73 \pm 0.04 \pm 0.02$}  & \multirow{2}{*}{$0.46 \pm 0.03 \pm 0.01$}  & \multirow{5}{*}{$0.46 \pm 0.04 \pm 0.02$} &  \\
$21-22$
 &   &   &   &  &  \\  \cline{2-4}
$22-23$
 & \multirow{2}{*}{$0.51 \pm 0.05 \pm 0.01$}  & \multirow{2}{*}{$0.46 \pm 0.03 \pm 0.04$}  & \multirow{2}{*}{$0.32 \pm 0.02 \pm 0.01$}  &  &  \\
$23-24$
 &   &   &   &  &  \\  \cline{2-4}
$24-25$
 & \multirow{2}{*}{$0.34 \pm 0.04 \pm 0.01$}  & \multirow{2}{*}{$0.30 \pm 0.02 \pm 0.01$}  & \multirow{2}{*}{$0.21 \pm 0.03 \pm 0.01$}  &  &  \\  \cline{5-5}
$25-26$
 &   &   &  &  &  \\  \cline{2-4}
$26-27$
 & \multirow{4}{*}{$0.52 \pm 0.05 \pm 0.02$}  & \multirow{2}{*}{$0.18 \pm 0.02 \pm 0.01$}  & \multirow{4}{*}{$0.20 \pm 0.02 \pm 0.01$} &  &  \\
$27-28$
 &   &   &  &  &  \\  \cline{3-3}
$28-29$
 &   & \multirow{2}{*}{$0.12 \pm 0.02 \pm 0.01$}  &  &  &  \\
$29-30$
 &   &   &  &  &  
    \end{tabular*}   
  \end{small}
\end{sidewaystable}

\begin{figure}[t]
  \setlength{\unitlength}{1mm}
  \centering
  \begin{picture}(150,60)
    %
    \put( 0,  0){ 
      \includegraphics*[width=75mm,height=60mm,%
      ]{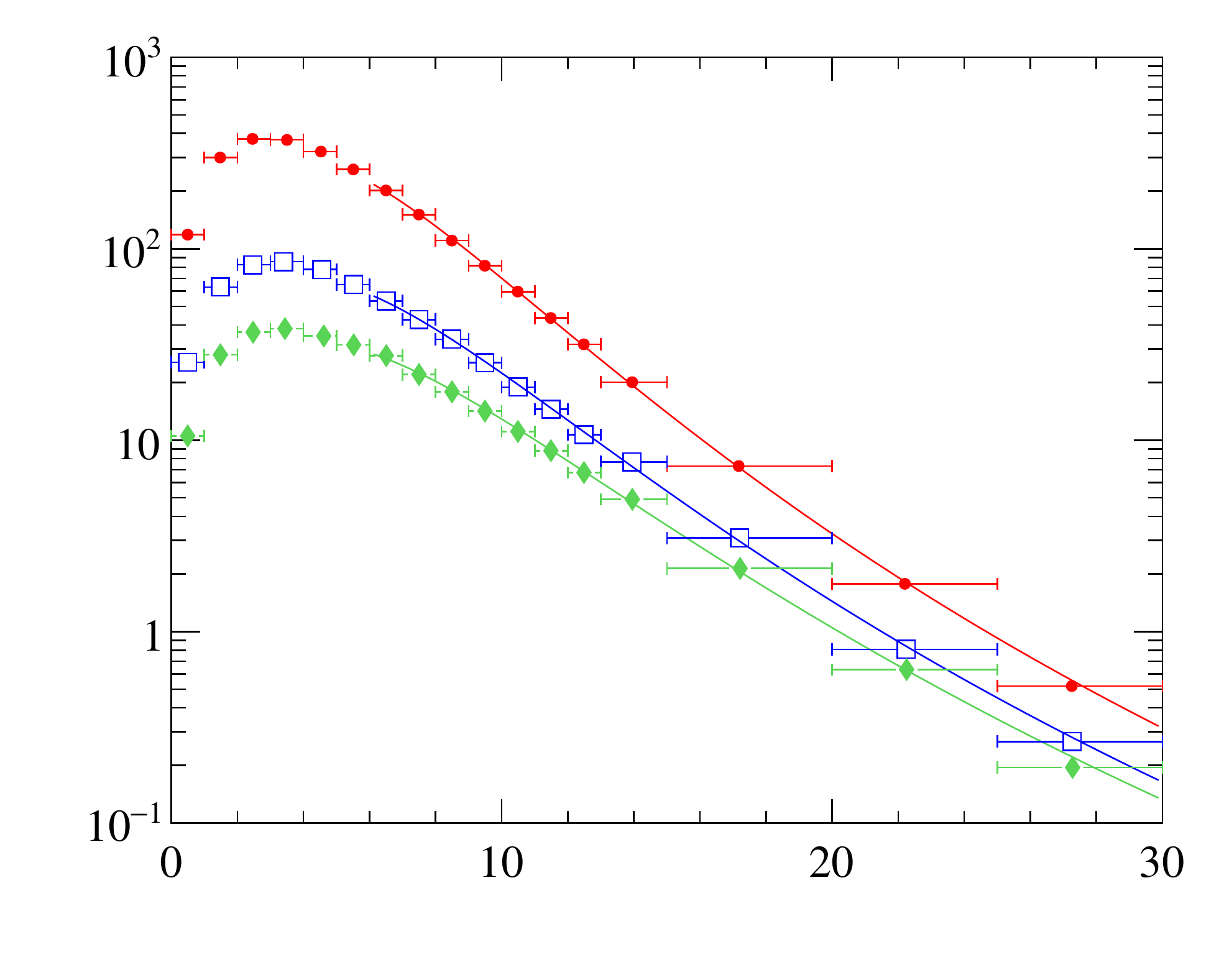}
    }
    \put(75,  0){ 
      \includegraphics*[width=75mm,height=60mm,%
      ]{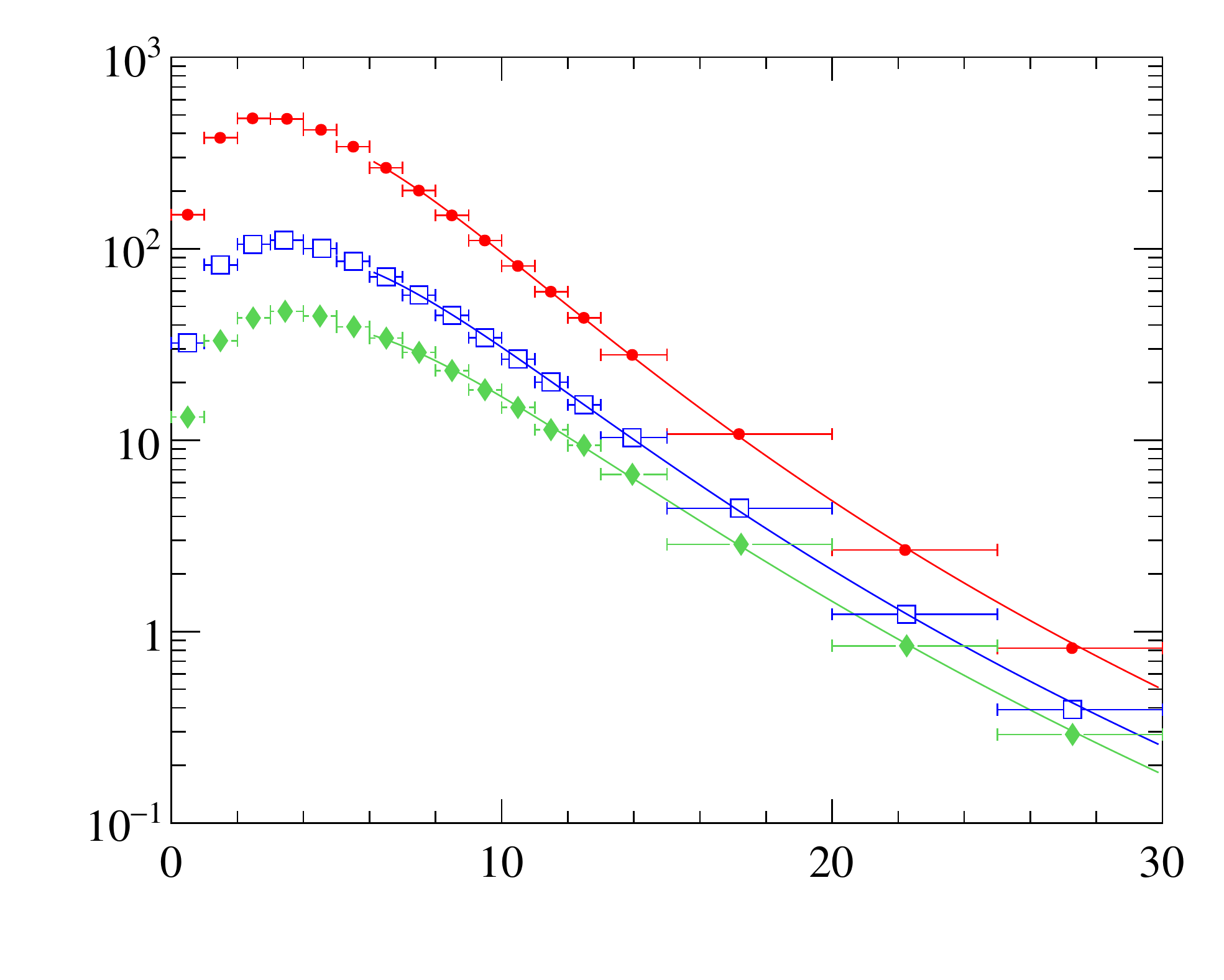}
    }
    \put( -2, 21) { \begin{sideways}
        $\frac{\deriv}{\deriv\pt}\upsigma^{\ups\to\mumu}~\left[\tfrac{\pb}{\!\gevc}\right]$
      \end{sideways}}
    \put( 73, 21) { \begin{sideways}
        $\frac{\deriv}{\deriv\pt}\upsigma^{\ups\to\mumu}~\left[\tfrac{\pb}{\!\gevc}\right]$
      \end{sideways}}
    \put( 34, 48) { $\begin{array}{r} \text{LHCb}~\sqrt{s}=7\tev \\  2.0<y<4.5  \end{array}$}
    \put(109, 48) { $\begin{array}{r} \text{LHCb}~\sqrt{s}=8\tev \\  2.0<y<4.5  \end{array}$}
    \put( 40,  2) { $\pt$} \put( 59,  2){ $\left[\!\gevc\right]$}
    \put(115,  2) { $\pt$} \put(134,  2){ $\left[\!\gevc\right]$}
    \put ( 53,36){\footnotesize $\begin{array}{cc}  
        {\color{red}\bullet}                    & \YoneS \\
        {\color{blue}\square}                   & \YtwoS \\
        {\color[rgb]{0.33,0.83,0.33}\blacklozenge} & \YthreeS \end{array}$}
    \put (128,36){\footnotesize $\begin{array}{cc}  
        {\color{red}\bullet}                    & \YoneS \\
        {\color{blue}\square}                   & \YtwoS \\
        {\color[rgb]{0.33,0.83,0.33}\blacklozenge} & \YthreeS \end{array}$}
  \end{picture}
  \caption { \small
    Differential cross-sections  
    $\frac{\deriv}{\deriv\pt}\upsigma^{\ups\to\mumu}$
    in the~range \mbox{$2.0<y<4.5$}
    for 
    (red solid circles)~\YoneS,
    (blue open squares)~\YtwoS
    and 
    (green solid diamonds)~\YthreeS~mesons
    for 
    (left)~\mbox{$\sqrt{s}=7\,\mathrm{TeV}$} and 
    (right)~\mbox{$\sqrt{s}=8\,\mathrm{TeV}$}~data.
    The~curves show the~fit results with 
    the~Tsallis~function
    in the~range \mbox{$6<\pt<30\gevc$}.    
    The data~points are positioned in 
    the~bins according to Eq.\,(6) in~Ref.~\cite{Lafferty:1994cj}.
  }
  \label{fig:diff_xsecPT} 
\end{figure}

\begin{figure}[t]
  \setlength{\unitlength}{1mm}
  \centering
  \begin{picture}(150,60)
    %
    \put( 0,  0){ 
      \includegraphics*[width=75mm,height=60mm,%
      ]{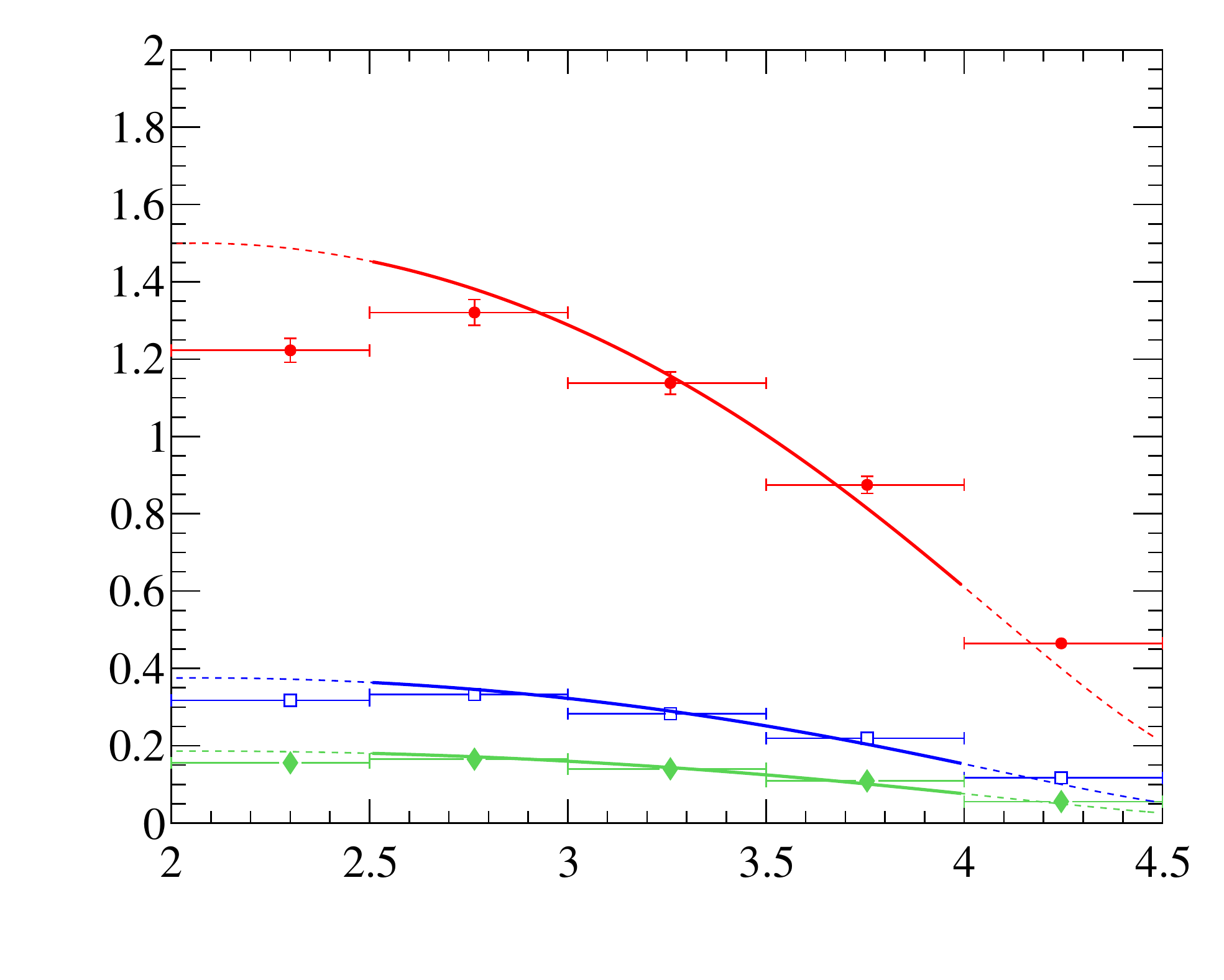}
    }
    \put(75,  0){ 
      \includegraphics*[width=75mm,height=60mm,%
      ]{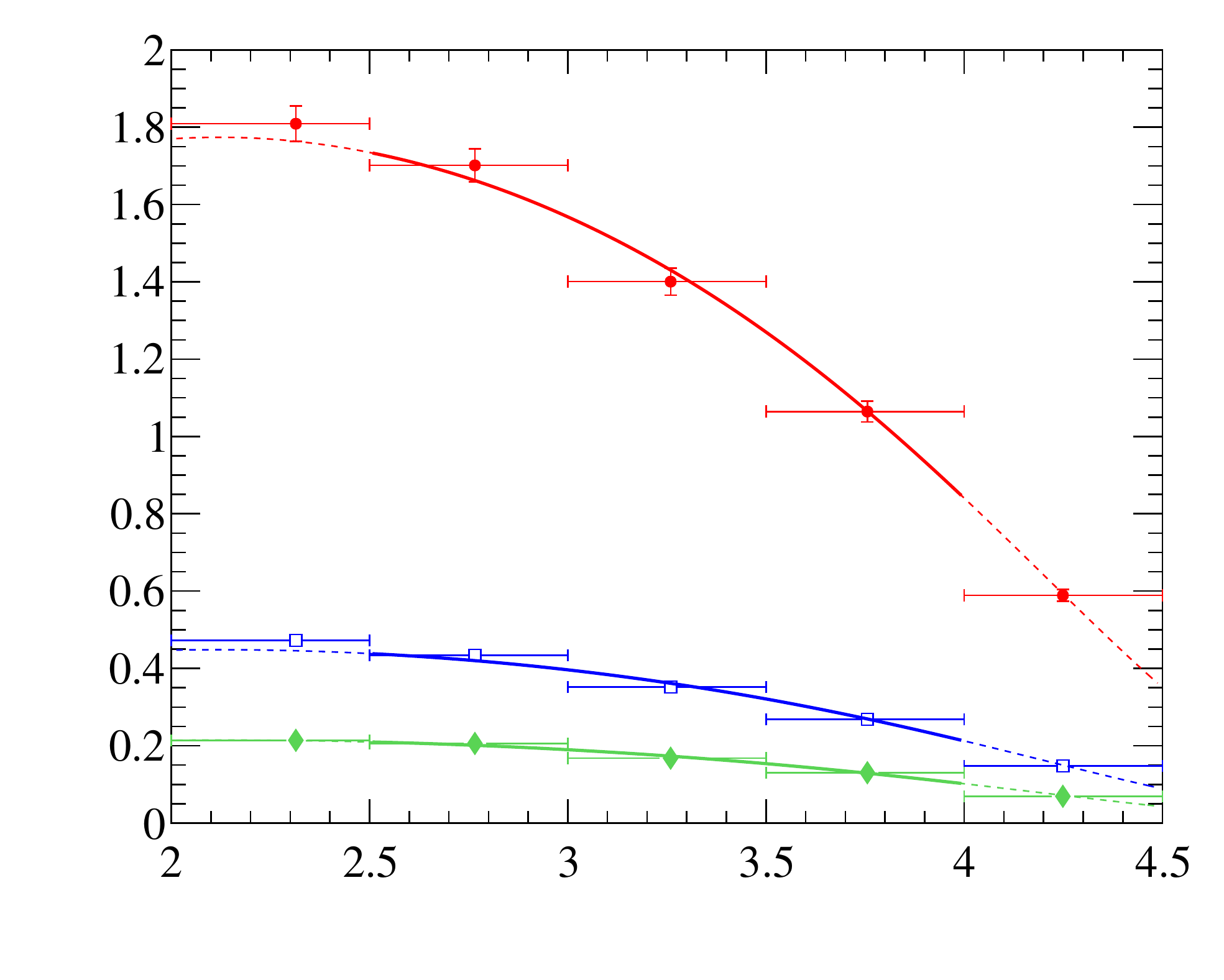}
    }
    \put( -1, 27) { \begin{sideways}
        $\frac{\deriv}{\deriv y}\upsigma^{\ups\to\mumu}~\left[\tfrac{\nb}{0.5}\right]$
      \end{sideways}}
    \put( 74, 27) { \begin{sideways}
        $\frac{\deriv}{\deriv y}\upsigma^{\ups\to\mumu}~\left[\tfrac{\nb}{0.5}\right]$
      \end{sideways}}
    \put( 35.5, 50.5) { $\text{LHCb}~\sqrt{s}=7\tev$}
    \put(110.5, 50.5) { $\text{LHCb}~\sqrt{s}=8\tev$}
    \put( 40,  2) { $y$} 
    \put(115,  2) { $y$} 
    \put ( 53,41){\footnotesize $\begin{array}{cc}  
        {\color{red}\bullet}                    & \YoneS \\
        {\color{blue}\square}                   & \YtwoS \\
        {\color[rgb]{0.33,0.83,0.33}\blacklozenge} & \YthreeS \end{array}$}
    \put (128,41){\footnotesize $\begin{array}{cc}  
        {\color{red}\bullet}                    & \YoneS \\
        {\color{blue}\square}                   & \YtwoS \\
        {\color[rgb]{0.33,0.83,0.33}\blacklozenge} & \YthreeS \end{array}$}
  \end{picture}
  \caption { \small
    Differential cross-sections  
    $\frac{\deriv}{\deriv y}\upsigma^{\ups\to\mumu}$
    in the~range \mbox{$\pt<30\gevc$}
    for 
    (red solid circles)~\YoneS,
    (blue open squares)~\YtwoS and 
    (green solid diamonds)~\YthreeS~mesons
    for 
    (left)~\mbox{$\sqrt{s}=7\,\mathrm{TeV}$} and 
    (right)~\mbox{$\sqrt{s}=8\,\mathrm{TeV}$}~data.
    Thick lines show fit results with 
    the~CO~model predictions
    from Ref.~\cite{ Kisslinger:2011fe,Kisslinger:2013mev,*Kisslinger:2014zga}
    in the~region~\mbox{$2.5<y<4.0$},
    and dashed lines show the~extrapolation 
    to the~full region~\mbox{$2.0<y<4.5$}.
    The~data points are positioned in 
    the~bins according to Eq.\,(6) in~Ref.~\cite{Lafferty:1994cj}.
  }
  \label{fig:diff_xsecY} 
\end{figure}

The double-differential production cross-sections multiplied by the~dimuon branching 
fractions for the~\ups~mesons are shown in Fig.~\ref{fig:diff_xsec}.
The corresponding production cross-section $\upsigma^{\ups\to\mumu}_{\mathrm{bin}}$
in $(\pt,y)$~bins are presented in 
Tables~\ref{tab:y1s7TeV}, \ref{tab:y2s7TeV} and~\ref{tab:y3s7TeV} for
$\sqs=7\tev$ and 
Tables~\ref{tab:y1s8TeV}, \ref{tab:y2s8TeV} and~\ref{tab:y3s8TeV} for
$\sqs=8\tev$.
The cross-sections 
integrated over~$y$~as a~function of $\pt$ 
and  integrated over~$\pt$~as a~function of rapidity 
are  shown in Figs.~\ref{fig:diff_xsecPT} 
and~\ref{fig:diff_xsecY}, 
respectively.

The transverse momentum spectra are fit using 
a~Tsallis~function~\cite{Tsallis:1987eu}
\begin{equation}
  \dfrac
  {\deriv\upsigma}
  {\pt\,\deriv\pt} \propto
  \left( 1 + \dfrac{ E^{\mathrm{kin}}_{\mathrm{T}}} {n\,T}\right)^{-n},
\end{equation}
where 
\mbox{$E^{\mathrm{kin}}_{\mathrm{T}}\equiv \sqrt{m_{\ups}^2 + p_{\mathrm{T}}^2}-m_{\ups}$}
is the~transverse kinetic energy, 
the~power $n$ and the~temperature parameter~$T$~are free parameters,
and $m_{\ups}$~is the~known mass of a~\ups~meson~\cite{PDG2014}.
This~function has a~power-law  
asymptotic behaviour 
$\propto p^{-n}_{\mathrm{T}}$ for high~\pt 
as~expected for hard scattering processes.
It~has been successfully applied to fit 
\pt~spectra~\cite{LHCb-PAPER-2015-032,Chatrchyan:2012xg,Zheng:2015tua,Marques:2015mwa} 
in wide ranges of
particle species, processes and kinematics.
A~fit with the~Tsallis~distribution 
for the~range $6<\pt<30\gevc$ 
is superimposed on the~differential cross-sections 
in Fig.~\ref{fig:diff_xsecPT}. The~fit quality is good for all cases. 
The~fitted values of the parameters $n$ and $T$ are 
listed in Table~\ref{tab:tsallis1}.
The~parameter $n$ for all cases is close to~8, 
compatible with 
the~high~\pt~asymptotic behaviour expected by 
the~CS model~\cite{Kartvelishvili:1978id,Berger:1980ni,Baier:1981uk,CSM,CSM1}.
The~temperature parameters $T$~show little dependence on $\sqs$ and increase
with the~mass of \ups~state.

\begin{table}[t]
  \centering
  \caption{\small 
    Results of the~fits to 
    the~transverse momentum spectra of \ups~mesons 
    using the~Tsallis function  in the~reduced range \mbox{$6<\pt<30\gevc$}.
  }\label{tab:tsallis1}
  \vspace*{3mm}
  \begin{tabular*}{0.75\textwidth}{@{\hspace{2mm}}c@{\extracolsep{\fill}}ccc@{\hspace{2mm}}}
    & $\sqs$  & $T~\left[\!\gev\right]$  & $n$
    \\[1mm]
    \hline 
    \\[-2mm]    
    \YoneS   
    & $\begin{array}{c} 7\tev         \\ 8\tev       \end{array}$
    & $\begin{array}{c} 1.19 \pm 0.04 \\ 1.20\pm0.04 \end{array}$  
    & $\begin{array}{c} 8.01 \pm 0.33 \\ 7.71\pm0.27 \end{array}$  
    \\
    \YtwoS   
    & $\begin{array}{c} 7\tev        \\ 8\tev       \end{array}$  
    & $\begin{array}{c} 1.33\pm0.05  \\ 1.37\pm0.05 \end{array}$  
    & $\begin{array}{c} 7.57\pm0.41  \\ 7.53\pm0.34 \end{array}$  
    \\
    \YthreeS   
    & $\begin{array}{c} 7\tev       \\ 8\tev       \end{array}$  
    & $\begin{array}{c} 1.53\pm0.07 \\ 1.63\pm0.06 \end{array}$  
    & $\begin{array}{c} 7.85\pm0.56 \\ 8.23\pm0.51 \end{array}$  
  \end{tabular*}   
\end{table}

The~shapes of the~rapidity spectra
are compared with the~CO model prediction
in the~region $2.5<y<4.0$ and are fitted using 
the~function given by Eq.~(1) of
Ref.~\cite{Kisslinger:2013mev,*Kisslinger:2014zga},
with free normalisation constants.
The~fit result, as well as the~extrapolation to  
the~full kinematic range $2.0<y<4.5$, is presented in 
Fig.~\ref{fig:diff_xsecY}.
The~quality of the fit is good for all~cases.

The~integrated production cross-sections multiplied by the~dimuon branching fractions 
in the~full range 
\mbox{$\pt <30\gevc$} and 
\mbox{$2.0<y<4.5$} at 
\mbox{$\sqs=7$} and~\mbox{$8\tev$}
are reported in~\mbox{Table}~\ref{tab:xsec_int}, where the~first 
uncertainties are statistical and the~second systematic.
The~same measurements are also shown integrated over 
the~reduced range \mbox{$\pt<15\gevc$} in the~same rapidity range, 
to allow the~comparison with previous 
measurements~\cite{LHCb-PAPER-2011-036,LHCb-PAPER-2013-016}.

\begin{table}[t]
  \centering
  \caption{ \small 
    The production cross-section
    $\upsigma^{\ups\to\mumu}$\,(in $\pb$) 
    for \ups~mesons 
    in the~full kinematic range~\mbox{$\pt<30\gevc$}\,(left two columns), 
    and reduced range~\mbox{$\pt<15\gevc$}\,(right two columns),  
    for~\mbox{$2.0<y<4.5$}.
    The first uncertainties are statistical 
    and the~second systematic. 
  } \label{tab:xsec_int}
  \vspace*{3mm}
  \begin{tabular*}{0.99\textwidth}{@{\hspace{1mm}}c@{\extracolsep{\fill}}cccc@{\hspace{1mm}}}
    & \multicolumn{2}{c}{$\pt<30\gevc$}
    & \multicolumn{2}{c}{$\pt<15\gevc$}
    \\[1mm]
    \cline {2-5}
    \\[-2mm]
    & $\sqs=7\tev$
    & $\sqs=8\tev$
    & $\sqs=7\tev$
    & $\sqs=8\tev$
    \\[1mm]
    \hline
    \\[-2mm]
    $\upsigma^{\YoneS\to\mumu}$
    &  $2510 \pm 3 \pm 80\phantom{0}$  
    &  $3280 \pm 3 \pm 100$ 
    &  $2460 \pm 3 \pm 80\phantom{0}$ 
    &  $3210 \pm 3 \pm 90\phantom{0}$  
    \\
    $\upsigma^{\YtwoS\to\mumu}$
    &  $\phantom{0}635 \pm 2 \pm 20\phantom{0}$  
    &  $\phantom{0}837 \pm 2 \pm 25\phantom{0}$  
    &  $\phantom{0}614 \pm 2 \pm 20\phantom{0}$ 
    &  $\phantom{0}807 \pm 2 \pm 24\phantom{0}$
    \\
    $\upsigma^{\YthreeS\to\mumu}$
    & $\phantom{0}313 \pm 2 \pm 10\phantom{0}$   
    & $\phantom{0}393 \pm 1 \pm 12\phantom{0}$ 
    & $\phantom{0}298 \pm 1 \pm 10\phantom{0}$  
    & $\phantom{0}373 \pm 1 \pm 11\phantom{0}$
  \end{tabular*}   
\end{table}

The ratios of integrated production cross-section 
$\mathscr{R}_{8/7}$  are presented in Table~\ref{tab:rsq}
for the~full\,\mbox{($\pt<30\gevc$)} and reduced\,\mbox{($\pt<15\gevc$)}~ranges.
The~results for the~reduced range are consistent 
with the~previous measurements,
confirming the~increase of the~bottomonium production cross-section
of approximately 30\% when the~centre-of-mass energy increases from  
$\sqs=7$~to~$8\tev$~\cite{LHCb-PAPER-2011-036,LHCb-PAPER-2013-016}.

\begin{table}[t]
  \centering
  \caption{ \small 
    The ratio of production cross-sections
    for \ups~mesons at \mbox{$\sqrt{s}=8$} to that 
    at \mbox{$\sqrt{s}=7\,\mathrm{TeV}$}
    in the~full kinematic range~\mbox{$\pt<30\gevc$}\,(left) 
    and reduced range~\mbox{$\pt<15\gevc$}\,(right)  
    for~\mbox{$2.0< y<4.5$}.
    The first uncertainties are statistical 
    and the second systematic. 
  } \label{tab:rsq}
  \vspace*{3mm}
  \begin{tabular*}{0.75\textwidth}{@{\hspace{2mm}}c@{\extracolsep{\fill}}cc@{\hspace{2mm}}}
    & $\pt<30\gevc$
    & $\pt<15\gevc$ 
    \\[1mm]
    \hline
    \\[-2mm]
    $\YoneS$
    & $1.307 \pm 0.002 \pm 0.025$ &  $1.304 \pm 0.002 \pm 0.024 $  
    \\
    $\YtwoS$
    & $1.319 \pm 0.005 \pm 0.025$ &  $1.315 \pm 0.005 \pm 0.024 $  
    \\
    $\YthreeS$
    & $1.258 \pm 0.007 \pm 0.024$ &  $1.254 \pm 0.007 \pm 0.023 $ 
  \end{tabular*}   
\end{table}

\begin{figure}[t]
  \setlength{\unitlength}{1mm}
  \centering
  \begin{picture}(150,60)
    \put(  0, 0){ 
      \includegraphics*[width=75mm,height=60mm,%
      ]{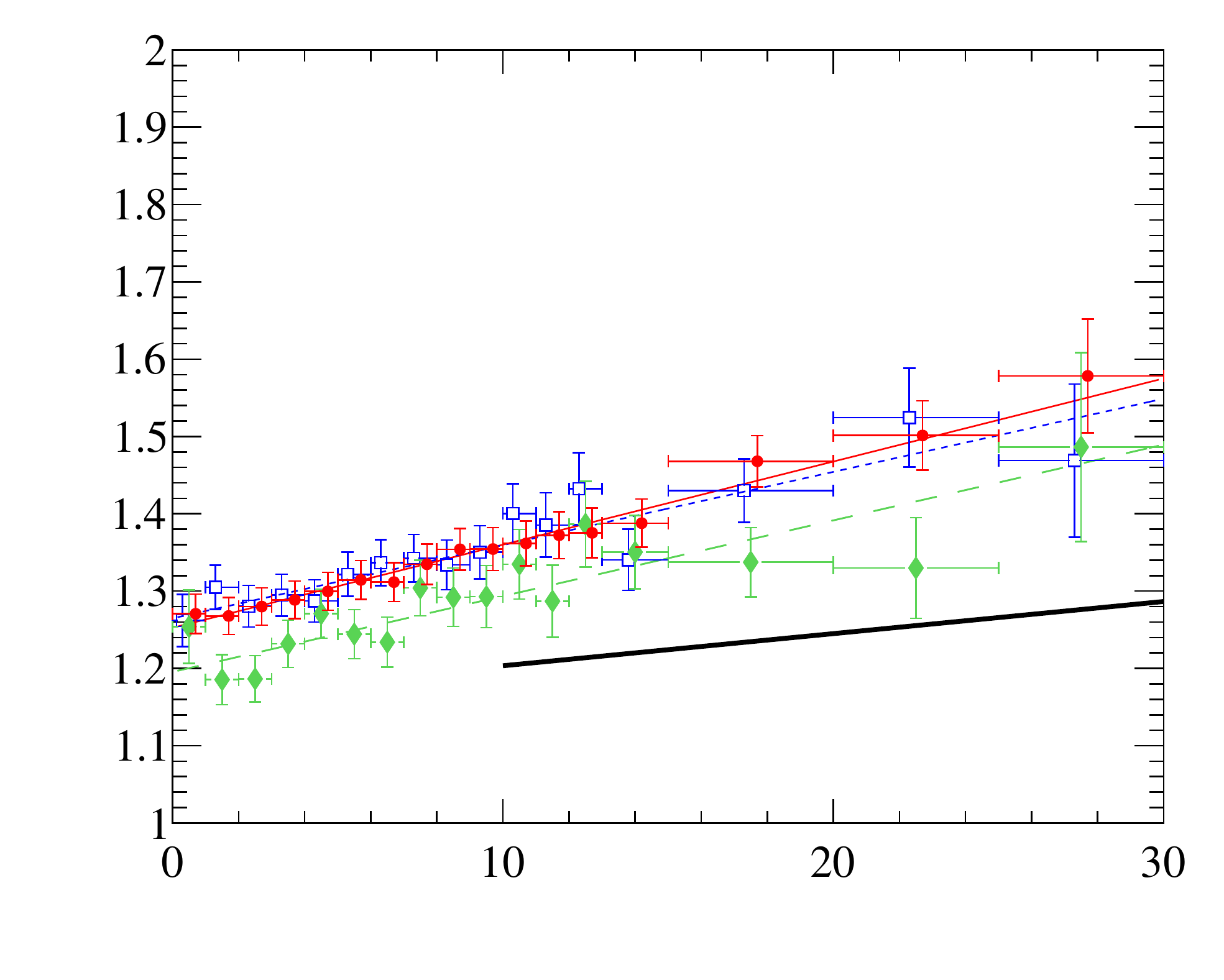}
    }
    \put( 75, 0){ 
      \includegraphics*[width=75mm,height=60mm,%
      ]{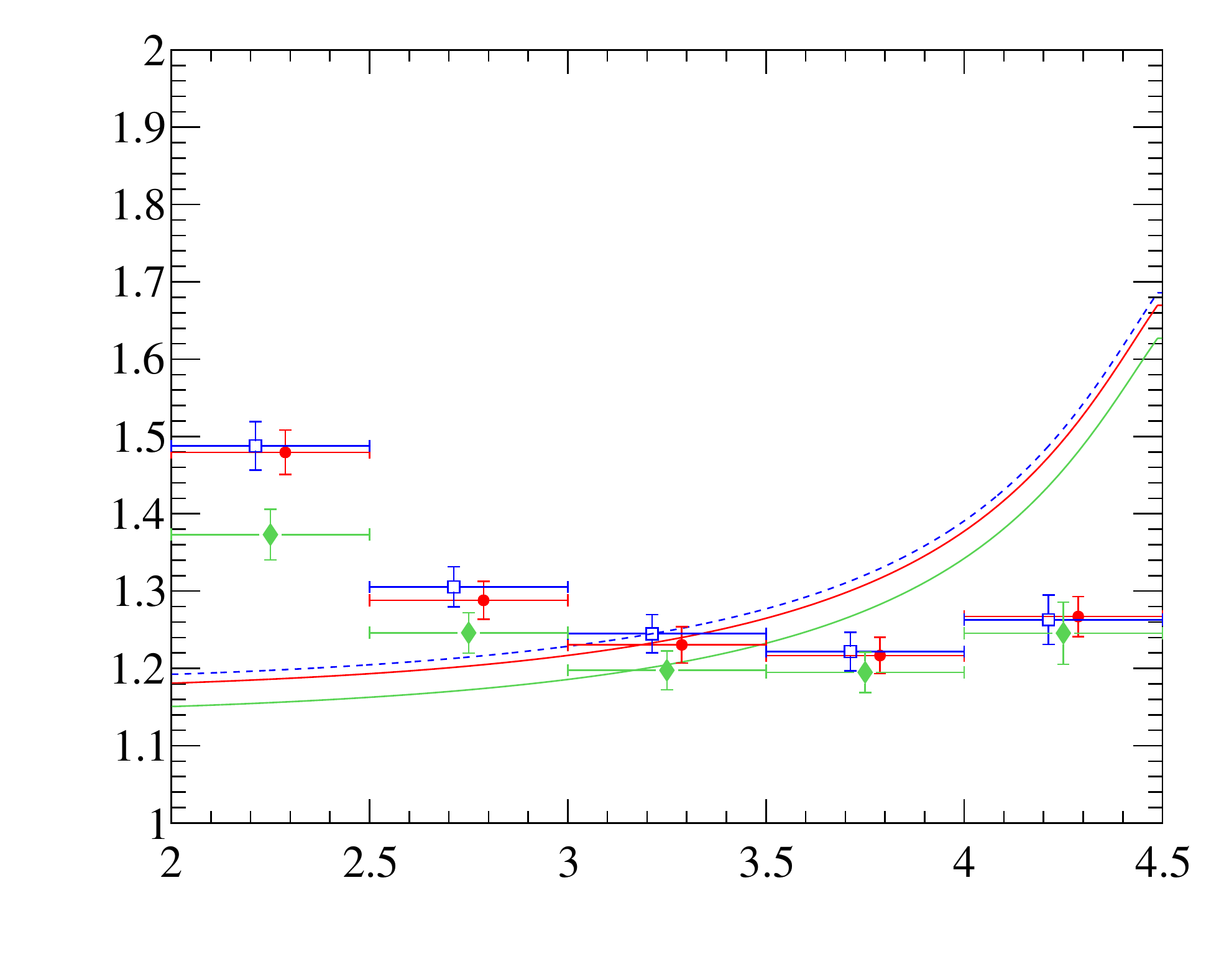}
    }
    \put(  1, 42) {\begin{sideways} $\mathscr{R}_{8/7}(\pt)$ \end{sideways} }
    \put( 76, 44) {\begin{sideways} $\mathscr{R}_{8/7}(y)$   \end{sideways} }
    \put( 14, 48) { $\begin{array}{l} \text{LHCb} \\  2.0<y<4.5  \end{array}$}
    \put( 89, 48) { $\begin{array}{l} \text{LHCb} \\ \pt<30\gevc \end{array}$}
    \put( 40,  2) { $\pt$} \put(59,  2){ $\left[\!\gevc\right]$}
    \put(115,  2) { $y$  } 
    \put ( 53,48){\footnotesize $\begin{array}{cl}  
        {\color{red}\bullet}                       & \YoneS \\
        {\color{blue}\square}                      & \YtwoS \\
        {\color[rgb]{0.33,0.83,0.33}\blacklozenge} & \YthreeS \end{array}$}
    \put (128,48){\footnotesize $\begin{array}{cc}  
        {\color{red}\bullet}                       & \YoneS \\
        {\color{blue}\square}                      & \YtwoS \\
        {\color[rgb]{0.33,0.83,0.33}\blacklozenge} & \YthreeS \end{array}$}
  \end{picture}
  \caption { \small
    Ratios of the differential cross-sections  
    (left)~\mbox{$\frac{\deriv}{\deriv\pt}\upsigma^{\ups\to\mumu}$} and 
    (right)~\mbox{$\frac{\deriv}{\deriv y}\upsigma^{\ups\to\mumu}$}
    at $\sqrt{s}=8$~and~$7\,\mathrm{TeV}$
    for
    (red solid circles)~\YoneS,
    (blue open squares)~\YtwoS and 
    (green solid diamonds)~\YthreeS.
    On~the~left hand plot, the~results of the~fit with a~linear function
    are shown with straight thin red solid, 
    blue dotted and green dashed lines.
    In~the~same plot, 
    the~next\nobreakdash-to\nobreakdash-leading order NRQCD theory predictions~\cite{Han:2014kxa}
    are shown as a~thick line.
    On~the~right hand plot, the~curved 
    red solid, 
    blue dotted and greed dashed 
    lines 
    show the~CO~model predictions~\cite{Kisslinger:2011fe,Kisslinger:2013mev,*Kisslinger:2014zga}
    with the~normalisation fixed from the~fits in~Fig.~\ref{fig:diff_xsecY}
    for \YoneS, \YtwoS and \YthreeS~mesons, respectively.
    Some data points are displaced from 
    the~bin centres to improve visibility.
  }
  \label{fig:resultsratio3}
\end{figure}

The~ratios $\mathscr{R}_{8/7}$ as a~function of \pt 
integrated over the~region~$2.0<y<4.5$ are shown 
in Fig.~\ref{fig:resultsratio3}a. 
The~ratios are fitted with a~linear function.
The~fit quality is good, with a~$p$-value exceeding
35\%~for all cases, and the~slopes are found 
to be 
\mbox{$10.8 \pm 0.6$}, 
\mbox{$9.5 \pm 1.2$} and 
\mbox{$9.8 \pm 1.6$}\,(in units of $10^{-3}/\left(\!\gevc\right)$)
for \YoneS, \YtwoS and \YthreeS, respectively.
The~measurements are compared with the NRQCD theory predictions~\cite{Han:2014kxa} 
in the~same kinematic range, where only uncertainties from the CO~long distance 
matrix elements are considered since most other uncertainties 
are expected to cancel in the~ratio.
The~theory predictions are independent on the~\ups~state 
and are consistently lower than the~measurements.

The ratio $\mathscr{R}_{8/7}$ as a~function of rapidity, 
integrated over the~region~\mbox{$\pt<30\gevc$} is shown 
in Fig.~\ref{fig:resultsratio3}b. 
The ratios are compared with the expectations 
from the CO~mechanism~\cite{Kisslinger:2011fe,Kisslinger:2013mev,*Kisslinger:2014zga}
with normalisation factors fixed from the~fits 
of~Fig.~\ref{fig:diff_xsecY}.
The~trend observed in data 
does not agree with the~pure CO~model.
It can be noted that also for open beauty hadrons
the~differential cross-sections exhibit 
a~larger rise as a~function of $\sqs$ 
at~smaller rapidities~\cite{LHCb-PAPER-2015-032}, 
while the FONLL calculations~\cite{Cacciari:1998it,*Cacciari:2001td,*Cacciari:2012ny}
predict this behaviour towards larger~rapidity.

\begin{figure}[p]
  \setlength{\unitlength}{1mm}
  \centering
  \begin{picture}(150,180)
    \put( 0,120){ 
      \includegraphics*[width=75mm,height=60mm,%
      ]{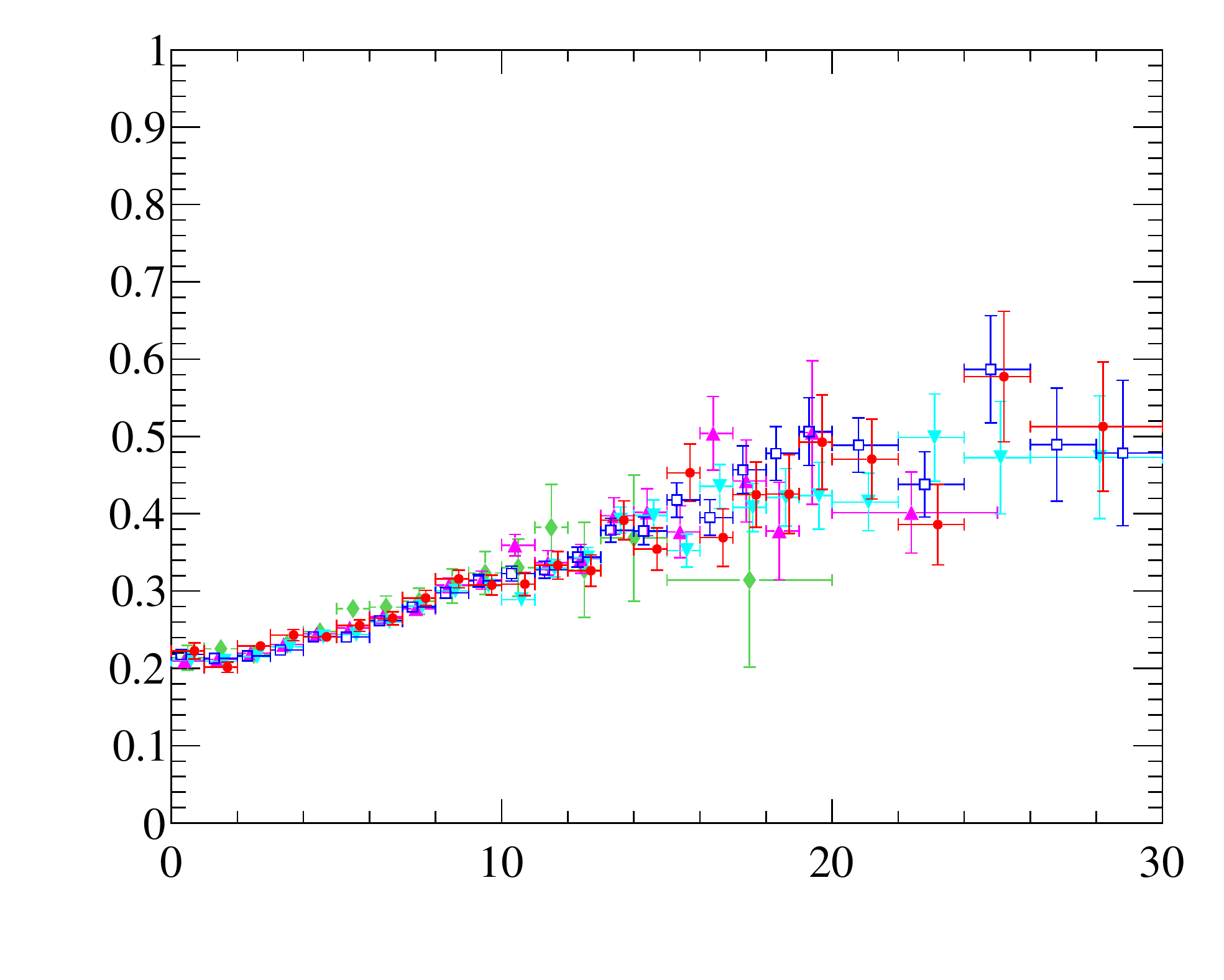}
    }
    \put(75,120){ 
      \includegraphics*[width=75mm,height=60mm,%
      ]{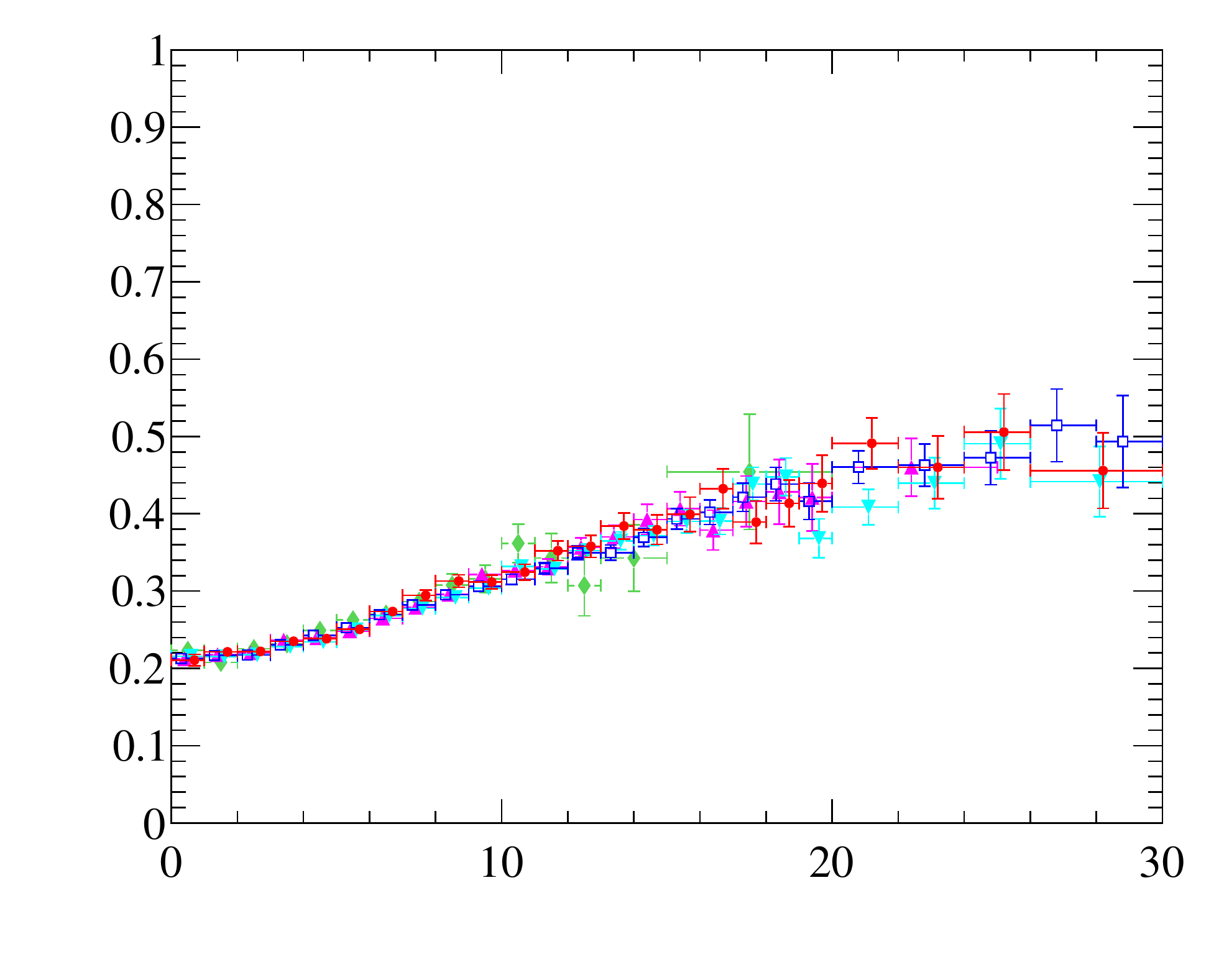}
    }
    \put( 0, 60){ 
      \includegraphics*[width=75mm,height=60mm,%
      ]{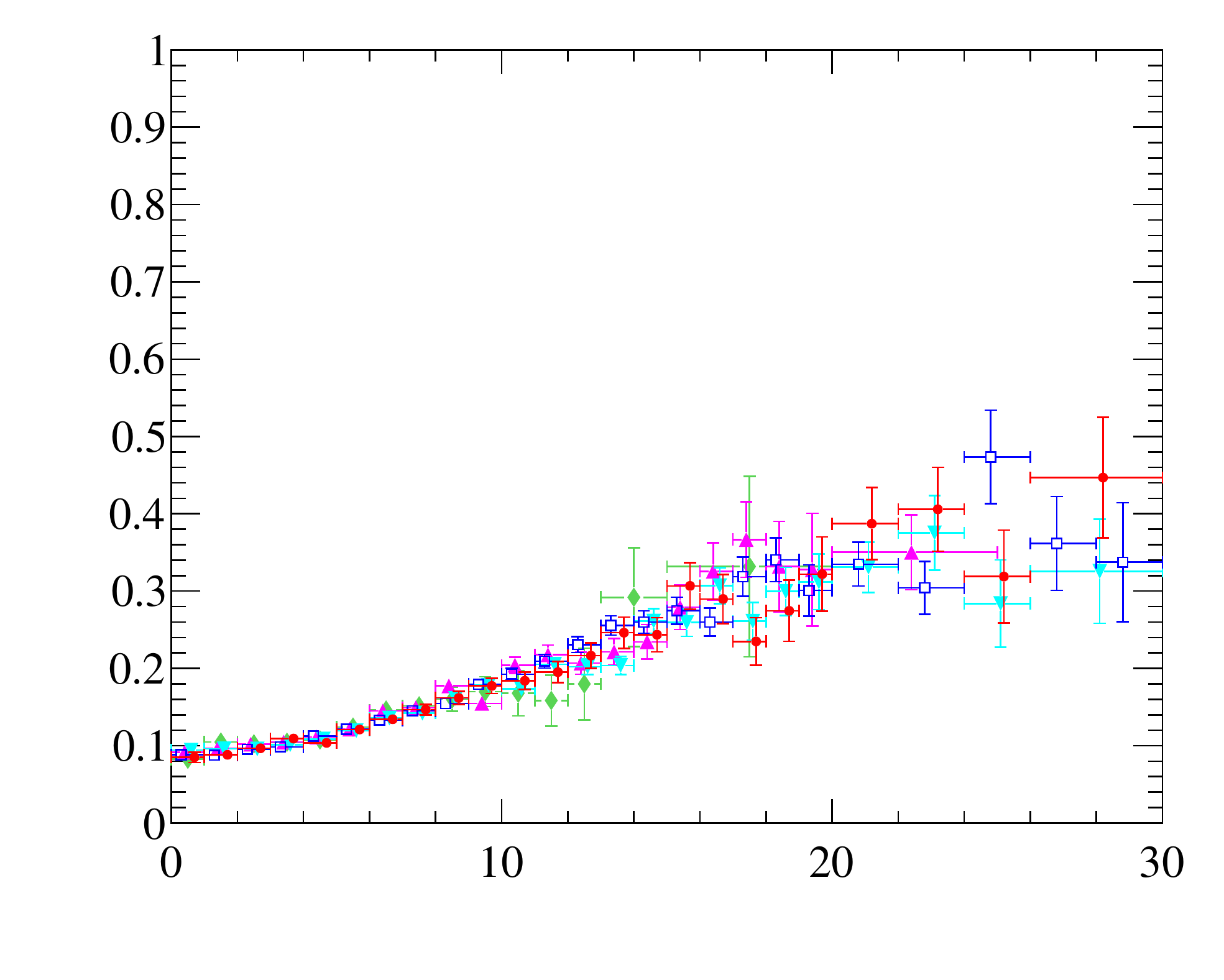}
    }
    \put(75, 60){ 
      \includegraphics*[width=75mm,height=60mm,%
      ]{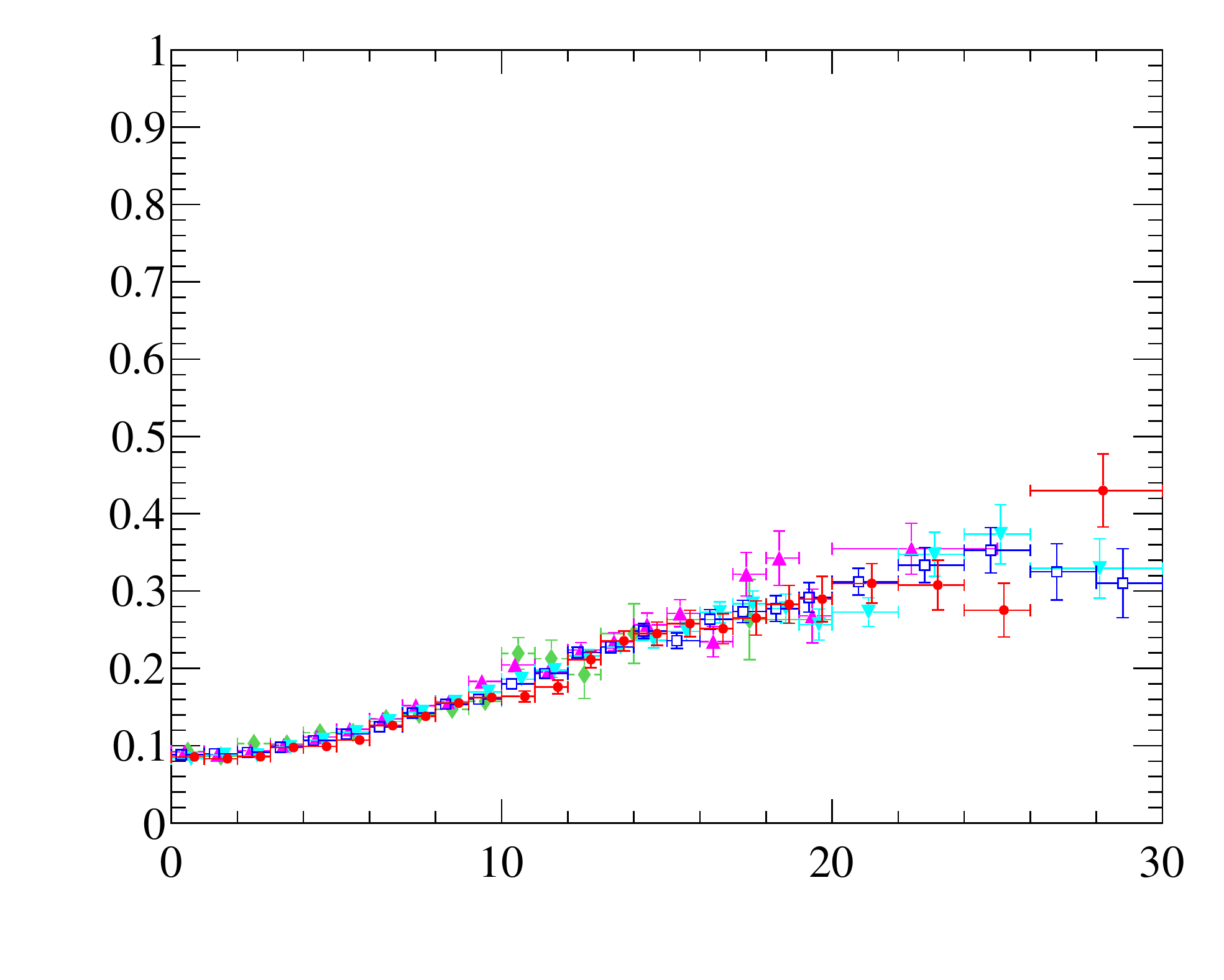}
    }
    \put( 0,  0){ 
      \includegraphics*[width=75mm,height=60mm,%
      ]{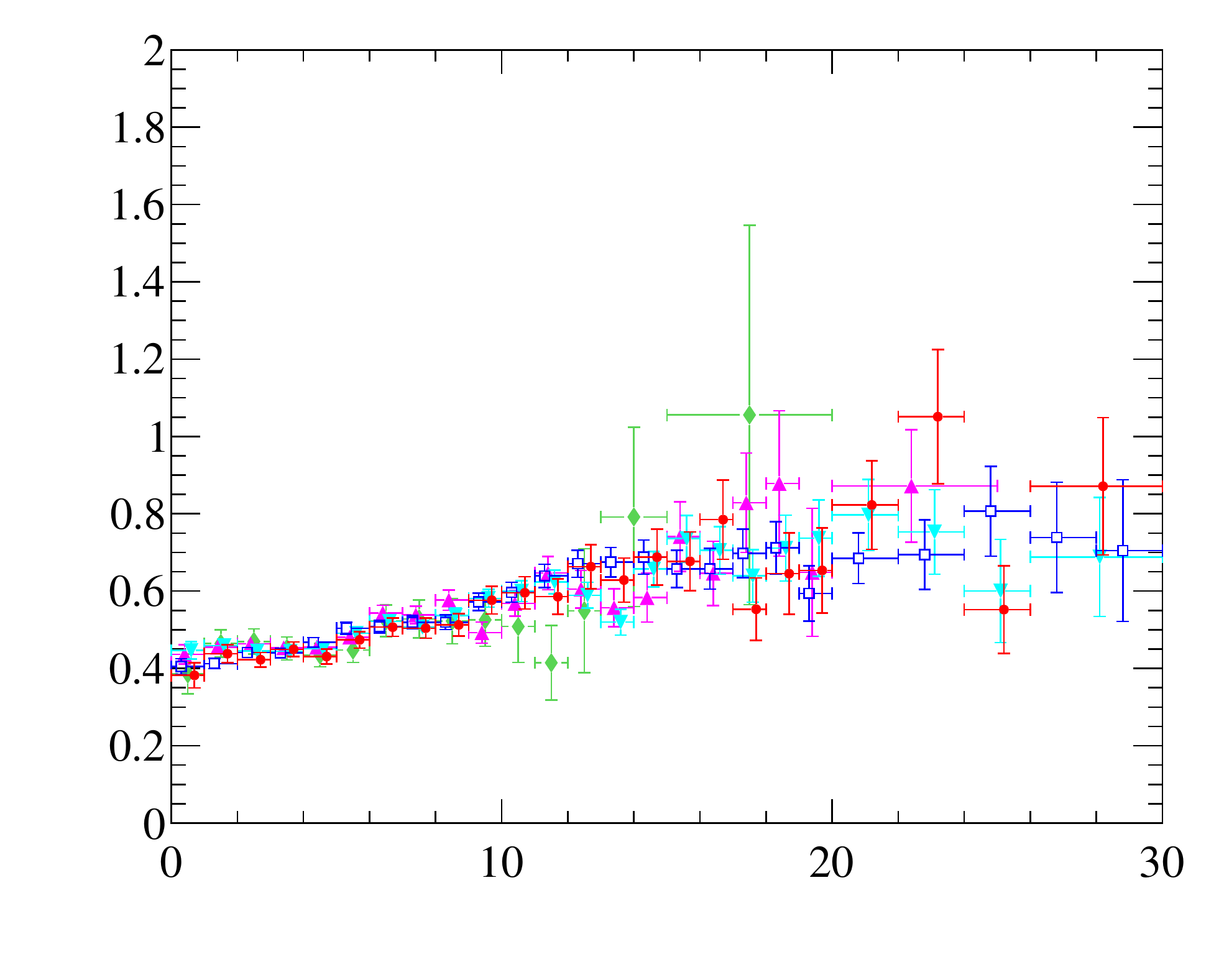}
    }
    \put(75,  0){ 
      \includegraphics*[width=75mm,height=60mm,%
      ]{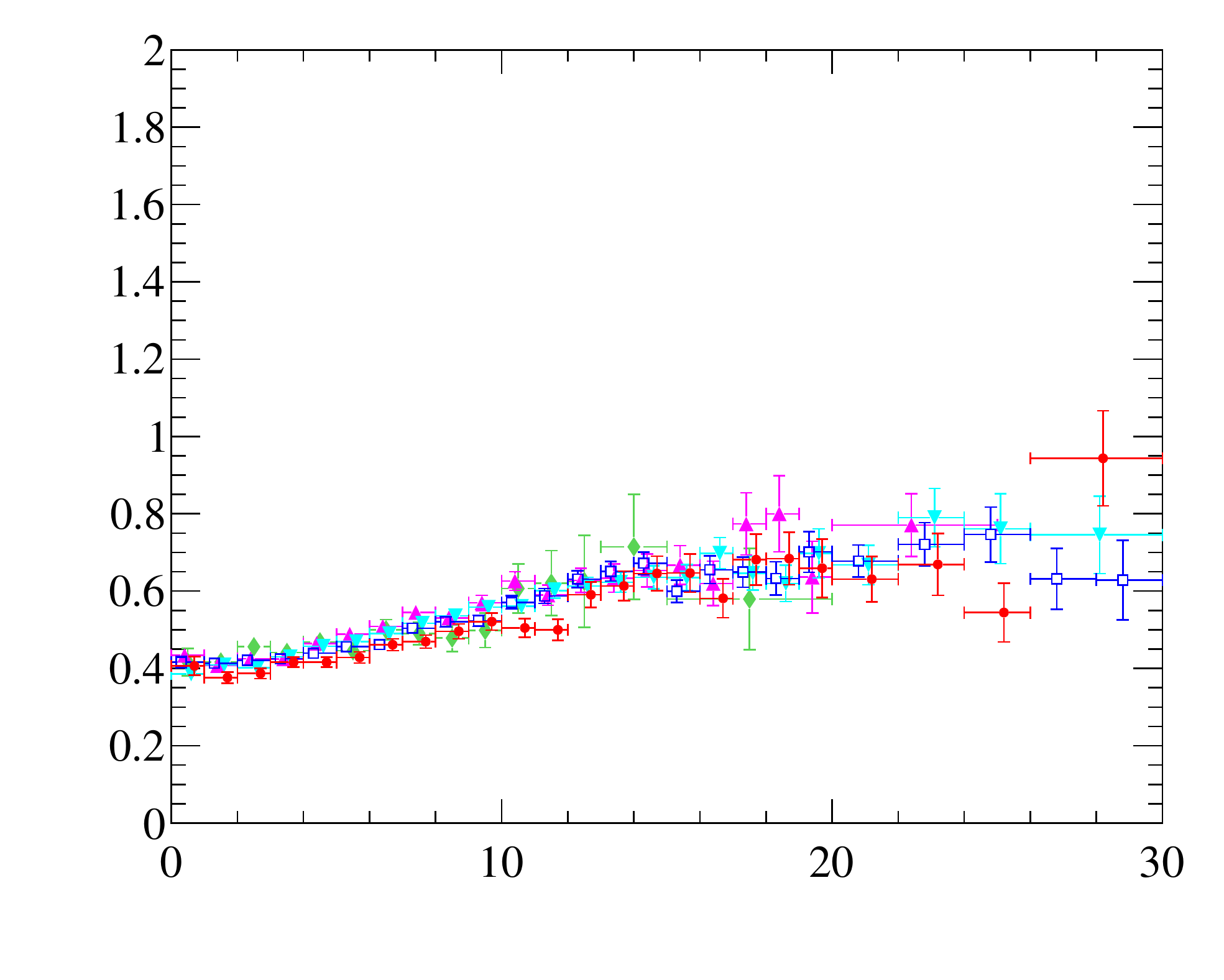}
    }
    \put(  2,170) { \begin{sideways} $\mathscr{R}_{2,1}$ \end{sideways}}
    \put( 77,170) { \begin{sideways} $\mathscr{R}_{2,1}$ \end{sideways}}
    \put(  2,109) { \begin{sideways} $\mathscr{R}_{3,1}$ \end{sideways}}
    \put( 77,109) { \begin{sideways} $\mathscr{R}_{3,1}$ \end{sideways}}
    \put(  2, 49) { \begin{sideways} $\mathscr{R}_{3,2}$ \end{sideways}} 
    \put( 77, 49) { \begin{sideways} $\mathscr{R}_{3,2}$ \end{sideways}} 
    \put (120,166){\scriptsize $\begin{array}{cc}  
        {\color{red}\bullet}                       & 2.0<y<2.5 \\
        {\color{blue}\square}                      & 2.5<y<3.0 \\
        {\color[rgb]{0,1,1}\blacktriangledown}     & 3.0<y<3.5 \\ 
        {\color[rgb]{1,0,1}\blacktriangle}         & 3.5<y<4.0 \\ 
        {\color[rgb]{0.33,0.83,0.33}\blacklozenge} & 4.0<y<4.5 \\ \end{array}$}
    \put (120,106){\scriptsize $\begin{array}{cc}  
        {\color{red}\bullet}                       & 2.0<y<2.5 \\
        {\color{blue}\square}                      & 2.5<y<3.0 \\
        {\color[rgb]{0,1,1}\blacktriangledown}     & 3.0<y<3.5 \\ 
        {\color[rgb]{1,0,1}\blacktriangle}         & 3.5<y<4.0 \\ 
        {\color[rgb]{0.33,0.83,0.33}\blacklozenge} & 4.0<y<4.5 \\ \end{array}$}
    \put (120, 46){\scriptsize $\begin{array}{cc}  
        {\color{red}\bullet}                       & 2.0<y<2.5 \\
        {\color{blue}\square}                      & 2.5<y<3.0 \\
        {\color[rgb]{0,1,1}\blacktriangledown}     & 3.0<y<3.5 \\ 
        {\color[rgb]{1,0,1}\blacktriangle}         & 3.5<y<4.0 \\ 
        {\color[rgb]{0.33,0.83,0.33}\blacklozenge} & 4.0<y<4.5 \\ \end{array}$}
    \put ( 45,166){\scriptsize $\begin{array}{cc}  
        {\color{red}\bullet}                       & 2.0<y<2.5 \\
        {\color{blue}\square}                      & 2.5<y<3.0 \\
        {\color[rgb]{0,1,1}\blacktriangledown}     & 3.0<y<3.5 \\ 
        {\color[rgb]{1,0,1}\blacktriangle}         & 3.5<y<4.0 \\ 
        {\color[rgb]{0.33,0.83,0.33}\blacklozenge} & 4.0<y<4.5 \\ \end{array}$}
    \put ( 45,106){\scriptsize $\begin{array}{cc}  
        {\color{red}\bullet}                       & 2.0<y<2.5 \\
        {\color{blue}\square}                      & 2.5<y<3.0 \\
        {\color[rgb]{0,1,1}\blacktriangledown}     & 3.0<y<3.5 \\ 
        {\color[rgb]{1,0,1}\blacktriangle}         & 3.5<y<4.0 \\ 
        {\color[rgb]{0.33,0.83,0.33}\blacklozenge} & 4.0<y<4.5 \\ \end{array}$}
    \put ( 45, 46){\scriptsize $\begin{array}{cc}  
        {\color{red}\bullet}                       & 2.0<y<2.5 \\
        {\color{blue}\square}                      & 2.5<y<3.0 \\
        {\color[rgb]{0,1,1}\blacktriangledown}     & 3.0<y<3.5 \\ 
        {\color[rgb]{1,0,1}\blacktriangle}         & 3.5<y<4.0 \\ 
        {\color[rgb]{0.33,0.83,0.33}\blacklozenge} & 4.0<y<4.5 \\ \end{array}$}
    \put( 15,168) { $\begin{array}{l} \text{LHCb}  \\ \sqrt{s}=7\tev \end{array}$}
    \put( 90,168) { $\begin{array}{l} \text{LHCb}  \\ \sqrt{s}=8\tev \end{array}$}
    \put( 15,108) { $\begin{array}{l} \text{LHCb}  \\ \sqrt{s}=7\tev \end{array}$}
    \put( 90,108) { $\begin{array}{l} \text{LHCb}  \\ \sqrt{s}=8\tev \end{array}$}
    \put( 15, 48) { $\begin{array}{l} \text{LHCb}  \\ \sqrt{s}=7\tev \end{array}$}
    \put( 90, 48) { $\begin{array}{l} \text{LHCb}  \\ \sqrt{s}=8\tev \end{array}$}
    \put( 40,122) { $\pt$} \put( 58,122){ $\left[\!\gevc\right]$}
    \put(115,122) { $\pt$} \put(133,122){ $\left[\!\gevc\right]$}
    \put( 40, 62) { $\pt$} \put( 58, 62){ $\left[\!\gevc\right]$}
    \put(115, 62) { $\pt$} \put(133, 62){ $\left[\!\gevc\right]$}
    \put( 40,  2) { $\pt$} \put( 58,  2){ $\left[\!\gevc\right]$}
    \put(115,  2) { $\pt$} \put(133,  2){ $\left[\!\gevc\right]$}
  \end{picture}
  \caption { \small
    The production ratios $\mathscr{R}_{\mathrm{i,j}}$
    for 
    (top)~\YtwoS to \YoneS, 
    (middle)~\YthreeS to \YoneS, 
    and 
    \mbox{(bottom)}~\YthreeS to \YtwoS, 
    measured with data collected at  
    (left)~\mbox{$\sqrt{s}=7\,\mathrm{TeV}$} and 
    (right)~\mbox{$\sqrt{s}=8\,\mathrm{TeV}$}.
    The~error bars indicate the~sum in quadrature 
    of the~statistical and systematic uncertainties.
    The~rapidity ranges 
    \mbox{$2.0< y<2.5$},
    \mbox{$2.5\le y<3.0$},
    \mbox{$3.0\le y<3.5$},
    \mbox{$3.5\le y<4.0$} and 
    \mbox{$4.0\le y<4.5$} are shown with 
    red circles, 
    blue squares, 
    cyan downward triangles, 
    magenta upward triangles 
    and green diamonds, respectively.
    Some data points are displaced from 
    the~bin centres to improve visibility.
  }
  \label{fig:results2}
\end{figure}

\begin{sidewaystable}[p]
  \centering
  \caption{\small 
    The ratio $\mathscr{R}_{\mathrm{2,1}}$
    for $\sqrt{s}=7\,\mathrm{TeV}$. 
    The~first uncertainties are statistical and the~second 
    are the~uncorrelated component of the~systematic uncertainties. 
    The~overall correlated systematic uncertainty is 0.7\% 
    and is not included in the~numbers in the~table.
    The~horizontal lines indicate bin boundaries.
  }\label{tab:R21at7TeV}
  \begin{small}
    \begin{tabular*}{0.99\textwidth}{@{\hspace{1mm}}c@{\extracolsep{\fill}}ccccc@{\hspace{1mm}}}
      $\pt\left[\!\gevc\right]$  
      &  $2.0<y<2.5$
      &  $2.5<y<3.0$
      &  $3.0<y<3.5$
      &  $3.5<y<4.0$
      &  $4.0<y<4.5$  
      \\
      \hline 
$0-1$
 & $0.223 \pm 0.010 \pm 0.002$  & $0.218 \pm 0.006 \pm 0.001$  & $0.211 \pm 0.006 \pm 0.001$  & $0.210 \pm 0.007 \pm 0.004$  & $0.214 \pm 0.015 \pm 0.005$ \\
$1-2$
 & $0.202 \pm 0.006 \pm 0.002$  & $0.213 \pm 0.004 \pm 0.001$  & $0.209 \pm 0.004 \pm 0.001$  & $0.212 \pm 0.004 \pm 0.001$  & $0.225 \pm 0.010 \pm 0.001$ \\
$2-3$
 & $0.229 \pm 0.006 \pm 0.001$  & $0.216 \pm 0.003 \pm 0.001$  & $0.215 \pm 0.003 \pm 0.001$  & $0.220 \pm 0.004 \pm 0.001$  & $0.218 \pm 0.009 \pm 0.004$ \\
$3-4$
 & $0.243 \pm 0.006 \pm 0.004$  & $0.224 \pm 0.003 \pm 0.001$  & $0.228 \pm 0.003 \pm 0.001$  & $0.231 \pm 0.004 \pm 0.001$  & $0.231 \pm 0.009 \pm 0.001$ \\
$4-5$
 & $0.241 \pm 0.006 \pm 0.001$  & $0.241 \pm 0.004 \pm 0.001$  & $0.241 \pm 0.004 \pm 0.001$  & $0.245 \pm 0.004 \pm 0.003$  & $0.247 \pm 0.010 \pm 0.002$ \\
$5-6$
 & $0.255 \pm 0.007 \pm 0.002$  & $0.241 \pm 0.004 \pm 0.001$  & $0.244 \pm 0.004 \pm 0.001$  & $0.252 \pm 0.005 \pm 0.001$  & $0.277 \pm 0.012 \pm 0.001$ \\
$6-7$
 & $0.265 \pm 0.008 \pm 0.003$  & $0.262 \pm 0.005 \pm 0.001$  & $0.260 \pm 0.005 \pm 0.002$  & $0.267 \pm 0.006 \pm 0.001$  & $0.279 \pm 0.014 \pm 0.004$ \\
$7-8$
 & $0.291 \pm 0.009 \pm 0.003$  & $0.279 \pm 0.006 \pm 0.002$  & $0.280 \pm 0.006 \pm 0.002$  & $0.277 \pm 0.007 \pm 0.003$  & $0.287 \pm 0.017 \pm 0.003$ \\
$8-9$
 & $0.316 \pm 0.011 \pm 0.002$  & $0.298 \pm 0.007 \pm 0.001$  & $0.300 \pm 0.007 \pm 0.002$  & $0.308 \pm 0.009 \pm 0.002$  & $0.307 \pm 0.021 \pm 0.005$ \\
$9-10$
 & $0.308 \pm 0.012 \pm 0.002$  & $0.313 \pm 0.008 \pm 0.001$  & $0.307 \pm 0.008 \pm 0.003$  & $0.314 \pm 0.011 \pm 0.004$  & $0.323 \pm 0.028 \pm 0.002$ \\
$10-11$
 & $0.309 \pm 0.014 \pm 0.003$  & $0.323 \pm 0.009 \pm 0.002$  & $0.289 \pm 0.009 \pm 0.001$  & $0.359 \pm 0.014 \pm 0.001$  & $0.33 \pm 0.04 \pm 0.01$ \\
$11-12$
 & $0.333 \pm 0.017 \pm 0.004$  & $0.328 \pm 0.011 \pm 0.003$  & $0.329 \pm 0.011 \pm 0.001$  & $0.337 \pm 0.015 \pm 0.002$  & $0.38 \pm 0.06 \pm 0.01$ \\
$12-13$
 & $0.326 \pm 0.019 \pm 0.005$  & $0.344 \pm 0.013 \pm 0.002$  & $0.343 \pm 0.013 \pm 0.001$  & $0.342 \pm 0.019 \pm 0.001$  & $0.33 \pm 0.06 \pm 0.01$ \\  \cline{6-6}
$13-14$
 & $0.392 \pm 0.025 \pm 0.005$  & $0.379 \pm 0.015 \pm 0.001$  & $0.392 \pm 0.017 \pm 0.001$  & $0.397 \pm 0.023 \pm 0.002$  & \multirow{2}{*}{$0.37 \pm 0.08 \pm 0.01$} \\
$14-15$
 & $0.354 \pm 0.026 \pm 0.007$  & $0.378 \pm 0.017 \pm 0.003$  & $0.398 \pm 0.020 \pm 0.005$  & $0.402 \pm 0.030 \pm 0.006$  &  \\  \cline{6-6}
$15-16$
 & $0.45 \pm 0.04 \pm 0.01$  & $0.418 \pm 0.022 \pm 0.002$  & $0.353 \pm 0.021 \pm 0.001$  & $0.377 \pm 0.033 \pm 0.004$  & \multirow{5}{*}{$0.31 \pm 0.11 \pm 0.01$} \\
$16-17$
 & $0.37 \pm 0.04 \pm 0.01$  & $0.395 \pm 0.023 \pm 0.004$  & $0.435 \pm 0.028 \pm 0.001$  & $0.50 \pm 0.05 \pm 0.01$  &  \\
$17-18$
 & $0.42 \pm 0.04 \pm 0.01$  & $0.457 \pm 0.031 \pm 0.001$  & $0.408 \pm 0.031 \pm 0.001$  & $0.44 \pm 0.05 \pm 0.01$  &  \\
$18-19$
 & $0.43 \pm 0.05 \pm 0.01$  & $0.478 \pm 0.035 \pm 0.002$  & $0.42 \pm 0.04 \pm 0.01$  & $0.38 \pm 0.06 \pm 0.01$  &  \\
$19-20$
 & $0.49 \pm 0.06 \pm 0.01$  & $0.51 \pm 0.04 \pm 0.01$  & $0.42 \pm 0.04 \pm 0.01$  & $0.51 \pm 0.09 \pm 0.01$  &  \\  \cline{2-6}
$20-21$
 & \multirow{2}{*}{$0.47 \pm 0.05 \pm 0.01$}  & \multirow{2}{*}{$0.489 \pm 0.035 \pm 0.002$}  & \multirow{2}{*}{$0.42 \pm 0.04 \pm 0.01$}  & \multirow{5}{*}{$0.40 \pm 0.05 \pm 0.01$} &  \\
$21-22$
 &   &   &   &  &  \\  \cline{2-4}
$22-23$
 & \multirow{2}{*}{$0.39 \pm 0.05 \pm 0.01$}  & \multirow{2}{*}{$0.44 \pm 0.04 \pm 0.01$}  & \multirow{2}{*}{$0.50 \pm 0.06 \pm 0.01$}  &  &  \\
$23-24$
 &   &   &   &  &  \\  \cline{2-4}
$24-25$
 & \multirow{2}{*}{$0.58 \pm 0.08 \pm 0.01$}  & \multirow{2}{*}{$0.59 \pm 0.07 \pm 0.01$}  & \multirow{2}{*}{$0.47 \pm 0.07 \pm 0.01$}  &  &  \\  \cline{5-5}
$25-26$
 &   &   &  &  &  \\  \cline{2-4}
$26-27$
 & \multirow{4}{*}{$0.51 \pm 0.08 \pm 0.01$}  & \multirow{2}{*}{$0.49 \pm 0.07 \pm 0.01$}  & \multirow{4}{*}{$0.47 \pm 0.08 \pm 0.02$} &  &  \\
$27-28$
 &   &   &  &  &  \\  \cline{3-3}
$28-29$
 &   & \multirow{2}{*}{$0.48 \pm 0.09 \pm 0.01$}  &  &  &  \\
$29-30$
 &   &   &  &  &  
\end{tabular*}   
\end{small}
\end{sidewaystable}

\begin{sidewaystable}[p]
  \centering
  \caption{\small 
    The~ratio $\mathscr{R}_{\mathrm{3,1}}$
    for $\sqrt{s}=7\,\mathrm{TeV}$. The~first uncertainties are statistical and the~second 
    are the~uncorrelated component of the~systematic uncertainties. 
    The~overall correlated systematic uncertainty is 0.7\% 
    and is not included in the~numbers in the~table.
    The~horizontal lines indicate bin boundaries.
  }\label{tab:R31at7TeV}
  \begin{small}
    \begin{tabular*}{0.99\textwidth}{@{\hspace{1mm}}c@{\extracolsep{\fill}}ccccc@{\hspace{1mm}}}
      $\pt\left[\!\gevc\right]$  
      &  $2.0<y<2.5$
      &  $2.5<y<3.0$
      &  $3.0<y<3.5$
      &  $3.5<y<4.0$
      &  $4.0<y<4.5$  
      \\
      \hline 
$0-1$
   & $0.085 \pm 0.007 \pm 0.001$  
   & $0.088 \pm 0.004 \pm 0.001$  
   & $0.094 \pm 0.004 \pm 0.002$  
   & $0.091 \pm 0.005 \pm 0.002$  
   & $0.083 \pm 0.010 \pm 0.003$ \\
$1-2$
   & $0.088 \pm 0.004 \pm 0.001$  
   & $0.088 \pm 0.003 \pm 0.001$  
   & $0.096 \pm 0.003 \pm 0.001$  
   & $0.096 \pm 0.003 \pm 0.001$  
   & $0.105 \pm 0.007 \pm 0.001$ \\
$2-3$
   & $0.097 \pm 0.004 \pm 0.001$  
   & $0.095 \pm 0.002 \pm 0.001$  
   & $0.096 \pm 0.002 \pm 0.001$  
   & $0.102 \pm 0.003 \pm 0.001$  
   & $0.102 \pm 0.006 \pm 0.004$ \\
$3-4$
   & $0.109 \pm 0.004 \pm 0.002$  
   & $0.099 \pm 0.002 \pm 0.001$  & $0.101 \pm 0.002 \pm 0.001$  & $0.105 \pm 0.003 \pm 0.001$  & $0.104 \pm 0.006 \pm 0.001$ \\
$4-5$
 & $0.104 \pm 0.004 \pm 0.001$  & $0.113 \pm 0.003 \pm 0.001$  & $0.109 \pm 0.003 \pm 0.001$  & $0.111 \pm 0.003 \pm 0.002$  & $0.108 \pm 0.007 \pm 0.001$ \\
$5-6$
 & $0.121 \pm 0.005 \pm 0.002$  & $0.121 \pm 0.003 \pm 0.001$  & $0.120 \pm 0.003 \pm 0.001$  & $0.121 \pm 0.004 \pm 0.001$  & $0.124 \pm 0.008 \pm 0.001$ \\
$6-7$
 & $0.134 \pm 0.006 \pm 0.002$  & $0.133 \pm 0.004 \pm 0.001$  & $0.136 \pm 0.004 \pm 0.001$  & $0.145 \pm 0.005 \pm 0.001$  & $0.146 \pm 0.010 \pm 0.004$ \\
$7-8$
 & $0.147 \pm 0.007 \pm 0.002$  & $0.145 \pm 0.004 \pm 0.001$  & $0.142 \pm 0.004 \pm 0.001$  & $0.149 \pm 0.005 \pm 0.002$  & $0.152 \pm 0.012 \pm 0.002$ \\
$8-9$
 & $0.162 \pm 0.008 \pm 0.002$  & $0.155 \pm 0.005 \pm 0.001$  & $0.161 \pm 0.005 \pm 0.002$  & $0.177 \pm 0.007 \pm 0.001$  & $0.160 \pm 0.016 \pm 0.002$ \\
$9-10$
 & $0.177 \pm 0.009 \pm 0.002$  & $0.179 \pm 0.006 \pm 0.001$  & $0.178 \pm 0.006 \pm 0.002$  & $0.155 \pm 0.008 \pm 0.003$  & $0.170 \pm 0.019 \pm 0.001$ \\
$10-11$
 & $0.184 \pm 0.011 \pm 0.001$  & $0.193 \pm 0.007 \pm 0.001$  & $0.173 \pm 0.007 \pm 0.001$  & $0.204 \pm 0.010 \pm 0.001$  & $0.168 \pm 0.026 \pm 0.013$ \\
$11-12$
 & $0.195 \pm 0.013 \pm 0.002$  & $0.209 \pm 0.008 \pm 0.003$  & $0.205 \pm 0.009 \pm 0.001$  & $0.218 \pm 0.012 \pm 0.002$  & $0.158 \pm 0.033 \pm 0.001$ \\
$12-13$
 & $0.217 \pm 0.016 \pm 0.003$  & $0.231 \pm 0.010 \pm 0.001$  & $0.202 \pm 0.010 \pm 0.001$  & $0.207 \pm 0.014 \pm 0.001$  & $0.18 \pm 0.05 \pm 0.01$ \\  \cline{6-6}
$13-14$
 & $0.246 \pm 0.019 \pm 0.005$  & $0.256 \pm 0.012 \pm 0.002$  & $0.204 \pm 0.012 \pm 0.001$  & $0.221 \pm 0.017 \pm 0.001$  & \multirow{2}{*}{$0.29 \pm 0.06 \pm 0.01$} \\
$14-15$
 & $0.244 \pm 0.022 \pm 0.003$  & $0.260 \pm 0.014 \pm 0.002$  & $0.261 \pm 0.015 \pm 0.004$  & $0.234 \pm 0.022 \pm 0.003$  &  \\  \cline{6-6}
$15-16$
 & $0.307 \pm 0.030 \pm 0.002$  & $0.275 \pm 0.017 \pm 0.001$  & $0.259 \pm 0.018 \pm 0.001$  & $0.279 \pm 0.028 \pm 0.003$  & \multirow{5}{*}{$0.33 \pm 0.12 \pm 0.01$} \\
$16-17$
 & $0.290 \pm 0.032 \pm 0.003$  & $0.260 \pm 0.018 \pm 0.002$  & $0.307 \pm 0.023 \pm 0.002$  & $0.33 \pm 0.04 \pm 0.01$  &  \\
$17-18$
 & $0.235 \pm 0.031 \pm 0.002$  & $0.319 \pm 0.025 \pm 0.002$  & $0.261 \pm 0.024 \pm 0.002$  & $0.37 \pm 0.05 \pm 0.01$  &  \\
$18-19$
 & $0.27 \pm 0.04 \pm 0.01$  & $0.340 \pm 0.028 \pm 0.001$  & $0.300 \pm 0.031 \pm 0.001$  & $0.33 \pm 0.06 \pm 0.01$  &  \\
$19-20$
 & $0.32 \pm 0.05 \pm 0.01$  & $0.301 \pm 0.032 \pm 0.006$  & $0.31 \pm 0.04 \pm 0.01$  & $0.33 \pm 0.07 \pm 0.01$  &  \\  \cline{2-6}
$20-21$
 & \multirow{2}{*}{$0.39 \pm 0.05 \pm 0.01$}  & \multirow{2}{*}{$0.335 \pm 0.028 \pm 0.002$}  & \multirow{2}{*}{$0.331 \pm 0.032 \pm 0.002$}  & \multirow{5}{*}{$0.35 \pm 0.05 \pm 0.01$} &  \\
$21-22$
 &   &   &   &  &  \\  \cline{2-4}
$22-23$
 & \multirow{2}{*}{$0.41 \pm 0.05 \pm 0.01$}  & \multirow{2}{*}{$0.304 \pm 0.034 \pm 0.002$}  & \multirow{2}{*}{$0.38 \pm 0.05 \pm 0.01$}  &  &  \\
$23-24$
 &   &   &   &  &  \\  \cline{2-4}
$24-25$
 & \multirow{2}{*}{$0.32 \pm 0.06 \pm 0.01$}  & \multirow{2}{*}{$0.47 \pm 0.06 \pm 0.01$}  & \multirow{2}{*}{$0.28 \pm 0.06 \pm 0.01$}  &  &  \\  \cline{5-5}
$25-26$
 &   &   &  &  &  \\  \cline{2-4}
$26-27$
 & \multirow{4}{*}{$0.45 \pm 0.08 \pm 0.01$}  & \multirow{2}{*}{$0.36 \pm 0.06 \pm 0.01$}  & \multirow{4}{*}{$0.33 \pm 0.06 \pm 0.02$} &  &  \\
$27-28$
 &   &   &  &  &  \\  \cline{3-3}
$28-29$
 &   & \multirow{2}{*}{$0.34 \pm 0.08 \pm 0.01$}  &  &  &  \\
$29-30$
 &   &   &  &  &  
\end{tabular*}   
\end{small}
\end{sidewaystable}

\begin{sidewaystable}[p]
  \centering
  \caption{\small 
    The ratio $\mathscr{R}_{\mathrm{2,1}}$
    for $\sqrt{s}=8\,\mathrm{TeV}$. The~first uncertainties are statistical and the~second 
    are the~uncorrelated component of the~systematic uncertainties. 
    The~overall correlated systematic uncertainty is 0.7\% 
    and is not included in the~numbers in the~table.    
    The~horizontal lines indicate bin boundaries.
  }\label{tab:R21at8TeV}
  \begin{small}
    \begin{tabular*}{0.99\textwidth}{@{\hspace{1mm}}c@{\extracolsep{\fill}}ccccc@{\hspace{1mm}}}
      $\pt\left[\!\gevc\right]$  
      &  $2.0<y<2.5$
      &  $2.5<y<3.0$
      &  $3.0<y<3.5$
      &  $3.5<y<4.0$
      &  $4.0<y<4.5$  
      \\
      \hline 
$0-1$
 & $0.211 \pm 0.007 \pm 0.003$  & $0.213 \pm 0.004 \pm 0.002$  & $0.216 \pm 0.004 \pm 0.001$  & $0.212 \pm 0.005 \pm 0.001$  & $0.223 \pm 0.010 \pm 0.002$ \\
$1-2$
 & $0.221 \pm 0.004 \pm 0.001$  & $0.217 \pm 0.003 \pm 0.001$  & $0.215 \pm 0.003 \pm 0.001$  & $0.218 \pm 0.003 \pm 0.001$  & $0.208 \pm 0.006 \pm 0.003$ \\
$2-3$
 & $0.222 \pm 0.004 \pm 0.001$  & $0.217 \pm 0.002 \pm 0.001$  & $0.218 \pm 0.002 \pm 0.001$  & $0.220 \pm 0.003 \pm 0.001$  & $0.225 \pm 0.006 \pm 0.001$ \\
$3-4$
 & $0.235 \pm 0.004 \pm 0.001$  & $0.231 \pm 0.002 \pm 0.001$  & $0.228 \pm 0.002 \pm 0.001$  & $0.237 \pm 0.003 \pm 0.002$  & $0.232 \pm 0.006 \pm 0.002$ \\
$4-5$
 & $0.238 \pm 0.004 \pm 0.001$  & $0.243 \pm 0.003 \pm 0.001$  & $0.234 \pm 0.002 \pm 0.001$  & $0.240 \pm 0.003 \pm 0.001$  & $0.249 \pm 0.007 \pm 0.005$ \\
$5-6$
 & $0.251 \pm 0.005 \pm 0.001$  & $0.253 \pm 0.003 \pm 0.001$  & $0.250 \pm 0.003 \pm 0.001$  & $0.249 \pm 0.003 \pm 0.002$  & $0.263 \pm 0.007 \pm 0.001$ \\
$6-7$
 & $0.274 \pm 0.005 \pm 0.003$  & $0.270 \pm 0.003 \pm 0.001$  & $0.268 \pm 0.003 \pm 0.002$  & $0.265 \pm 0.004 \pm 0.002$  & $0.270 \pm 0.009 \pm 0.001$ \\
$7-8$
 & $0.294 \pm 0.006 \pm 0.002$  & $0.282 \pm 0.004 \pm 0.002$  & $0.278 \pm 0.004 \pm 0.002$  & $0.279 \pm 0.005 \pm 0.002$  & $0.287 \pm 0.010 \pm 0.002$ \\
$8-9$
 & $0.313 \pm 0.007 \pm 0.002$  & $0.295 \pm 0.004 \pm 0.002$  & $0.292 \pm 0.004 \pm 0.001$  & $0.296 \pm 0.006 \pm 0.001$  & $0.308 \pm 0.014 \pm 0.006$ \\
$9-10$
 & $0.312 \pm 0.008 \pm 0.002$  & $0.306 \pm 0.005 \pm 0.001$  & $0.304 \pm 0.005 \pm 0.001$  & $0.322 \pm 0.007 \pm 0.002$  & $0.316 \pm 0.018 \pm 0.001$ \\
$10-11$
 & $0.324 \pm 0.010 \pm 0.002$  & $0.315 \pm 0.006 \pm 0.002$  & $0.332 \pm 0.006 \pm 0.002$  & $0.327 \pm 0.008 \pm 0.003$  & $0.362 \pm 0.025 \pm 0.004$ \\
$11-12$
 & $0.352 \pm 0.012 \pm 0.004$  & $0.329 \pm 0.007 \pm 0.002$  & $0.328 \pm 0.007 \pm 0.004$  & $0.331 \pm 0.010 \pm 0.003$  & $0.343 \pm 0.032 \pm 0.004$ \\
$12-13$
 & $0.358 \pm 0.014 \pm 0.004$  & $0.350 \pm 0.008 \pm 0.002$  & $0.352 \pm 0.009 \pm 0.001$  & $0.357 \pm 0.012 \pm 0.002$  & $0.31 \pm 0.04 \pm 0.01$ \\  \cline{6-6}
$13-14$
 & $0.384 \pm 0.016 \pm 0.003$  & $0.350 \pm 0.009 \pm 0.001$  & $0.365 \pm 0.010 \pm 0.004$  & $0.370 \pm 0.015 \pm 0.002$  & \multirow{2}{*}{$0.34 \pm 0.04 \pm 0.01$} \\
$14-15$
 & $0.379 \pm 0.018 \pm 0.005$  & $0.370 \pm 0.011 \pm 0.003$  & $0.372 \pm 0.012 \pm 0.001$  & $0.393 \pm 0.018 \pm 0.008$  &  \\  \cline{6-6}
$15-16$
 & $0.399 \pm 0.021 \pm 0.005$  & $0.393 \pm 0.013 \pm 0.002$  & $0.390 \pm 0.015 \pm 0.003$  & $0.407 \pm 0.022 \pm 0.003$  & \multirow{5}{*}{$0.45 \pm 0.07 \pm 0.01$} \\
$16-17$
 & $0.432 \pm 0.025 \pm 0.002$  & $0.402 \pm 0.016 \pm 0.002$  & $0.390 \pm 0.017 \pm 0.002$  & $0.379 \pm 0.024 \pm 0.008$  &  \\
$17-18$
 & $0.389 \pm 0.027 \pm 0.003$  & $0.421 \pm 0.018 \pm 0.001$  & $0.439 \pm 0.021 \pm 0.001$  & $0.416 \pm 0.032 \pm 0.005$  &  \\
$18-19$
 & $0.414 \pm 0.030 \pm 0.001$  & $0.438 \pm 0.021 \pm 0.003$  & $0.448 \pm 0.024 \pm 0.001$  & $0.43 \pm 0.04 \pm 0.01$  &  \\
$19-20$
 & $0.44 \pm 0.04 \pm 0.01$  & $0.416 \pm 0.023 \pm 0.001$  & $0.368 \pm 0.024 \pm 0.007$  & $0.42 \pm 0.04 \pm 0.01$  &  \\  \cline{2-6}
$20-21$
 & \multirow{2}{*}{$0.491 \pm 0.033 \pm 0.002$}  & \multirow{2}{*}{$0.460 \pm 0.021 \pm 0.003$}  & \multirow{2}{*}{$0.409 \pm 0.022 \pm 0.005$}  & \multirow{5}{*}{$0.46 \pm 0.04 \pm 0.01$} &  \\
$21-22$
 &   &   &   &  &  \\  \cline{2-4}
$22-23$
 & \multirow{2}{*}{$0.46 \pm 0.04 \pm 0.01$}  & \multirow{2}{*}{$0.463 \pm 0.027 \pm 0.002$}  & \multirow{2}{*}{$0.440 \pm 0.032 \pm 0.004$}  &  &  \\
$23-24$
 &   &   &   &  &  \\  \cline{2-4}
$24-25$
 & \multirow{2}{*}{$0.51 \pm 0.05 \pm 0.01$}  & \multirow{2}{*}{$0.473 \pm 0.035 \pm 0.001$}  & \multirow{2}{*}{$0.49 \pm 0.05 \pm 0.01$}  &  &  \\  \cline{5-5}
$25-26$
 &   &   &  &  &  \\  \cline{2-4}
$26-27$
 & \multirow{4}{*}{$0.46 \pm 0.05 \pm 0.01$}  & \multirow{2}{*}{$0.51 \pm 0.05 \pm 0.01$}  & \multirow{4}{*}{$0.44 \pm 0.04 \pm 0.01$} &  &  \\
$27-28$
 &   &   &  &  &  \\  \cline{3-3}
$28-29$
 &   & \multirow{2}{*}{$0.49 \pm 0.06 \pm 0.01$}  &  &  &  \\
$29-30$
 &   &   &  &  &  
    \end{tabular*}   
  \end{small}
\end{sidewaystable}

\begin{sidewaystable}[p]
  \centering
  \caption{\small 
    The ratio $\mathscr{R}_{\mathrm{3,1}}$
    for $\sqrt{s}=8\,\mathrm{TeV}$. The~first uncertainties are statistical and the~second 
    are the~uncorrelated component of the~systematic uncertainties. 
    The~overall correlated systematic uncertainty is 0.7\% 
    and is not included in the~numbers in the~table.    
    The~horizontal lines indicate bin boundaries.
  }\label{tab:R31at8TeV}
  \begin{small}
    \begin{tabular*}{0.99\textwidth}{@{\hspace{1mm}}c@{\extracolsep{\fill}}ccccc@{\hspace{1mm}}}
      $\pt\left[\!\gevc\right]$  
      &  $2.0<y<2.5$
      &  $2.5<y<3.0$
      &  $3.0<y<3.5$
      &  $3.5<y<4.0$
      &  $4.0<y<4.5$  
      \\
      \hline 
$0-1$
 & $0.086 \pm 0.004 \pm 0.001$ 
 & $0.089 \pm 0.003 \pm 0.001$  
 & $0.083 \pm 0.003 \pm 0.001$  
 & $0.092 \pm 0.003 \pm 0.001$  
 & $0.093 \pm 0.007 \pm 0.001$ \\
$1-2$
 & $0.083 \pm 0.003 \pm 0.001$ 
 & $0.090 \pm 0.002 \pm 0.001$ 
 & $0.088 \pm 0.002 \pm 0.001$ 
 & $0.089 \pm 0.002 \pm 0.001$  
 & $0.087 \pm 0.004 \pm 0.002$ \\
$2-3$
 & $0.086 \pm 0.003 \pm 0.001$  
 & $0.091 \pm 0.002 \pm 0.001$  
 & $0.087 \pm 0.002 \pm 0.001$  
 & $0.094 \pm 0.001 \pm 0.001$  
 & $0.103 \pm 0.004 \pm 0.001$ \\
$3-4$
 & $0.098 \pm 0.003 \pm 0.001$  
 & $0.098 \pm 0.002 \pm 0.001$  
 & $0.098 \pm 0.002 \pm 0.001$  
 & $0.100 \pm 0.002 \pm 0.001$  
 & $0.102 \pm 0.004 \pm 0.001$ \\
$4-5$
 & $0.099 \pm 0.003 \pm 0.001$  
 & $0.107 \pm 0.002 \pm 0.001$  
 & $0.107 \pm 0.002 \pm 0.001$  
 & $0.111 \pm 0.002 \pm 0.001$  
 & $0.117 \pm 0.005 \pm 0.004$ \\
$5-6$
 & $0.107 \pm 0.003 \pm 0.001$  
 & $0.115 \pm 0.002 \pm 0.001$  
 & $0.117 \pm 0.002 \pm 0.001$  
 & $0.121 \pm 0.003 \pm 0.001$  
 & $0.117 \pm 0.005 \pm 0.001$ \\
$6-7$
 & $0.126 \pm 0.004 \pm 0.001$  
 & $0.125 \pm 0.002 \pm 0.001$  
 & $0.132 \pm 0.002 \pm 0.001$  
 & $0.135 \pm 0.003 \pm 0.002$  
 & $0.135 \pm 0.006 \pm 0.001$ \\
$7-8$
 & $0.138 \pm 0.004 \pm 0.001$ 
 & $0.142 \pm 0.003 \pm 0.001$  
 & $0.144 \pm 0.003 \pm 0.001$  
 & $0.152 \pm 0.004 \pm 0.001$  
 & $0.141 \pm 0.007 \pm 0.001$ \\
$8-9$
 & $0.155 \pm 0.005 \pm 0.001$  
 & $0.154 \pm 0.003 \pm 0.001$  
 & $0.156 \pm 0.003 \pm 0.001$  
 & $0.157 \pm 0.004 \pm 0.001$  
 & $0.147 \pm 0.009 \pm 0.003$ \\
$9-10$
 & $0.162 \pm 0.006 \pm 0.001$  
 & $0.160 \pm 0.004 \pm 0.001$  
 & $0.170 \pm 0.004 \pm 0.001$  
 & $0.183 \pm 0.005 \pm 0.002$  
 & $0.157 \pm 0.012 \pm 0.001$ \\
$10-11$
 & $0.164 \pm 0.007 \pm 0.001$  
 & $0.180 \pm 0.005 \pm 0.002$  
 & $0.186 \pm 0.005 \pm 0.001$  
 & $0.205 \pm 0.007 \pm 0.002$  
 & $0.220 \pm 0.019 \pm 0.009$ \\
$11-12$
 & $0.176 \pm 0.008 \pm 0.003$  
 & $0.193 \pm 0.005 \pm 0.001$  
 & $0.198 \pm 0.005 \pm 0.003$  
 & $0.195 \pm 0.007 \pm 0.001$  
 & $0.213 \pm 0.024 \pm 0.004$ \\
$12-13$
 & $0.211 \pm 0.010 \pm 0.002$  
 & $0.221 \pm 0.006 \pm 0.001$  
 & $0.216 \pm 0.007 \pm 0.001$  
 & $0.224 \pm 0.009 \pm 0.002$  
 & $0.192 \pm 0.031 \pm 0.002$ \\  \cline{6-6}
$13-14$
 & $0.236 \pm 0.013 \pm 0.001$  
 & $0.228 \pm 0.007 \pm 0.001$  
 & $0.227 \pm 0.008 \pm 0.003$  
 & $0.235 \pm 0.011 \pm 0.002$  
 & \multirow{2}{*}{$0.245 \pm 0.040 \pm 0.008$} \\
$14-15$
 & $0.245 \pm 0.015 \pm 0.003$  
 & $0.248 \pm 0.009 \pm 0.003$  
 & $0.236 \pm 0.010 \pm 0.001$  
 & $0.257 \pm 0.014 \pm 0.005$  &  \\  \cline{6-6}
$15-16$
 & $0.258 \pm 0.017 \pm 0.002$  
 & $0.236 \pm 0.010 \pm 0.002$  
 & $0.248 \pm 0.011 \pm 0.002$  
 & $0.271 \pm 0.017 \pm 0.001$  
 & \multirow{5}{*}{$0.263 \pm 0.050 \pm 0.002$} \\
$16-17$
 & $0.251 \pm 0.019 \pm 0.003$  
 & $0.263 \pm 0.012 \pm 0.002$  
 & $0.272 \pm 0.014 \pm 0.001$  
 & $0.235 \pm 0.019 \pm 0.006$  &  \\
$17-18$
 & $0.265 \pm 0.022 \pm 0.002$  
 & $0.274 \pm 0.014 \pm 0.001$  
 & $0.283 \pm 0.017 \pm 0.001$  
 & $0.322 \pm 0.028 \pm 0.002$  &  \\
$18-19$
 & $0.283 \pm 0.024 \pm 0.002$  
 & $0.277 \pm 0.017 \pm 0.002$  
 & $0.278 \pm 0.018 \pm 0.001$  
 & $0.343 \pm 0.035 \pm 0.004$  &  \\
$19-20$
 & $0.290 \pm 0.029 \pm 0.001$  
 & $0.292 \pm 0.019 \pm 0.001$  
 & $0.257 \pm 0.020 \pm 0.003$  
 & $0.268 \pm 0.034 \pm 0.007$  &  \\  \cline{2-6}
$20-21$
 & \multirow{2}{*}{$0.310 \pm 0.025 \pm 0.002$}  
 & \multirow{2}{*}{$0.312 \pm 0.017 \pm 0.004$}  
 & \multirow{2}{*}{$0.273 \pm 0.018 \pm 0.005$}  
 & \multirow{5}{*}{$0.355 \pm 0.032 \pm 0.009$} &  \\
$21-22$
 &   &   &   &  &  \\  \cline{2-4}
$22-23$
 & \multirow{2}{*}{$0.308 \pm 0.032 \pm 0.002$}  
 & \multirow{2}{*}{$0.334 \pm 0.023 \pm 0.001$}  
 & \multirow{2}{*}{$0.348 \pm 0.028 \pm 0.002$}  &  &  \\
$23-24$
 &   &   &   &  &  \\  \cline{2-4}
$24-25$
 & \multirow{2}{*}{$0.275 \pm 0.035 \pm 0.001$}  
 & \multirow{2}{*}{$0.353 \pm 0.029 \pm 0.002$}  
 & \multirow{2}{*}{$0.374 \pm 0.040 \pm 0.002$}  &  &  \\  \cline{5-5}
$25-26$
 &   &   &  &  &  \\  \cline{2-4}
$26-27$
 & \multirow{4}{*}{$0.430 \pm 0.050 \pm 0.005$}  
 & \multirow{2}{*}{$0.325 \pm 0.040 \pm 0.001$}  
 & \multirow{4}{*}{$0.329 \pm 0.040 \pm 0.005$} &  &  \\
$27-28$
 &   &   &  &  &  \\  \cline{3-3}
$28-29$
 &   & \multirow{2}{*}{$0.310 \pm 0.040 \pm 0.004$}  &  &  &  \\
 $29-30$
 &   &   &  &  & 
\end{tabular*}   
  \end{small}
\end{sidewaystable}

\begin{figure}[t]
  \setlength{\unitlength}{1mm}
  \centering
  \begin{picture}(150,120)
    \put( 0, 60){ 
      \includegraphics*[width=75mm,height=60mm,%
      ]{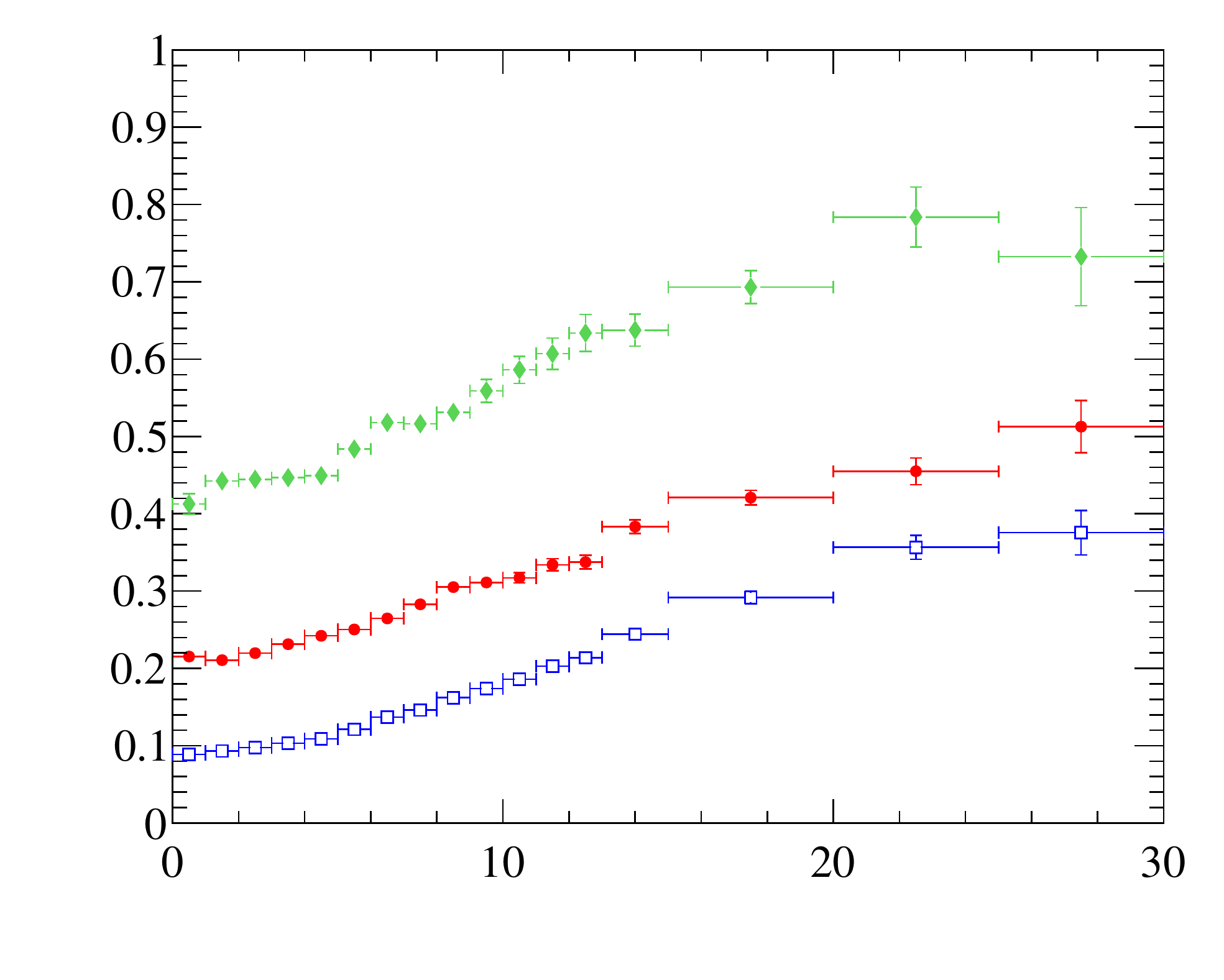}
    }
    \put(75, 60){ 
      \includegraphics*[width=75mm,height=60mm,%
      ]{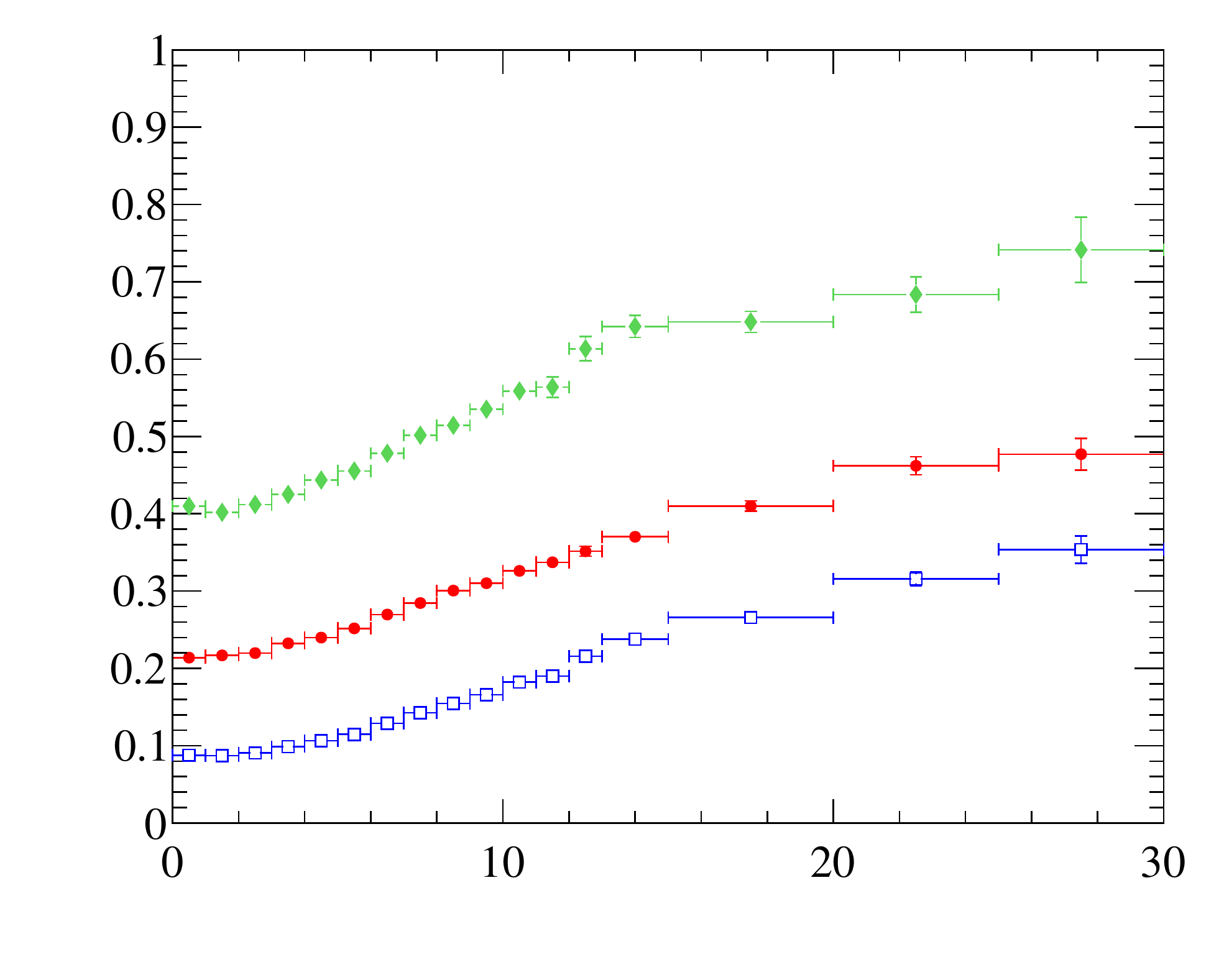}
    }
    \put( 0,  0){ 
      \includegraphics*[width=75mm,height=60mm,%
      ]{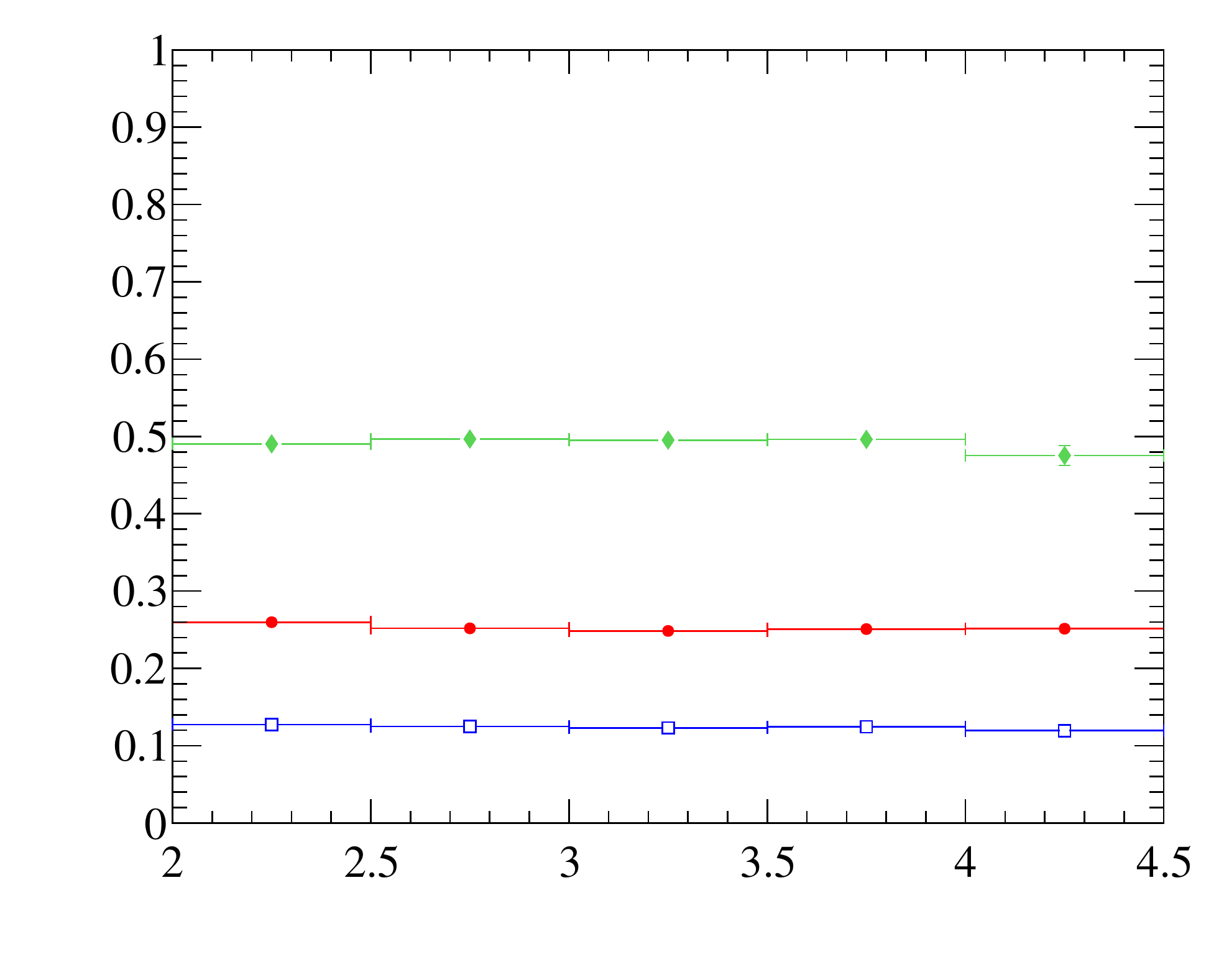}
    }
    \put(75,  0){ 
      \includegraphics*[width=75mm,height=60mm,%
      ]{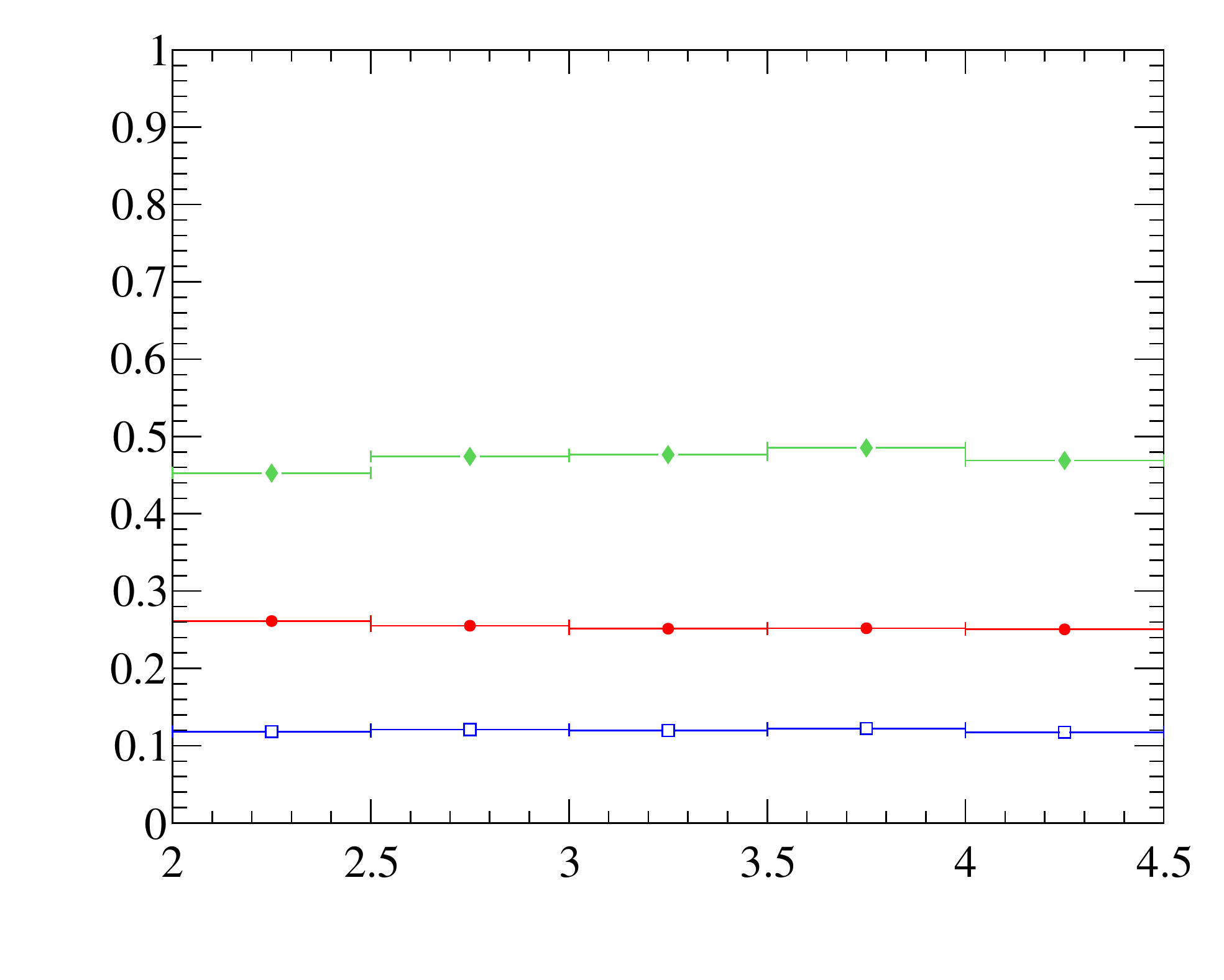}
    }
    \put(  1,103) { \begin{sideways} $\mathscr{R}_{\mathrm{i,j}}(\pt)$ \end{sideways}}
    \put( 76,103) { \begin{sideways} $\mathscr{R}_{\mathrm{i,j}}(\pt)$ \end{sideways}}
    \put(  1, 44) { \begin{sideways} $\mathscr{R}_{\mathrm{i,j}}(y)$ \end{sideways}} 
    \put( 76, 44) { \begin{sideways} $\mathscr{R}_{\mathrm{i,j}}(y)$ \end{sideways}} 
    \put( 14,108) { $\begin{array}{l} \text{LHCb}~\sqrt{s}=7\tev \\ 2.0<y<4.5   \end{array}$}
    \put( 89,108) { $\begin{array}{l} \text{LHCb}~\sqrt{s}=8\tev \\ 2.0<y<4.5   \end{array}$}
    \put( 14, 48) { $\begin{array}{l} \text{LHCb}~\sqrt{s}=7\tev \\ \pt<30\gevc \end{array}$}
    \put( 89, 48) { $\begin{array}{l} \text{LHCb}~\sqrt{s}=8\tev \\ \pt<30\gevc \end{array}$}
    \put( 40, 62) { $\pt$} \put( 58, 62){ $\left[\!\gevc\right]$}
    \put(115, 62) { $\pt$} \put(133, 62){ $\left[\!\gevc\right]$}
    \put( 40,  2) { $y$} 
    \put(115,  2) { $y$} 
    \put (130,76){\footnotesize $\begin{array}{cc}  
        {\color{red}\bullet}                       & \mathscr{R}_{2,1} \\
        {\color{blue}\square}                      & \mathscr{R}_{3,1} \\
        {\color[rgb]{0.33,0.83,0.33}\blacklozenge} & \mathscr{R}_{3,2} \end{array}$}
    \put ( 55,76){\footnotesize $\begin{array}{cc}  
        {\color{red}\bullet}                       & \mathscr{R}_{2,1} \\
        {\color{blue}\square}                      & \mathscr{R}_{3,1} \\
        {\color[rgb]{0.33,0.83,0.33}\blacklozenge} & \mathscr{R}_{3,2} \end{array}$}
    \put (130,48){\footnotesize $\begin{array}{cc}  
        {\color{red}\bullet}                       & \mathscr{R}_{2,1} \\
        {\color{blue}\square}                      & \mathscr{R}_{3,1} \\
        {\color[rgb]{0.33,0.83,0.33}\blacklozenge} & \mathscr{R}_{3,2} \end{array}$}
    \put ( 55,48){\footnotesize $\begin{array}{cc}  
        {\color{red}\bullet}                       & \mathscr{R}_{2,1} \\
        {\color{blue}\square}                      & \mathscr{R}_{3,1} \\
        {\color[rgb]{0.33,0.83,0.33}\blacklozenge} & \mathscr{R}_{3,2} \end{array}$}
  \end{picture}
  \caption { \small
    The production ratios 
    (red solid circles)~$\mathscr{R}_{2,1}$,
    (blue open squares)~$\mathscr{R}_{3,1}$ and     
    (green solid diamonds)~$\mathscr{R}_{3,2}$
    for (left)~$\sqrt{s}=7\,\mathrm{TeV}$ and 
    (right)~$\sqrt{s}=8\,\mathrm{TeV}$~data, 
    integrated over the~(top)~$2.0<y<4.5$~region
    and (bottom)~$\pt<30\gevc$~region.
  }
  \label{fig:results3}
\end{figure}

The~ratios  $\mathscr{R}_{\mathrm{i,j}}$  at 
\mbox{$\sqs=7$}~and~$8\tev$ are reported in Fig.~\ref{fig:results2} 
and Tables~\ref{tab:R21at7TeV},
\ref{tab:R31at7TeV}, 
\ref{tab:R21at8TeV}~and~\ref{tab:R31at8TeV}
as a~function of $\pt$ for different 
rapidity bins.
The~same ratios 
as a~function of \pt integrated over rapidity,  
and as a~function of $y$ integrated over \pt,
are shown in~Fig.~\ref{fig:results3}.
The~ratios $\mathscr{R}_{\mathrm{i,j}}$~show
little dependence on rapidity
and increase as a~function of~$\pt$, 
in agreement with previous observations 
by LHCb~\cite{LHCb-PAPER-2011-036,LHCb-PAPER-2013-016},
ATLAS~\cite{AtlasUpsilon} and CMS~\cite{CmsUpsilon} 
at~\mbox{$\sqs=7\tev$}.
The~ratios  of integrated cross-sections $\mathscr{R}_{\mathrm{i,j}}$ 
at $\sqs=7$ and $8\tev$ are reported in Table~\ref{tab:rin_int},
for the~full and the~reduced \pt~kinematic regions.
All~ratios $\mathscr{R}_{\mathrm{i,j}}$ agree with 
previous LHCb measurements.
The~ratio $\mathscr{R}_{\mathrm{2,1}}$~agrees with 
the~estimates 
of $0.27$~from~Refs.
~\cite{Kisslinger:2009pw,
  *Kisslinger:2012tu,
  *Kisslinger:2012np,Kisslinger:2013mev,*Kisslinger:2014zga},
while 
$\mathscr{R}_{\mathrm{3,1}}$~significantly exceeds
the~expected value of~$0.04$~\cite{Kisslinger:2009pw,
  *Kisslinger:2012tu,
  *Kisslinger:2012np,Kisslinger:2013mev,*Kisslinger:2014zga}
but agrees with the~range~\mbox{$0.14-0.22$}, 
expected for the~hypothesis 
of a~large admixture of  
a~hybrid quarkonium state in the~\YthreeS~meson state~\cite{Kisslinger:2009pw,
  *Kisslinger:2012tu,
  *Kisslinger:2012np}.

\begin{table}[t]
  \centering
  \caption{ \small 
    The ratios $\mathscr{R}_{\mathrm{i,j}}$
    in the~full kinematic range~\mbox{$\pt<30\gevc$}
    and in the reduced range~\mbox{$\pt<15\gevc$}
    for~\mbox{$2.0< y<4.5$}.
    The first uncertainties are statistical 
    and the~second systematic. 
  } \label{tab:rin_int}
  \vspace*{3mm}
  \begin{tabular*}{0.70\textwidth}{@{\hspace{2mm}}c@{\extracolsep{\fill}}cc@{\hspace{2mm}}}
    & $\sqs=7\tev$
    & $\sqs=8\tev$
    \\[1mm]
    \hline 
    \\[-2mm]    
    & \multicolumn{2}{c}{$\pt<30\gevc$}
    \\[1mm]
    \hline
    \\[-2mm]
    $\mathscr{R}_{\mathrm{2,1}}$
    & $0.253 \pm 0.001 \pm 0.004$
    & $0.255 \pm 0.001 \pm 0.004$
    \\
    $\mathscr{R}_{\mathrm{3,1}}$
    & $0.125 \pm 0.001 \pm 0.002$
    & $0.120 \pm 0.000 \pm 0.002$
    \\
    $\mathscr{R}_{\mathrm{3,2}}$
    & $0.493 \pm 0.003 \pm 0.007$
    & $0.470 \pm 0.002 \pm 0.007$
    \\[1mm]
    \cline {2-3}
    \\[-2mm]    
    & \multicolumn{2}{c}{$\pt<15\gevc$}
    \\[1mm]
    \cline {2-3}
    \\[-2mm]
    $\mathscr{R}_{\mathrm{2,1}}$
    & $0.249 \pm 0.001 \pm 0.004$
    & $0.251 \pm 0.001 \pm 0.004$
    \\
    $\mathscr{R}_{\mathrm{3,1}}$
    & $0.121 \pm 0.001 \pm 0.002$
    & $0.116 \pm 0.000 \pm 0.002$
    \\
    $\mathscr{R}_{\mathrm{3,2}}$
    & $0.485 \pm 0.003 \pm 0.007$
    & $0.463 \pm 0.002 \pm 0.007$
 \end{tabular*}   
\end{table}

\section{Summary} 
\label{sec:summary}

The~forward  production of \ups~mesons is studied
in $\proton\proton$~collisions at
centre-of-mass energies of~$7$ and~$8\tev$
using~data samples corresponding to integrated luminosities 
of~\mbox{$1\invfb$}~and  \mbox{$2\invfb$}
respectively, collected with the~LHCb detector. 
The~double differential production cross\nobreakdash-sections  
are measured as a~function of meson transverse momenta 
and rapidity for 
the~range $\pt<30\gevc$, $2.0<y<4.5$.
The~measured increase in the~production cross\nobreakdash-sections
of \ups~mesons between $\sqs=8$ and $7\tev$ significantly 
exceeds theory expectations and confirms the~previous 
LHCb observations~\cite{LHCb-PAPER-2011-036,LHCb-PAPER-2013-016}.
For the~region $\pt<15\gevc$ the~results agree with 
the~previous measurements~\cite{LHCb-PAPER-2011-036,LHCb-PAPER-2013-016},
and supersede them.


\section*{Acknowledgements}

\noindent
We thank 
K.-T.~Chao,
H.~Han
and 
H.-S.~Shao for providing the~theory 
predictions for our measurements.
We also would like to thank  
S.~P.~Baranov,
L.~S.~Kisslinger,
J.-P.~Lansberg, 
A.~K.~Likhoded
and 
A.~V.~Luchinsky  
for interesting and stimulating 
discussions on quarkonia production.
We express our gratitude to our colleagues in the~CERN
accelerator departments for the~excellent performance of the~LHC. 
We~thank the technical and administrative staff at the~LHCb
institutes. 
We~acknowledge support from CERN and from the~national
agencies: CAPES, CNPq, FAPERJ and FINEP\,(Brazil); 
NSFC\,(China);
CNRS/IN2P3\,(France); 
BMBF, DFG, HGF and MPG\,(Germany); 
INFN\,(Italy); 
FOM and NWO\,(The Netherlands); 
MNiSW and NCN\,(Poland); 
MEN/IFA\,(Romania); 
MinES and FANO\,(Russia); 
MinECo\,(Spain); 
SNSF and SER\,(Switzerland); 
NASU\,(Ukraine); 
STFC\,(United Kingdom); 
NSF\,(USA).
The~Tier1 computing centres are supported by 
IN2P3\,(France), 
KIT and BMBF\,(Germany), 
INFN\,(Italy), 
NWO and SURF\,(The~Netherlands), 
PIC\,(Spain), 
GridPP\,(United Kingdom).
We~are indebted to the~communities behind the~multiple open 
source software packages on which we depend. 
We~are also thankful for the~computing resources and 
the~access to software R\&D tools provided 
by Yandex~LLC\,(Russia).
Individual groups or members have received support from 
EPLANET, Marie Sk\l{}odowska\nobreakdash-Curie Actions and 
ERC\,(European Union), 
Conseil g\'{e}n\'{e}ral de Haute\nobreakdash-Savoie, 
Labex ENIGMASS and OCEVU, 
R\'{e}gion Auvergne\,(France), 
RFBR\,(Russia), XuntaGal and GENCAT\,(Spain), 
Royal Society and Royal
Commission for the Exhibition of 1851\,(United Kingdom).



\clearpage
\addcontentsline{toc}{section}{References}
\setboolean{inbibliography}{true}
\bibliographystyle{LHCb}
\bibliography{local,main,LHCb-PAPER,LHCb-CONF,LHCb-DP,LHCb-TDR}

\newpage

\centerline{\large\bf LHCb collaboration}
\begin{flushleft}
\small
R.~Aaij$^{38}$, 
B.~Adeva$^{37}$, 
M.~Adinolfi$^{46}$, 
A.~Affolder$^{52}$, 
Z.~Ajaltouni$^{5}$, 
S.~Akar$^{6}$, 
J.~Albrecht$^{9}$, 
F.~Alessio$^{38}$, 
M.~Alexander$^{51}$, 
S.~Ali$^{41}$
% ...

{\footnotesize \it
% ...
$ ^{5}$Clermont Universit\'{e}, Universit\'{e} Blaise Pascal, CNRS/IN2P3, LPC, Clermont-Ferrand, France\\
$ ^{6}$CPPM, Aix-Marseille Universit\'{e}, CNRS/IN2P3, Marseille, France\\
% ...
$ ^{9}$Fakult\"{a}t Physik, Technische Universit\"{a}t Dortmund, Dortmund, Germany\\
% ...
$ ^{37}$Universidad de Santiago de Compostela, Santiago de Compostela, Spain\\
$ ^{38}$European Organization for Nuclear Research (CERN), Geneva, Switzerland\\
% ...
$ ^{41}$Nikhef National Institute for Subatomic Physics, Amsterdam, The Netherlands\\
% ...
$ ^{46}$H.H. Wills Physics Laboratory, University of Bristol, Bristol, United Kingdom\\
% ...
$ ^{51}$School of Physics and Astronomy, University of Glasgow, Glasgow, United Kingdom\\
$ ^{52}$Oliver Lodge Laboratory, University of Liverpool, Liverpool, United Kingdom\\
% ...
}
\end{flushleft}

\end{document}